\documentclass[authoryear,review]{elsarticle}
\usepackage{appendix}
\usepackage{setspace}
\usepackage[font=small,  margin=2cm]{caption}
\usepackage[tbtags]{amsmath}
\usepackage{amsthm,amssymb,amsfonts}
\usepackage{natbib}
\usepackage{multirow}
\usepackage{lscape}
\journal{Journal of Econometrics \ Templates}
\usepackage{subfigure}
\usepackage{makecell}
\usepackage{exscale}
\usepackage{booktabs}
\usepackage{array}
\usepackage{fullpage}
\usepackage{url}
\usepackage{algorithm}
\usepackage{algorithmic}
\usepackage{bm}
\usepackage{smile}
\usepackage{mathtools}
\usepackage{wrapfig}
\usepackage{lipsum}
\usepackage{mathrsfs}
\usepackage{relsize}
\usepackage{dsfont}
\usepackage{multirow}
\usepackage{apalike}
\usepackage{chngcntr}
\usepackage[usenames,dvipsnames,svgnames,table]{xcolor}
\usepackage[colorlinks=true,
linkcolor=blue,
urlcolor=blue,
citecolor=blue]{hyperref}

\numberwithin{equation}{section}
\numberwithin{theorem}{section}
\numberwithin{corollary}{section}
\counterwithout{asmp}{section}

\numberwithin{definition}{section}

\begin{document}
	
\begin{frontmatter}
\title{ Subgroup Identification with Latent Factor Structure}	

\author[myfirstaddress]{Yong He
}\address[myfirstaddress]{Institute for Financial Studies, Shandong University, Jinan, 250100, China}

\author[mysecondaddress]{Dong Liu}\address[mysecondaddress]{Shanghai University of Finance and Economics, Shanghai, 200433, China}

\author[mythirdaddress]{Fuxin Wang}\address[mythirdaddress]{University of Wisconsin-Madison, Madison, WI 53706, USA}	

\author[myfourthaddress]{Mingjuan Zhang
}\address[myfourthaddress]{School of Statistics and Mathematics, Shanghai Lixin University of Accounting and Finance, Shanghai 201209, China}	

\author[myfiveaddress]{Wen-Xin Zhou}\address[myfiveaddress]{ Department of Information and Decision Sciences, University of Illinois at Chicago, Chicago, IL 60607, USA}

\begin{abstract}
Subgroup analysis has garnered increasing attention for its ability to identify meaningful subgroups within heterogeneous populations, thereby enhancing predictive power. However, in many fields such as social science and biology, covariates are often highly correlated due to common factors. This correlation poses significant challenges for subgroup identification, an issue that is often overlooked in existing literature. In this paper, we aim to address this gap in the ``diverging dimension"
regime by proposing a center-augmented subgroup identification method within the Factor Augmented (sparse) Linear Model framework. This method bridges dimension reduction and sparse regression. Our proposed approach is adaptable to the high cross-sectional dependence among covariates and offers computational advantages with a complexity of $O(nK)$, compared to the $O(n^2)$ complexity of the conventional pairwise fusion penalty method in the literature, where $n$ is the sample size and $K$ is the number of subgroups.
We also investigate the asymptotic properties of the oracle estimators under conditions on the minimal distance between group centroids. To implement the proposed approach, we introduce a Difference of Convex functions-based Alternating Direction Method of Multipliers (DC-ADMM) algorithm and demonstrate its convergence to a local minimizer in a finite number of steps. We illustrate the superiority of the proposed method through extensive numerical experiments and a real macroeconomic data example. An \texttt{R} package, \texttt{SILFS}, implementing the method is also available on CRAN\footnote{\url{https://cran.r-project.org/web/packages/SILFS/index.html}}.
\end{abstract}
\end{frontmatter}

\section{Introduction}
With advancements in data collection and information technology, the dimensionality of data has exponentially increased across various research and application fields. Concurrently, it is believed that meaningful subgroups exist within heterogeneous populations in many real datasets. In macroeconomics, \cite{phillips2007transition}  analyzes the convergence in the cost of living indices among 19 U.S. metropolitan cities, demonstrating that cities within different groups behave quite differently. In portfolio allocations, stocks within the same industries exhibit similar characteristics \citep{Livingston1977INDUSTRYMO}. In precision medicine, patients may belong to various latent groups, and those within the same group can be considered together for making treatment decisions; see, for example,  \cite{ma2017concave} and \cite{Chen2021Simultaneous}. \cite{wang2021identifying} also points out that geographic adjacency is a natural criterion for grouping when analyzing international trade and economic geography datasets. These examples underline the importance of identifying latent group structures prior to conducting statistical inference, as this approach significantly enhances statistical efficiency.

Statisticians and econometricians typically characterize group structures using group-specific parameters in statistical models. This topic is closely related to concepts such as integrative analysis, transfer learning, and multi-task representation learning.
For integrative analysis, \cite{gertheiss2012regularization} and \cite{ollier2017regression} assume that the coefficients of the $k$-th group, denoted as $\bbeta_k$, can be decomposed into $\bomega$ and $\bzeta_k$, where $\bomega$ is a common parameter shared by all groups, and $\bzeta_k$ is the group-specific parameter representing the heterogeneity of group structures.
Similarly, in the context of transfer learning or multi-task representation learning, \cite{li2022transfer} and \cite{tian2023learning} also use group-specific parameters to characterize the heterogeneity of group structures.  However, it is worth noting that the aforementioned works presume the group membership is known a priori, which is not the case in many real applications.
In the context of subgroup analysis, \cite{ma2017concave}, \cite{zhang2019robust} and \cite{he2022center} employ group-specific regression coefficients while assuming unknown group membership. They aim to cluster the samples and conduct statistical inference simultaneously. In this paper, we also adopt the unknown group membership framework and consider the following linear model with group-specific intercepts:
\begin{equation}
Y_{i}=\alpha_{i}+\bm{x}_{i}^{\top}\bm{\beta}+\epsilon_{i},\ i=1,2,\dots,n,\ \text{or equivalently,}\ \ \bm{Y} = \bm{\alpha}+\bm{X}\bm{\beta}+\bm{\epsilon},
\label{equ:Subgroup1}
\end{equation}
where $\bm{Y}=(Y_{1},Y_{2},\dots,Y_{n})^{\top}$ is the response vector, $\bm{X}=(\bm{x}_{1},\bm{x}_{2},\dots,\bm{x}_{n})^{\top}$ denotes the design matrix, $\bepsilon=(\epsilon_{1},\epsilon_{2},\dots,\epsilon_{n})^{\top}$ is the noise vector, and the unknown vector of regression coefficients $\bbeta\in\RR^p$ is sparse.
The group structure is characterized by the intercept parameters $\bm{\alpha} = (\alpha_{1}, \alpha_{2}, \dots, \alpha_{n})^{\top}$, meaning that individuals within the same group share the same group-specific intercept parameter. Specifically, let the number of subgroups be $K$, and define $\mathcal{G} = \cbr{\mathcal{G}_{k}}_{k=1}^K$ as a partition of $\cbr{1,2,\cdots,n}$, satisfying $\mathcal{G}_{k} \cap \mathcal{G}_{k^{\prime}} = \emptyset$ for $k \neq k^{\prime}$. It is assumed that $\alpha_i=\gamma_k$ for $i$ in $\cG_k$, and $\bgamma=(\gamma_1,\gamma_2,\cdots,\gamma_K)^{\top}$ represents the centroid of the group-specific parameters. It is worth noting that $\alpha_i$ may reflect heterogeneity driven by latent variables $\bz_i$, that is, $\alpha_i=\alpha+\bz_i^{\top}\btheta$. For example, in microeconomics, the impact of education on wages is a widely studied topic. However,  individuals may have different backgrounds, such as family background and work experience, which can be modeled as $\bz_i$. See, for example, \cite{f82aa4c5-5616-3bf8-a501-20c3a9982c3d}, \cite{https://doi.org/10.1111/obes.12237} and \cite{connolly2006differences}.  More generally, the coefficients for $\bz_i$ can be subject-specific, as the same family background and work experience may have different impacts for different individuals. In this case, the model would become
\begin{equation}\label{equ:Subgroup11}
Y_{i}=\alpha+\bz_i^{\top}\btheta_i+\bm{x}_{i}^{\top}\bm{\beta}+\epsilon_{i},\ i=1,2,\dots,n,
\end{equation}
Throughout this article, we focus on the model (\ref{equ:Subgroup1}), assuming that the heterogeneity arises from unobserved covariates $\bz_i$. In fact, model (\ref{equ:Subgroup11}) is similar to model (\ref{equ:Subgroup1}), and our proposed algorithms and the associated theoretical properties for model (\ref{equ:Subgroup1}) can be extended to model (\ref{equ:Subgroup11}) with slight modifications.

\begin{figure}[ht]
	\hspace{-8.5mm}
\includegraphics[width=18cm]{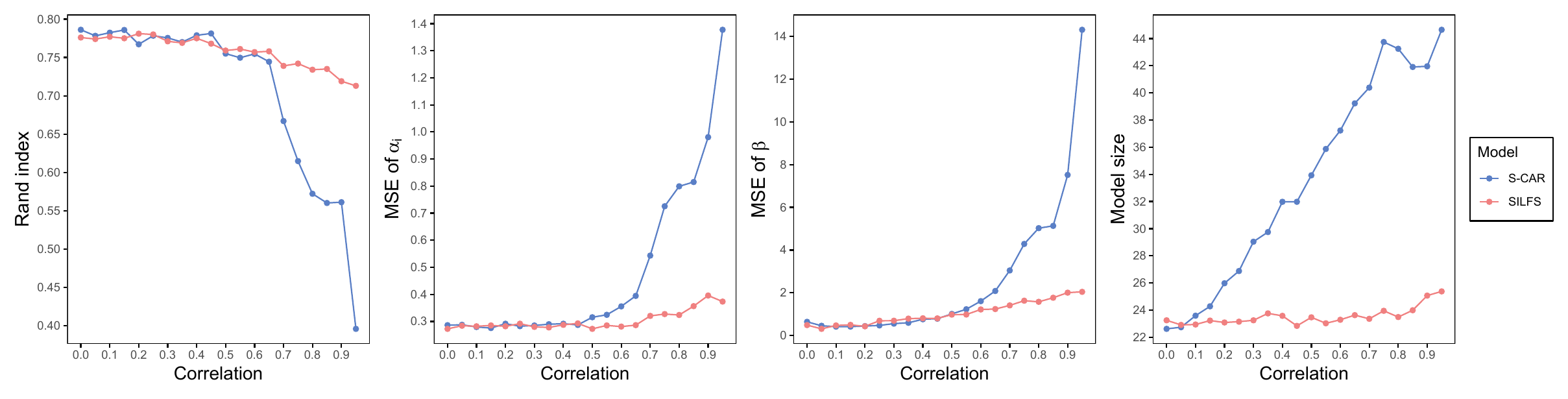}
        \hspace{-5.2mm}
	\caption{Comparison of line charts of estimation results for S-CAR and SILFS models based on 100 replications. The model size is defined as the number of non-zero components in $\hat{\bbeta}$.}
	\label{fig:1}
\end{figure}

In datasets from social science and biology, high-dimensional covariates are often highly correlated, possibly due to the existence of
common factors. See, for example, \cite{stock2002forecasting}, \cite{bai2002determining}, \cite{fan2020factor}, \cite{vatcheva2016multicollinearity}, \cite{porcu2019mendelian} and \cite{he2022statistical}. The high cross-sectional dependence among covariates can lead to unsatisfactory results in various statistical inference problems, including the subgroup identification problem discussed in this paper.  In the following, we present a toy example to
illustrate the impacts of high collinearity on the subgroup identification problem.
For model (\ref{equ:Subgroup1}), each group-specific $\alpha_i$ takes value of either $1$ or $-1$ with equal probability. We generate the residuals $\bm{\epsilon}$ from $\cN(0,0.01\bm{I}_{p})$, and let $n=p=100$. Set $\bm{\beta}=(\beta_1,\dots,\beta_{10},\bm{0}^{\top}_{p-10})^{\top}$, where the nonzero coefficients $\beta_i$ $(1\le i\le 10)$ are drawn from $\cU(2,5)$. Let $\bm{X}=(\bm{x}_1,\dots,\bm{x}_n)^\top$ be $n$ independent copies from multivariate normal distribution $\mathcal{N}(\bm{0},\bm{\Xi})$, where $\bm{\Xi}\in \mathbb{R}^{p\times p}$ is a covariance matrix with off-diagonal elements being $\rho$ and diagonal elements being $1$. In this toy example, the parameter  $\rho$ is varied over the interval [0, 0.95] with a step size of 0.05 to illustrate different strengths of cross-sectional dependency. {For each given $\rho$, we consider the  Center-augmented Regularizer (CAR) method proposed by \cite{he2022center}, referred to as Standard CAR (S-CAR). We report the averages of model size, estimation errors (MSE), and Rand Index over 100 replications in Figure \ref{fig:1}. As illustrated in the Figure \ref{fig:1} (the blue lines), both the clustering performance and the estimation accuracy of S-CAR deteriorate as the cross-sectional dependency strength, $\rho$, increases. The CAR method shows limited power in detecting group structures under high cross-sectional dependency, especially when $\rho>0.7$.}

To the best of our knowledge, there exists no literature addressing subgroup analysis under conditions of high collinearity among covariates, due to both computational and statistical challenges. In this work, we propose to address the high cross-sectional dependency among covariates using a factor model approach, inspired by the Factor Augmented (sparse linear) Regression Model  (abbreviated as FARM) introducted in \cite{fan2020factor} and \cite{fan2023latent}.
Factor model is widely used as an effective tool for dimension reduction and can capture various levels of cross-sectional dependence. It is well recogonized that many macroeconomic and financial datasets exhibit high  cross-sectional dependence. For instance,  \cite{fama1992cross} and \cite{fama2015five} provide evidence that average stock returns, characterized by high cross-sectional dependence, are driven by Fama-French factors. Additionally, \cite{mccracken2016fred} confirm that the well-studied microeconomic dataset FRED-MD also exhibits latent factor structures. \cite{johnstone2018pca} investigate numerous datasets showing spiked structures in covariance matrices across various fields, including microarrays, satellite images, medical shapes, climate data, and signal detection. Therefore, it is reasonable to mitigate high collinearity and achieve dimension reduction simultaneously by assuming a latent factor model. In this paper, we adopt the FARM framework to address the challenges posed by high collinearity and high-dimensionality in the subgroup analysis problem. More specifically, we assume the vector of regression coefficients $\bbeta$ is sparse with a support set $\cS$, and the observed $\bx_i$'s satisfy  the following factor structure:
\begin{equation}\label{equ:Fac}
	\bx_{i}=\bB\bbf_i+\bu_i, \ i=1,2,\dots,n,\ \text{or in matrix form, }\bX=\bF\bB^{\top}+\bU,
\end{equation}
where $\bbf_i\in\RR^r$, $\bm{F}=(\bm{f}_{1},\bm{f}_{2},\dots,\bm{f}_{n})^{\top}$ is the $n\times r$ factor score matrix, $\bm{B}$ is the $p\times r$ factor loading matrix, and $\bm{U}=(\bm{u}_{1},\bm{u}_{2},\dots,\bm{u}_{n})^{\top}$ is the matrix of idiosyncratic errors of dimension $n\times p$. To illustrate the advantage of the FARM framework in the presence of high collinearity, we substitute (\ref{equ:Fac}) into (\ref{equ:Subgroup1}) to form the following sparse regression model
\begin{equation}\label{equ:Subgroup2}
	Y_{i}=\alpha_{i}+\bbf^{\top}_i\btheta+\bu^{\top}_i\bbeta+\epsilon_i,\ i=1,2,\cdots,n, ~~\text{or in matrix form, } \bY=\balpha+\bF\btheta+\bU\bbeta+\bepsilon,
\end{equation}
where $\btheta=\bB^{\top}\bbeta$ and $\bF \in\RR^{n\times r}$ extract essential information from $\bX$, and $\bU\in\RR^{n\times p}$ contributes additional sparse information to the response. The identifiability condition on $\bB$ and  $\bF$  will be discussed in Section \ref{sec:2.1}. This construction augments the highly cross-dependent covariates $\bX\in\RR^{n\times p}$ to weakly dependent covariates ${\bF, \bU} \in \RR^{n\times (p+r)}$. Subsequently, we adopt the FARM framework to capture the cross-sectional dependence of covariates in  subgroup analysis. Our goal is to identify latent groups characterized by the intercept parameters $\bm{\alpha} = (\alpha_{1}, \alpha_{2}, \dots, \alpha_{n})^{\top}$ and achieve variable selection simultaneously under model (\ref{equ:Subgroup2}). Hereafter, we treat the group number $K$ and factor number $r$ as fixed, while allowing the dimension $p$ to increase with the sample size $n$.

\subsection{Literature Review}

A closely related line of research focuses on sparse linear regression and variable selection. Over the last two decades, numerous regularization-based methods have been proposed, including well-known approaches such as LASSO \citep{tibshirani1996regression}, SCAD \citep{fan2001variable}, elastic net \citep{zou2005regularization} and adaptive LASSO \citep{zou2006adaptive}. The effectiveness of regularized regression methods often relies on constraints imposed on the covariance matrix of the covariates and limitations on collinearity levels. However, these constraints may be inappropriate in real macroeconomic or financial applications.


Another closely related line of research focuses on large-dimensional approximate factor models. \cite{bai2003inferential} explored the theory of large-dimensional approximate factor models, and theoretical guarantees for determining the number of factors can be found in works such as \cite{bai2002determining} and \cite{ahn2013eigenvalue}. Robust factor analysis has also gained increasing attention in recent years, as evidenced by studies like \cite{chen2021quantile}, \cite{he2022large} and \cite{he2023huber}. Furthermore, \cite{stock2002forecasting} demonstrates the consistency of Ordinary Least Squares (OLS) estimates in factor-augmented regressions, treating latent factors as predictors. \cite{ando2011quantile} extends this approach to quantile factor-augmented regressions, with further extensions discussed in \cite{wang2019quantile} and \cite{fan2021augmented}. In high-dimensional regression settings, latent factors often cannot fully explain the response, particularly when their effect is weak. \cite{fan2020factor} proposes the Factor Augmented (sparse linear) Regression Model, which assumes sparsity in the coefficient vector and integrates both latent factors and idiosyncratic components into the linear regression framework. \cite{fan2023latent} introduces hypothesis tests for assessing the adequacy of latent factor regression models. Additionally, \cite{tu2024consistent} present a new class of information criteria aimed at achieving consistent factor and covariate selection jointly in factor-augmented regression.


The last closely related line of research focuses on subgroup analysis. Typical approaches for subgroup identification can generally be categorized into two distinct types.
The first type is the finite mixture model (FMM), which employs model-based clustering techniques. Representative works include but are not limited to \cite{banfield1993model}, \cite{hastie1996discriminant}, and \cite{wei2013latent} for mixture models under Gaussian distributions; \cite{shen2010grouping} for logistic-normal regression; \cite{yao2014robust} for the Student 
$t$ distribution; and \cite{song2014robust} for the Laplace distribution. For finite mixture models, \cite{khalili2007variable} propose penalized likelihood methods for variable selection and establish their consistency. In high-dimensional cases, references include \cite{zhang2020estimation} and \cite{wang2023statistical}. Although the EM algorithm is effective and powerful in finding maximum likelihood estimators with latent group variables in FMM, it suffers from significant computational burdens in high dimensions. The second type is regularization methods, which use fusion penalties to enforce sparsity in pairwise differences among group-specific parameters to achieve clustering. Examples include \cite{hocking2011clusterpath}, \cite{pan2013cluster}, \cite{chi2015splitting}, and \cite{wu2016new}. For subgroup analysis of linear models, \cite{ma2017concave} propose a concave Pairwise Fusion Penalty (PFP) capable of recovering latent groups without prior information on the number of groups. Many subsequent studies adopt this penalty for its favorable statistical properties, including extensions to quantile regressions \citep{zhang2019robust} and functional regressions \citep{zhang2022subgroup} for group pursuits. From a computational perspective,  the PFP, which pairs $n$ samples, has a computational complexity of $O(n^2)$. To mitigate this, \cite{wang2024homogeneity}  introduce a threshold PFP method to reduce computational time and enhance group detection accuracy. Moreover, \cite{he2022center} propose a more efficient Center-augmented Regularization (CAR) method, reducing computational complexity to $O(nK)$ by incorporating group centroids as parameters. However, due to the concavity of PFP and CAR, their performance is sensitive to the selection of initial values. Therefore, choosing appropriate initial values remains a challenging task.

\subsection{Contributions and Structure of the Paper}

In this work, we propose a clustering method within the FARM framework, coined Subgroup Identification with Latent Factor Structures (SILFS). SILFS demonstrates superior performance compared to existing methods in two critical aspects: identifying group memberships and simultaneously recovering the support of regression coefficients in the presence of high collinearity among covariates. Returning to Figure \ref{fig:1} (the red lines),  we observe that SILFS exhibits robust performance across varying levles of cross-sectional dependence parameterized by $\rho$. Notably, for $\rho>0.5$, SILFS significantly outperforms the S-CAR method, highlighting the necessity of accounting for high cross-sectional dependency among covariates in subgroup identification problems and the novelty of our approach. Additionally, the SILFS method draws inspiration from Center-augmented Regularization, offering computational advantages over pairwise fusion penalization methods. Theoretically, we establish properties of the local minima for the optimization problem under an $\ell_1$-type CAR penalty. We provide, to the best of our knowledge, the first rigorous theoretical guarantee that CAR can effectively identify subgroups under the $\ell_1$ distance. To achieve these local minima, we propose a DC-ADMM algorithm that fully exploits the mathematical structure of the CAR penalty. In implementing the DC decomposition of the CAR penalty under inequality constraints, we convert these inequalities into equivalent equalities by introducing slack variables and incorporating indicator functions into the objective function. We derive explicit iteration formulas for the equivalent optimization problem and demonstrate that our algorithm converges to a Karush-Kuhn-Tucker (KKT) point of the objective function after a finite number of iterations.

The remainder of this paper is organized as follows. Section \ref{sec:2} introduces the optimization problem of SILFS and presents a two-step algorithm to solve it. In Section \ref{sec:3}, we establish the consistency of the oracle estimators and demonstrate that they serve as local minimizers of the proposed optimization problem with probability approaching one. Section \ref{sec:4} presents extensive simulation studies that illustrate the superior performance of the SILFS method in both clustering and variable selection. Furthermore, computational speed comparisons with various methods highlight the computational advantages of our proposed approach. In Section \ref{sec:5}, we apply the SILFS method to a large panel of China's commodity trade data involving all countries worldwide from 2019. We conclude with a discussion on potential future research directions and summarize our findings in Section \ref{sec:6}. The supplementary material includes proofs for all theorems and additional details on algorithm implementations.

To conclude this section, we introduce the notations that will be used throughout the remainder of this paper. For a matrix $\bM=(M_{ij})_{i\in p,j\in q}$, denote $\left\|\bm{M}\right\|_{\mathbb{F}}$ and $\left\|\bm{M}\right\|_{2}$ as its Frobenius norm and operator norm, respectively. Write $\left\|\bm{M}\right\|_{1}=\mathop{\text{max}}_{j\in\{1,\dots,q\}}\sum_{i=1}^{p}\left|M_{ij}\right|$, $\left\|\bm{M}\right\|_{\infty}=\mathop{\text{max}}_{i\in\{1,\dots,p\}}\sum_{j=1}^{q}\left|M_{ij}\right|$ and $\left\|\bm{M}\right\|_{\text{max}}=\max_{i,j}|M_{ij}|$. Let $\cS_1$ and $\cS_2$ be two subsets of $\cbr{1,2,\cdots,p}$. We denote $\bM_{\cS_1 \cS_2}$ as the submatrix of $\bM$ with row indices in $\cS_1$ and column indices in $\cS_2$. The matrix $\bm{M}_{\mathcal{S}}$ denotes the submatrix of $\mathbf{M}$ consisting of columns whose indices belong to the set $\mathcal{S}$. Here, $|\mathcal{S}_1|$ denotes the cardinality of $\mathcal{S}_1$. Moreover, $\bI_n$ represents the $n$-dimensional identity matrix, and $\bm{0}_{(p\times q)}$ denotes the zero-matrix of size of $p\times q$. $\bm{1}_p$ and $\bm{0}_p$ denote the $p$-dimensional vectors with all elements being $1$ and $0$, respectively. The function $\text{sgn}(a)$ is defined such that $\text{sgn}(a)=1$ if $a>0$, $\text{sgn}(a)=-1$ if $a<0$, and $\text{sgn}(0) = 0$. For a set of indices $\mathcal{S}$ and a function $f(\bx)$ where $\mathbf{x} \in \mathbb{R}^n$,  $\nabla _{\cS} f(\bx)$ denotes the sub-gradient of $f(\bx)$ with respect to $\bx_{\cS}$, where $\bx_{\cS}$ consists of elements of $\bx$ indexed by $\cS$. 
For two sequences of real numbers $\{a_{n}\}_{n=1}^{\infty}$ and $\{ b_{n} \}_{n=1}^{\infty}$, we use the notation $a_{n} = o(b_{n})$ or $b_{n} \gg a_{n}$ to indicate that $a_{n}/b_{n} \rightarrow 0$ as $n \rightarrow \infty$, and $a_{n} = O(b_{n})$ if there exist a positive integer $N$ and a positive constant $C$ such that $|a_{n}| \leq C|b_{n}|$ for all $n \geq N$.

\section{Methodology}\label{sec:2}

In this section, we introduce a two-step procedure to estimate the parameters $\bbeta_i$ and $\balpha_i$ in model (\ref{equ:Subgroup2}). In Section \ref{sec:2.1}, we first estimate the unobserved $\bF$ and $\bU$. Subsequently, in Section \ref{sec:2.2}, we formulate a center-augmented regularization optimization problem using plug-in estimators to achieve simultaneous clustering and variable selection. Section \ref{sec:2.3} discusses implementation details of the algorithm, including the selection of initial values, the number of subgroups $K$, and the shrinkage parameters $\lambda_{1}$ and $\lambda_{2}$.

\subsection{Estimation of Factor Model}\label{sec:2.1}

Given the presence of an unknown factor structure in model (\ref{equ:Subgroup1}), our first step is to estimate these factors from the observed predictors $\bX$. To achieve this, we employ Principal Component Analysis (PCA) to estimate the latent factors $\bm{F}$ and the factor loading matrix $\bm{B}$. Specifically, \cite{fan2013large} demonstrate that the PCA method is equivalent to the following constrained least squares:
\begin{equation*}
	\begin{aligned}
		\left(\hat{\bm{F}},\hat{\bm{B}}\right) &= \mathop{\text{argmin}}_{\bm{F}\in \mathbb{R}^{n\times r}, \bm{B}\in \mathbb{R}^{p\times r }} \|\bm{X}-\bm{F}\bm{B}^{\top}\|_{\mathbb{F}}^{2},\\
		\text{s.t.} &\qquad \frac{1}{n}\bm{F}^{\top}\bm{F}=\bm{I}_{r}, \quad \bm{B}^{\top}\bm{B} \ \text{is diagonal},
	\end{aligned}
\end{equation*}
where the normalization constraints address identifiability concerns \citep{bai2003inferential}.  Through simple algebra, the estimated factors $\widehat{\bF}/\sqrt{n}$ correspond to the eigenvectors associated with the largest $r$ eigenvalues of $\bX\bX^{\top}$, and $\widehat{\bB}=\widehat{\bF}^{\top}\bX/n$. The idiosyncratic components can be computed straightforwardly as $\widehat{\bU}=\bX-\widehat{\bF}\widehat{\bB}^{\top}$.

However, the number of latent factors $r$ is unknown in practice. There are various methods in the literature to derive consistent estimators of $r$, such as the Information Criteria (IC) proposed by \cite{bai2002determining} and the Eigenvalue-Ratio (ER) criterion proposed by \cite{lam2012factor} and \cite{ahn2013eigenvalue}. Since the estimation of $r$ is often conducted separately, we assume that the number of factors $r$ is given.

\subsection{Clustering Procedure}\label{sec:2.2}

To recovery the group membership and estimate the regression coefficients $\bbeta$ simultaneously, we plug in the PCA estimators $\widehat{\bF}$ and $\widehat{\bU}$ and formulate the following  optimization problem with a given number of subgroups $K$:
\begin{equation}\label{opt:CAFARM}
	\argmin_{\bTheta}\ Z(\bm{\Theta})=\frac{1}{2n}\big\|\bm{Y}-\bm{\alpha}-\hat{\bm{F}}\bm{\theta}-\hat{\bm{U}}\bm{\beta}\big\|_2^{2}
+\lambda_{1}g(\balpha,\bgamma)+\lambda_{2}\left\|\bm{\beta}\right\|_{1},
\end{equation}
where $\bTheta=(\balpha^{\top},\bbeta^{\top},\btheta^{\top},\bgamma^{\top})^{\top}$, $\lambda_1$ and $\lambda_2$ are tuning parameters, and $g(\balpha,\bgamma)$ is a penalty for group pursuits. The $\ell_1$ penalty, $\left\|\bm{\beta}\right\|_{1}$, encourages sparsity for variable selection.

In this paper, we choose $g(\balpha,\bgamma)$ as the CAR penalty for group pursuit. The CAR method, initially proposed by \cite{he2022center}, aims to estimate both the group-specific parameter $\balpha$ and the group centroids $\bgamma$ simultaneously. This is achieved by clustering each subject $i$ into group $k$ based on the nearest distance. Specifically, $$g(\balpha,\bgamma)=\sum_{i=1}^{n}\text{min}\{d\left(\alpha_{i},\gamma_{1}\right),\cdots,d\left(\alpha_{i},\gamma_{K}\right)\},$$
where $d(x,y)$ is a distance function between $x$ and $y$.  This approach is analogous to the $K$-means method when using $\ell_2$-type distance ($d(x, y) = (x - y)^2$) or the $K$-median method when employing $\ell_1$-type distance ($d(x, y) = |x - y|$). From an optimization perspective, CAR exhibits a computational complexity of $O(nK)$, whereas pairwise penalties used in \cite{ma2017concave} and \cite{zhang2019robust}, such as $\sum_{i < j} p_{\lambda}(|\alpha_i - \alpha_j|)$ with penalty function $p_{\lambda}(\cdot)$, require $O(n^2)$ computational complexity. Hence, CAR significantly reduces computational burdens by incorporating centroid parameters. Furthermore, as CAR is a non-convex penalty, it presents challenges for both theoretical analysis and optimization. It is worth noting that CAR does not directly offer clustering results. If $\hat{\gamma}_k$ represents the estimated group centroids, our clustering rule for the $i$-th subject is defined as $i \in \mathcal{G}_{k^*}$ if and only if $k^* = \argmin_{k} d(\hat{\alpha}_{i}, \hat{\gamma}_{k})$ for $k \le K$.

Recall that the optimization problem outlined in \eqref{opt:CAFARM} depends on the choice of the distance function. In the following, we introduce the DC-ADMM  and Cyclic Coordinate Descent (CCD) algorithm to solve the problem for $\ell_1$-type and $\ell_2$-type distances, respectively.

\subsubsection{DC-ADMM Algorithm}\label{sec:2.2.1}

In this section, we propose  the DC-ADMM algorithm to solve (\ref{opt:CAFARM}) under the $\ell_1$-type distance. The corresponding theoretical results are elaborated in Theorem \ref{thm:A1}. The optimization problem is as follows:
\begin{equation}\label{Aopt:CAFARM}
\argmin_{\bTheta}\ Z(\bm{\Theta})=\frac{1}{2n}\big\|\bm{Y}-\bm{\alpha}-\widehat{\bm{F}}\bm{\theta}-\widehat{\bm{U}}\bm{\beta}\big\|_2^{2}
+\lambda_{1}\sum_{i=1}^{n}\text{min}\{d\left(\alpha_{i},\gamma_{1}\right),\cdots,d\left(\alpha_{i},\gamma_{K}\right)\}+\lambda_{2}\left\|\bm{\beta}\right\|_{1}.
\end{equation}
This optimization problem is challenging due to the identifiability issue with the parameter $\bgamma$ and its non-convex nature. For example, when $K = 2$ and $(\tilde{\gamma}_1, \tilde{\gamma}_2)$ minimizes the loss function in \eqref{opt:CAFARM}, it is evident that $(\tilde{\gamma}_2, \tilde{\gamma}_1)$ also serves as a minimizer. To address the identifiability issue, we impose the condition that $\gamma_{1} \leq \gamma_{2} \leq \cdots \leq \gamma_{K}$, which is also crucial for the following optimization. Motivated by the works of \cite{an2005dc} and \cite{wu2016new} on Difference of Convex (DC) programming to address non-convex optimization problems, we reformulate the CAR penalty $g(\balpha,\bgamma)$ as a difference of convex functions under the identifiability condition of $\bgamma$.
Specifically, we have
$g(\balpha,\bgamma)=g_1(\balpha,\bgamma)-g_2(\balpha,\bgamma)$ with
$$
g_1(\balpha,\bgamma)=\sum_{i=1}^{n}\sum_{k=1}^{K}d(\alpha_{i},\gamma_{k}),\quad g_2(\balpha,\bgamma)=\sum_{i=1}^{n}\sum_{k=2}^{K}\text{max}\{d(\alpha_{i},\gamma_{k-1}),d(\alpha_{i},\gamma_{k})\}.
$$
We reparameterize the variables by letting $\delta_{ik}=\alpha_{i}-\gamma_{k}$ for $i=1,\cdots,n \ \text{and} \ k=1,\cdots,K$. Now, both $g_1(\bdelta)$ and $g_2(\bdelta)$ are convex functions with respect to $\bdelta_i$, where $\bdelta=(\bdelta_{1}^\top,\cdots,\bdelta_{n}^\top)^{\top}$ with $\bdelta_{i}=(\delta_{i1},\cdots,\delta_{iK})^{\top}$.

Therefore, the original problem is equivalent to
\begin{align*}
\argmin_{\bTheta}\ &Z(\bm{\Theta},\bdelta)=\frac{1}{2n}\big\|\bm{Y}-\bm{\alpha}-\widehat{\bm{F}}\bm{\theta}-\widehat{\bm{U}}\bm{\beta}\big\|_2^{2} +\lambda_{1}g_1(\bdelta)-\lambda_1g_2(\bdelta)+\lambda_{1}\left\|\bm{\beta}\right\|_{1}\\
&\text { subject to}\quad \delta_{ik}=\alpha_{i}-\gamma_{k},\ i=1,\cdots,n,\ k=1,\cdots,K,\ \text{and}\ \gamma_{1}\leq\gamma_{2}\leq\cdots\leq\gamma_{K}.
\end{align*}
We can then define a sequence of lower approximations of $g_2(\bdelta)$ as
$$g_2^{(m)}(\bdelta)=g_2(\widehat{\bm{\delta}}^{(m-1)})+(\nabla g_{2}(\widehat{\bm{\delta}}^{(m-1)}))^{\top}(\bm{\delta}-\widehat{\bm{\delta}}^{(m-1)}),$$
where $\nabla g_{2}(\bdelta)$ is the sub-gradient of $g_{2}(\bdelta)$ and $\widehat{\bm{\delta}}^{(m-1)}$ is the estimator of $\bdelta$ from the $(m-1)$-th iteration. More specifically, let $\nabla_{ik}g_{2}(\bdelta)$ be the sub-gradient of $g_{2}(\bdelta)$ with respect to $\delta_{ik}$. Under the $\ell_1$-type distance for $1<k<K$, we have
$$
\nabla_{ik}g_{2}(\bm{\delta})=\text{sgn}\big(\delta_{ik}\big)\mathbb{I}\left(\big|\delta_{ik}\big|>\big|\delta_{i(k-1)}\big|\right)+\text{sgn}\big(\delta_{ik}\big)\mathbb{I}\left(\big|\delta_{ik}\big|>\big|\delta_{i(k+1)}\big|\right),
$$
and otherwise we have
$$
\nabla_{i1}g_{2}(\bm{\delta})=\text{sgn}\big(\delta_{i1}\big)\mathbb{I}\rbr{\big|\delta_{i1}\big|>\big|\delta_{i2}\big|},\quad \nabla_{iK}g_{2}(\bm{\delta})=\text{sgn}\big(\delta_{iK}\big)\mathbb{I}\rbr{\big|\delta_{iK}\big|>\big|\delta_{i(K-1)}\big|}. $$

The DC programming approach encourages us to optimize an upper approximation in the $(m+1)$-th iteration, which is
\begin{equation}
	\begin{aligned}\label{Aopt:UpperA1}
		\argmin_{\bm{\Theta},\bm{\delta}}&\ Z^{(m+1)}(\bm{\Theta},\bdelta)=\frac{1}{2n}\big\|\bm{Y}-\bm{\alpha}-\widehat{\bm{F}}\bm{\theta}-\widehat{\bm{U}}\bm{\beta}\big\|_2^{2} + \lambda_{1}\sum_{i=1}^{n} \sum_{k=1}^{K}d(\delta_{ik})-\lambda_1(\nabla g_{2}(\widehat{\bm{\delta}}^{(m)}))^{\top}(\bm{\delta}-\widehat{\bm{\delta}}^{(m)})+\lambda_{2}\|\bm{\beta}\|_{1}\\
		&\text{subject to}\quad \delta_{ik}=\alpha_{i}-\gamma_{k}\quad i=1,\cdots,n, \   k=1,\cdots,K, \ \text{and}\ \gamma_{1}\leq\gamma_{2}\leq\cdots\leq\gamma_{K}.
	\end{aligned}
\end{equation}

Clearly, the optimization problem (\ref{Aopt:UpperA1}) is convex with equality and inequality constraints. We denote the global minimizer in the $m$-th iteration as 
$\cbr{\widehat{\bTheta}^{(m+1)},\widehat{\bdelta}^{(m+1)}}$. Inspired by \cite{giesen2019combining}, who employed slack variables for inequality constraints in the standard ADMM, we construct the slack vector $\bm{y}\in\RR^{K-1}$ with $y_k=\gamma_k-\gamma_{k+1}$ for $k<K-1$. To ensure $y_k<0$, we define a loss function $\mathbb{I}_{\infty}(y_k>0)$, where $\mathbb{I}_{\infty}(y_k>0)=\infty$ when $y_k>0$ and $0$ otherwise. By incorporating this loss function, $\mathbb{I}_{\infty}(y_k>0)$, (\ref{Aopt:UpperA1}) transforms into the following equivalent optimization problem without inequality constraints:
\begin{equation}
\begin{aligned}\label{Aopt:Upper1}
\argmin_{\bm{\Theta},\bm{\delta}}\ &\frac{1}{2n}\big\|\bm{Y}-\bm{\alpha}-\widehat{\bm{F}}\bm{\theta}-\widehat{\bm{U}}\bm{\beta}\big\|_2^{2} + \lambda_{1}\sum_{i=1}^{n} \sum_{k=1}^{K}d(\delta_{ik})-\lambda_1(\nabla g_{2}(\widehat{\bm{\delta}}^{(m)}))^{\top}\bm{\delta}+\lambda_{2}\|\bm{\eta}\|_{1}+\sum_{k=1}^{K-1}\mathbb{I}_{\infty}(y_k>0)\\
		&\text{subject to}\quad \bm{\delta} = \bm{C}_{1}\bm{\alpha}-\bm{C}_{2}\bm{\gamma},\ \  \bm{D}\bm{\gamma}=\bm{y} \ \ \text{and} \ \ \bbeta=\bm{\eta},
\end{aligned}
\end{equation}
where
$$
\bm{C}_{1} =  \begin{pmatrix} \bm{1}_{K} & \bm{0}_{K}&\cdots &\bm{0}_{K}\\
		\bm{0}_{K} &\bm{1}_{K} & \cdots & \bm{0}_{K}\\
		\vdots&\vdots&\cdots&\vdots\\
		\bm{0}_{K}&\bm{0}_{K}&\cdots&\bm{1}_{K}\end{pmatrix}_{((n\times K) \times n)},\ \ \bm{C}_{2} = \begin{pmatrix} \bm{I}_{K} \\
		\bm{I}_{K} \\
		\vdots\\
		\bm{I}_{K} \end{pmatrix}_{((n\times K) \times K)}, \ \ \bm{D} = \begin{pmatrix} 1 & -1& 0 &\cdots & 0\\
		0 &1 & -1 & \cdots &0\\
		\vdots&\vdots&\vdots&\cdots&\vdots\\
		0&\cdots&0&1&-1\end{pmatrix}_{((K-1) \times K)}.
$$
The constraint $\bbeta=\bm{\eta}$ is imposed to separate the $\ell_1$-norm of $\bbeta$ from the quadratic loss. Finally, we form the scaled Lagrangian problem as 
\begin{equation}
\begin{aligned}
\argmin_{\bm{\Theta},\bm{\delta},\bm{y},\bm{u},\bm{v},\bm{w}}&\frac{1}{2n}\big\|\bm{Y}-\bm{\alpha}-\widehat{\bm{F}}\bm{\theta}-\widehat{\bm{U}}\bm{\beta}\big\|_2^{2} + \lambda_{1}\sum_{i=1}^{n} \sum_{k=1}^{K}d(\delta_{ik})-\lambda_1(\nabla g_{2}(\widehat{\bm{\delta}}^{(m)}))^{\top}\bm{\delta}+\lambda_{2}\|\bm{\eta}\|_{1}\notag \\
&+\sum_{k=1}^{K-1}\mathbb{I}_{\infty}(y_k>0)
+\frac{\rho_1}{2}(\|\bm{\delta}-\bm{C}_1\bm{\alpha}+\bm{C}_{2}\bm{\gamma}
+\bm{u}\|^{2}_2)+\frac{\rho_2}{2}(\|\bm{D}\bm{\gamma}-\bm{y}
+\bm{v}\|^{2}_2)+\frac{\rho_3}{2}(\|\bm{\eta}-\bm{\bbeta}+\bw\|^{2}_2),
\notag
\end{aligned}
\end{equation}
where $\bm{u}=(\bu_{1}^\top,\cdots,\bu_{n}^\top)^{\top}$ with $\bu_{i}=(u_{i1},\cdots,u_{iK})^{\top}$,   $\bm{v}=(v_1,\cdots,v_{K-1})^{\top}$ and $\bm{w}=(w_1,\cdots,w_p)^{\top}$
are the Lagrangian multipliers with corresponding augmented parameters denoted as $\rho_1$, $\rho_2$ and $\rho_3$. The standard ADMM procedures can be expressed as
\begin{equation}\label{Aopt:update}
\begin{aligned}
&\begin{aligned}\widehat{\bm{\Theta}}^{s+1} = \argmin_{\displaystyle \bm{\Theta}}&\frac{1}{2n}\big\|\bm{Y}-\bm{\alpha}-\widehat{\bm{F}}\bm{\theta}-\widehat{\bm{U}}\bm{\beta}\big\|_2^{2} +\frac{\rho_1}{2}\|\widehat{\bm{\delta}}^{s}-\bm{C}_1\bm{\alpha}+\bm{C}_{2}\bm{\gamma} +\widehat{\bm{u}}^{s}\|^{2}_2+\frac{\rho_2}{2}\|\bm{D}\bm{\gamma}-\widehat{\bm{y}}^{s}+\widehat{\bm{v}}^{s}\|^{2}_2\\
&+\frac{\rho_3}{2}(\|\widehat{\bm{\eta}}^{s}-\bm{\bbeta}+\widehat{\bw}^{s}\|^{2}_2), \end{aligned}\\
&\widehat{\bm{\delta}}^{s+1} = \argmin_{\bm{\delta}}\frac{\rho_1}{2}\big\|\bm{\delta}-\bm{C}_1\widehat{\bm{\alpha}}^{s+1}+\bm{C}_2\widehat{\bm{\gamma}}^{s+1}+\widehat{\bm{u}}^{s}\big\|_2^{2} - \lambda_1(\nabla g_{2}(\widehat{\bm{\delta}}^{(m-1)}))^\top(\bm{\delta}-\widehat{\bm{\delta}}^{(m-1)})+ \lambda_{1}\|\bm{\delta}\|_{1},\\
&\widehat{\bm{y}}^{s+1} = \argmin_{\bm{y}}\frac{\rho_2}{2}\|\bm{D}\widehat{\bm{\gamma}}^{s+1}-\bm{y}+\widehat{\bm{v}}^{s}\|^{2}_2 + \sum_{k=1}^{K-1}\mathbb{I}_{\infty}(y_k>0),\\
&\widehat{\bm{\eta}}^{s+1}=\argmin_{\bm{\eta}}\frac{\rho_3}{2}(\|\bm{\eta}-\widehat{\bm{\bbeta}}^{s+1}+\widehat{\bw}^{s}\|^{2}_2)+\lambda_{2}\|\bm{\eta}\|_{1},\\
&\widehat{\bm{u}}^{s+1} = \widehat{\bm{u}}^{s}+\widehat{\bm{\delta}}^{s+1}-\bC_1\widehat{\balpha}^{s+1}+\bC_2\widehat{\bgamma}^{s+1},\\
&\widehat{\bm{v}}^{s+1} = \widehat{\bm{v}}^{s}+\bD\widehat{\bgamma}^{s+1}-\widehat{\bm{y}}^{s+1},\\
&\widehat{\bm{w}}^{s+1} = \widehat{\bm{w}}^{s}+\widehat{\bm{\eta}}^{s+1}-\widehat{\bbeta}^{s+1},
\end{aligned}
\end{equation}
where the subscript $s$ denotes the $s$-th iteration in the standard ADMM algorithm.

Note that the first optimization in (\ref{Aopt:update}) is a quadratic form, and we can find the global minimizer by setting its derivative to zero. To update $\bdelta$ and $\bbeta$, we apply the soft-thresholding operator to obtain the explicit forms:
$$\widehat{\bm{\delta}}^{s+1} = \text{ST}\Big((\bm{C}_1\widehat{\bm{\alpha}}^{s+1}-\bm{C}_2\widehat{\bm{\gamma}}^{s+1}-\widehat{\bm{u}}^{s+1}+\lambda_1\nabla g_{2}(\widehat{\bm{\delta}}^{(m)})/\rho_1),\lambda_1/\rho_1\Big),\quad
\widehat{\bm{\eta}}^{s+1} = \text{ST}\Big((\widehat{\bm{\bbeta}}^{s+1}-\widehat{\bw}^{s}),\lambda_2/\rho_3\Big),
$$
where $\text{ST}_t(\bx)$ is the soft-thresholding operator applied to each element of $\bx$, defined as $\text{ST}_t(u)=\text{sgn}(u)(|u|-t)$. By some simple algebra, the updating formula for the slack vector $\by$ is given by
\begin{equation}
    \widehat{y}_k^{s+1} = \min\big\{0,\bm{D}\widehat{\bm{\gamma}}^{s+1}+\widehat{v}_k^{s}\big\}.
\label{Aopt:yupdate}
\end{equation}

In the above DC-ADMM algorithm, the parameters $\rho_1$, $\rho_2$ and $\rho_3$ influence the convergence speed \citep{boyd2011distributed,wu2016new}. For practical implementation, we typically set $\rho_1=\rho_2=\rho_3=1/2$. The complete DC-ADMM algorithm is summarized in Algorithm \ref{alg:1}.

\begin{algorithm}[hbtp]
	\caption{DC-ADMM for the Plug-in Optimization}
	\label{alg:1}
	\begin{algorithmic}[1]
		\STATE {\textbf{Input:} Datasets $\bm{Y}$ and $\bm{X}$; PCA estimators $\hat{\bm{F}}$ and $\hat{\bm{U}}$; tuning parameters $\lambda_{1}$, $\lambda_{2}$ and tolerance $\epsilon_{1},\ \epsilon_{2}>0$ for inner and outer layer algorithms.}
		\STATE {\textbf{Global Initialize:} Set $m=0$ and DCA-initial value $\hat{\bm{\alpha}}^{(m)}$, $\hat{\bm{\gamma}}^{(m)}$. For inner ADMM, the initializations  of the dual variables $\bm{u}^{0}$, $\bm{v}^{0}$, $\bm{w}^{0}$
 are chosen to be zero vectors and $\hat{\bm{\eta}}^{0}=\bm{1}_{p}$.}
		\WHILE {$m=0$ or $|Z(\hat{\bm{\Theta}}^{(m)},\hat{\bm{\delta}}^{(m)})-Z(\hat{\bm{\Theta}}^{(m-1)},\hat{\bm{\delta}}^{(m-1)})|> \epsilon_{1}$}
		\STATE {\textbf{ADMM Initialize:} Find the upper bound function $Z^{(m+1)}(\bm{\Theta},\bm{\delta})$ in the form of (\ref{Aopt:UpperA1}) at $(\hat{\bm{\Theta}}^{(m)},\hat{\bm{\delta}}^{(m)})$. Initialize $\hat{\bm{\alpha}}^{0}=\hat{\bm{\alpha}}^{(m)}$ and  $\hat{\bm{\gamma}}^{0}=\hat{\bm{\gamma}}^{(m)}$. $\hat{\bm{y}}^{0}$ is calculated by the equation (\ref{Aopt:yupdate}) with $\hat{\bm{\gamma}}^{(m)}$ and $\bm{v}^{0}$.}
        \WHILE {$s=0$ or $|Z^{(m+1)}(\hat{\bm{\Theta}}^{s},\hat{\bm{\delta}}^{s})-Z^{(m+1)}(\hat{\bm{\Theta}}^{s-1},\hat{\bm{\delta}}^{s-1})|> \epsilon_{2}$}
		\STATE Update $\hat{\bm{\Theta}}^{s+1},\hat{\bm{\delta}}^{s+1},\hat{\bm{y}}^{s+1},\hat{\bm{u}}^{s+1},\hat{\bm{v}}^{s+1} $ and $\hat{\bm{w}}^{s+1}$ by (\ref{Aopt:update}).
        \STATE $s \leftarrow s+1$.
        \ENDWHILE
		\STATE {\textbf{ADMM Output:} Updated variables $(\hat{\bm{\Theta}}^{(m+1)},\hat{\bm{\delta}}^{(m+1)})$.}
		\STATE $m \leftarrow m+1$.\\
		\ENDWHILE
		\STATE {\textbf{Global Output:} Estimations of intercept $\hat{\bm{\alpha}}$, group centroids $\hat{\bm{\gamma}}$  and regression coefficients $\hat{\bm{\theta}}$ and $\hat{\bm{\beta}}$.}
	\end{algorithmic}
\end{algorithm}

In the following theorem, we establish that the proposed DC-ADMM algorithm guarantees convergence to a local minimizer in finite steps.

\begin{theorem}\label{thm:A1}
In the DC-ADMM algorithm with $\ell_{1}$-type distance, the sequence $\{Z(\widehat{\bm{\Theta}}^{(m)})\}$ decreases with $m$ and converges in finite steps. Specifically, there exists an $m^{\star}<\infty$ such that $Z(\widehat{\bm{\Theta}}^{(m)})=Z(\widehat{\bm{\Theta}}^{(m^{\star})})$ for all $m \geq m^{\star}$. Furthermore, $\widehat{\bm{\Theta}}^{(m^{\star})}$ is a KKT point of $Z(\bm{\Theta})$.
\end{theorem}

Although the ADMM algorithm typically ensures convergence to a global minimizer, the DC-ADMM guarantees only a KKT point due to the nonconvex nature of the CAR optimization problem. A variant DC algorithm proposed by \cite{breiman1993deterministic} may achieve a global minimizer, but it is often criticized for its slow convergence speed. In Section \ref{sec:3}, we also discuss the statistical properties of the local minimizer.

It is worth noting that the DC-ADMM can also be extended to handle $\ell_2$-type distance by modifying the updating formula of $\bdelta$ in (\ref{Aopt:update}). However, the computational burden of DC-ADMM under $\ell_2$-type distance can be prohibitive. Therefore, for efficiency, we propose an alternative approach using the Cyclic Coordinate Descent (CCD) algorithm to optimize under $\ell_2$-type distance.

\subsubsection{Cyclic Coordinate Decent Algorithm}\label{sec:2.2.2}

In this section, we introduce the CCD algorithm tailored to optimize the problem (\ref{opt:CAFARM}) with $\ell_2$-type distance. Without loss of generality, the superscript $m$ denotes the $m$-th step estimator in the coordinate descent algorithm. For instance, $\hat{\balpha}^{(m)}$ and $\hat{\bgamma}^{(m)}$ represent the $m$-th step estimators of $\balpha$ and $\bgamma$, respectively. Given the estimation process, it is clear that $\hat{\bF}^{\top}\hat{\bU}=\bm{0}$. Consequently, we update $\bm{\theta}$ and $\bbeta$ as follows:
\begin{align}
	\hat{\bm{\theta}}^{(m+1)}&=\hat{\bm{F}}^{\top}(\bm{Y}-\hat{\bm{\alpha}}^{(m)})/n,
	\label{equ:UpdateTheta}\\
	\hat{\bbeta}^{(m+1)}&= \argmin_{\bm{\beta}}\frac{1}{2n}\big\|\bm{Y}-\hat{\bm{\alpha}}^{(m)}-\hat{\bm{F}}\hat{\bm{\theta}}^{(m+1)}-\hat{\bm{U}}\bm{\beta}\big\|_2^{2} + \lambda_{2}\|\bm{\beta}\|_{1}.
	\label{equ:Lasso}
\end{align}
Clearly, the optimization problem in (\ref{equ:Lasso}) is a standard LASSO problem,  and many existing packages, such as the \texttt{R} package \texttt{glmnet}, can be used to solve it. Given $\{\hat{\bm{\theta}}^{(m+1)},\hat{\bm{\beta}}^{(m+1)},\hat{\bm{\gamma}}^{(m)}\}$, updating $\bm{\alpha}$ is equivalent to solve the following optimization problem:
\begin{equation}\label{equ:UpdateAlpha}
	\hat{\balpha}^{(m+1)}=\argmin_{\balpha}\ \frac{1}{2n}\big\|\bm{Y}-\bm{\alpha}-\hat{\bm{F}}\hat{\bm{\theta}}^{(m+1)}-\hat{\bm{U}}\hat{\bm{\beta}}^{(m+1)}\big\|_2^{2}
	+\lambda_{1}g(\balpha,\hat{\bgamma}^{(m)}).
\end{equation}

Following procedures similar to the DC algorithm introduced in Section  \ref{sec:2.2.1}, it is straightforward to verify that both $g_1(\balpha,\bgamma)$ and $g_2(\balpha,\bgamma)$ are convex functions with respect to $\balpha$. Therefore, the corresponding optimization in (\ref{equ:UpdateAlpha}), based on the lower approximation of $g(\balpha,\bgamma)$, yields
\begin{equation}\label{equ:UpdateAlpha1}
	\begin{aligned}
		\hat{\balpha}^{(m+1)}=&\argmin_{\balpha}\ \frac{1}{2n}\big\|\bm{Y}-\bm{\alpha}-\hat{\bm{F}}\hat{\bm{\theta}}^{(m+1)}-\hat{\bm{U}}\hat{\bm{\beta}}^{(m+1)}\big\|_2^{2}\\
		&+\lambda_{1}\sum_{i=1}^{n}\sum_{k=1}^{K}(\alpha_{i}-\hat{\gamma}_{k}^{(m)})^2
		-\lambda_{1}\rbr{\nabla g(\hat{\bm{\alpha}}^{(m)},\hat{\bm{\bgamma}}^{(m)})}^{\top}(\bm{\alpha}-\bm{\alpha}^{(m)}),
	\end{aligned}
\end{equation}
where $\nabla g_{2}(\balpha^{(m)},\bm{\gamma}^{(m)})$ is the sub-gradient of $g_2(\balpha^{(m)},\bgamma^{(m)})$ with respect to $\balpha^{(m)}$,
$$\nabla g_2(\balpha^{(m)},\bgamma^{(m)})=\left\{2\sum_{i=1}^n\sum_{k=2}^{K}\max\{|\alpha^{(m)}_{i}-\gamma^{(m)}_{k-1}|,|\alpha^{(m)}_{i}-\gamma^{(m)}_{k}|\}\text{sgn}(\alpha^{(m)}_{i}-(\gamma^{(m)}_{k-1}+\gamma^{(m)}_{k})/2)\right\}.$$
Note that the objective function in (\ref{equ:UpdateAlpha1}) is quadratic with respect to $\balpha$, and the explicit form of $\hat{\balpha}^{(m+1)}$ can be derived by setting its first derivative equal to zero:
\begin{equation}\label{equ:UpdateAlpha2}
	\hat{\bm{\alpha}}^{(m+1)} = \frac{n}{1+2\lambda_{1}Kn}\left[\frac{1}{n}(\bm{Y}-\hat{\bm{F}}\hat{\bm{\theta}}^{(m+1)}-\hat{\bm{U}}\hat{\bm{\beta}}^{(m+1)})+2\lambda_{1}\Big(\sum_{k=1}^{K}\hat{\gamma}^{(m)}_{k}\Big)\bm{1}_{n}+\lambda_{1}\rbr{\nabla g(\hat{\bm{\alpha}}^{(m)},\hat{\bm{\bgamma}}^{(m)})}\right].
\end{equation}
The updating formula for $\bm{\gamma}$ is defined as
\begin{equation}\label{equ:UpdateGamma}
\hat{\bm{\gamma}}^{(m+1)}=\argmin_{\bm{\gamma}}\sum_{i=1}^{n}\min\{(\hat{\alpha}^{(m+1)}_i-\gamma_{1})^2,\cdots,(\hat{\alpha}^{(m+1)}_i-\gamma_{K})^2\}.
\end{equation}
This degenerates into the standard $K$-means clustering procedure and can be implemented using existing \texttt{R} packages such as \texttt{Ckmeans.1d.dp}.

Once the main updating rules are determined, compute $Z(\hat{\bm{\Theta}})$ and update the main parameters at each iteration, repeating until the stopping criterion is met. The stopping criterion is defined such that either $Z(\hat{\bm{\Theta}}^{(m)})$ is sufficiently close to $Z(\hat{\bm{\Theta}}^{(m-1)})$ or the maximum number of iterations is reached. The summarized CCD algorithm for optimizing the problem with $\ell_2$-type distance is outlined in Algorithm \ref{alg:2}.

\begin{algorithm}[hbtp]
	\caption{Cyclic Coordinate Descent Algorithm for the Optimization Problem with  $\ell_2$-type Distance}
	\label{alg:2}
	\begin{algorithmic}[1]
		\STATE {\textbf{Input:} Datasets $\bm{Y}$ and $\bm{X}$; PCA estimators $\hat{\bm{F}}$ and $\hat{\bm{U}}$; tuning parameters $\lambda_{1}$, $\lambda_{2}$ and tolerance $\epsilon>0$.}
		\STATE {\textbf{Initialize:} Set $m=0$ and initial value $\hat{\bm{\alpha}}^{(0)}$, $\hat{\bm{\gamma}}^{(0)}$.}
		\WHILE {$m=0$ or $|Z(\hat{\bm{\Theta}}^{(m)})-Z(\hat{\bm{\Theta}}^{(m-1)})|> \epsilon$}
		\STATE Update $\hat{\btheta}^{(m+1)}$ by (\ref{equ:UpdateTheta}) and update $\hat{\bbeta}^{(m+1)}$ in (\ref{equ:Lasso}) by the standard LASSO optimization.
		\STATE Update $\hat{\balpha}^{(m+1)}$ by (\ref{equ:UpdateAlpha2}).
		\STATE Update $\hat{\bgamma}^{(m+1)}$ in (\ref{equ:UpdateGamma}) by K-means algorithms.
		\STATE $m \leftarrow m+1$.\\
		\ENDWHILE
		\STATE {\textbf{Output:} Estimations of intercept $\hat{\bm{\alpha}}$, group centroids $\hat{\bm{\gamma}}$  and regression coefficients $\hat{\bm{\theta}}$ and $\hat{\bm{\beta}}$.}
	\end{algorithmic}
\end{algorithm}

Note that if the $\ell_2$-type distance in Algorithm \ref{alg:2} is replaced with the $\ell_1$-type distance, the optimization problem in (\ref{equ:UpdateGamma}) becomes a standard $K$-median algorithm, implementable using the \texttt{R} package \texttt{Ckmeans.1d.dp}. However, updating $\balpha$ in (\ref{equ:UpdateAlpha1}) would lack an explicit form, necessitating complex algorithms such as ADMM to obtain iterative solutions. Moreover, it remains an open question  whether the cyclic coordinate descent algorithm converges to a local minimizer for the non-convex optimization problem. Simulation results show no significant difference in performances between Algorithm \ref{alg:1} and Algorithm \ref{alg:2}, despite the latter bearing greater computational burden. Hence, we recommend using Algorithm \ref{alg:2} for $\ell_2$-type distance due to its computational efficiency, reserving DC-ADMM for solving (\ref{opt:CAFARM}) with $\ell_1$-type distance.

\subsection{Practical Implementation Details}\label{sec:2.3}

We first investigate the method for selecting  the number of subgroups $K$ and the shrinkage parameters $\lambda_1$ and $\lambda_2$. To alleviate computational burdens, we use the Bayesian Information Criterion (BIC) to determine the group number $K$. Specifically, let
$$
\text{BIC}(K)=\log{\big[\sum_{i=1}^{n}(Y_{i}-\hat{Y}_{i})^{2}/n\big]}+a_{n}(\hat{S}+K)\frac{\log{(n)}}{n},
$$
where $\hat{Y}_{i}=\sum_{k=1}^{K}\mathbb{I}(i\in \hat{\cG}_{k})\hat{\gamma}_{k}+\hat{\bm{f}}_{i}^{\top}\hat{\bm{\theta}}+\hat{\bm{u}}_{i}^{\top}\hat{\bm{\beta}}$ and $\hat{S}=\sum_{j=1}^{p}\mathbb{I}(\hat{\beta}_{j}\neq0)$. For better practical performance, we set $a_{n}=2\log(nK+p)$.
As $\lambda_{1}$ tends to infinity, it results in $\hat{\alpha}_i=\hat{\gamma}_k$ for $i\in\hat{\cG}_k$. Similarly, the sparsity of $\hat{\bbeta}$ increases as $\lambda_2$ grows.
To minimize differences $|\hat{\alpha}_i-\hat{\gamma}_k|$ within each group, we initially set $\lambda_{1}$ to a relatively large value and $\lambda_{2}$ to a relatively small value to control the bias. Subsequently, we determine the optimal group number by minimizing the BIC.
Similar strategies are also implemented in \cite{ma2017concave} and \cite{he2022center}. With the estimated group number $K$, we further adopt the generalized cross-validation (GCV) method  by \cite{tang2021individualized}, which is an approximate version of the method by \cite{pan2013cluster}, to select $\lambda_1$ and $\lambda_2$. Specifically, the GCV is defined as
$$
\text{GCV}(\lambda_{1},\lambda_{2}) = \frac{\sum_{i=1}^{n}(Y_{i}-\hat{Y}_{i})^{2}}{(n-\text{df})^2},
$$
where $\text{df}=\sum_{j=1}^{p}\mathbb{I}(\hat{\beta}_{j}\neq0)$. Alternatively, one can simultaneously select the optimal $K$, $\lambda_{1}$ and $\lambda_{2}$ by minimizing the BIC. However, this approach is computationally intensive due to the large size of the grid search.

Next, we focus on the selection of initial values  ($\hat{\balpha}^{(0)}$, $\hat{\bgamma}^{(0)}$) in Algorithm \ref{alg:1}. This step is more challenging because the algorithm may converge to a local minimizer due to the non-convexity of the objective function. Inspired by the ridge regression estimators used for initializing values by \cite{he2022center}, we incorporate the primary factor structure and fit the following ridge regression to obtain $\hat{\balpha}^{(0)}$:
\begin{equation}\label{equ:Init}
	\hat{\balpha}^{(0)}=\argmin_{\balpha}\ \frac{1}{2n}\big\|\bm{Y}-\bX^*\bbeta^*\big\|_2^{2}
	+\lambda^*\left\|\bm{\beta}^*\right\|_{2}^2,
\end{equation}
where $\bX^*=(\hat{\bF},\bI_n)$ and $\bbeta^*=(\btheta^{\top},\balpha^{\top})^{\top}$. The tuning parameter $\lambda^*$ is determined by cross-validation. Given $\hat{\balpha}^{(0)}$, $\hat{\bgamma}^{(0)}$ can be obtained using the $K$-means (or $K$-median) algorithm. The rationale behind using  ridge regression is that  the pseudo $\bX^*$ has a dimension of $r+n$, which is larger than the sample size $n$. We exclude $\hat{\bU}$ from the regression for two reasons: firstly, due to the latent factor structure, the common factors $\bbf_i\in\RR^r$ are considered to contain the most crucial information; secondly,  incorporating the idiosyncratic errors $\bu_{i}\in\RR^p$ into ridge regression would introduce significant computational challenges due to its high dimensionality. Indeed, our simulation results indicate that the proposed strategy performs well and offers computational advantages.

\section{Theoretical Properties}\label{sec:3}

In this section, we investigate the statistical properties of our proposed estimators under mild conditions. First, we study the convergence rate of the ``oracle" estimators, assuming prior knowledge of group memberships. Subsequently, we establish the asymptotic relationship between the oracle estimator and the local minimizer of the optimization problem (\ref{opt:CAFARM}).

Throughout the statement of the following theorems, we denote the true values of $\balpha$, $\bgamma$ and $\bbeta$ as $\balpha_0$, $\bgamma_0$ and $\bbeta_0$, respectively, and define $\btheta_0=\bB^\top\bbeta_0$. Additionally, we use $\left|\mathcal{G}_{\text{min}}\right|$ and $\left|\mathcal{G}_{\text{max}}\right|$ to represent the minimum and maximum cardinalities of $\mathcal{G}_{k}$, respectively. Let $\bm{\Omega}=(\Omega_{ik})$ be the matrix indicating the true group membership, where $\Omega_{ik}=1$ implies that subject $i$ belongs to group $\mathcal{G}_{k}$ and $\Omega_{ik}=0$ otherwise.  Hence, it is straightforward to verify that $\balpha_0=\bOmega\bgamma_0$. In the ideal scenario where $\bOmega$ is known in advance, the oracle estimator can be defined as 
\begin{equation}\label{equ:oracle}(\hat{\bm{\gamma}}^{or},\hat{\bm{\theta}}^{or},\hat{\bm{\beta}}^{or})=\mathop{\text{argmin}}_{\bm{\gamma},\bm{\theta},\bm{\beta}}\frac{1}{2n}\big\|\bm{Y}-\bm{\Omega}\bm{\gamma}-\hat{\bm{F}}\bm{\theta}-\hat{\bm{U}}\bm{\beta}\big\|_2^{2} + \lambda_{2}\left\|\bm{\beta}\right\|_{1} .
\end{equation}
Write $\hat{\bm{\alpha}}^{or}=\bm{\Omega}\hat{\bm{\gamma}}^{or}$. Note that the oracle estimator only incorporates prior grouping information $\bOmega$, without prior knowledge of the sparsity of $\bbeta_0$.

In the following, we first introduce some assumptions related to the estimation of factor models. These technical assumptions are commonly used in the literature on large-dimensional factor models, such as \cite{fan2013large}, \cite{fan2020factor} and \cite{fan2023latent}.
\begin{asmp}\label{asmp:1}
	(Tail Probability) The sequence $\left\{\bm{f}_{i},\bm{u}_{i}\right\}_{-\infty}^{\infty}$ is strictly stationary with $\mathbb{E}\left(f_{ij}\right)=\mathbb{E}\left(u_{is}\right)=\mathbb{E}\left(f_{ij}u_{is}\right)=0$ for all $s\le p$, $j\le r$ and $i\le n$. Furthermore, we assume $\bm{f}_{i}$ and $\bm{u}_{i}$ satisfy exponential-type tail probability:  $$\mathbb{P}\Big(\frac{\left|\bm{a}^{\top}\bm{f}_{i}\right|}{\left\|\bm{a}\right\|_{2}}\geq t\Big)\leq\exp{\Big(-\Big(\displaystyle\frac{t}{V_{1}}\Big)^{l_{1}}\Big)}\quad \text{and}\quad \mathbb{P}\Big(\frac{\left|\bm{b}^{\top}\bm{u}_{i}\right|}{\left\|\bm{b}\right\|_{2}}\ge t\Big)\leq\exp{\Big(-\Big(\displaystyle\frac{t}{V_{2}}\Big)^{l_{2}}\Big)},$$
	where $\bm{a}\in \mathbb{R}^{r},\bm{b}\in \mathbb{R}^{p}$. $V_{1},V_{2},l_{1}$ and $l_{2}$ are positive constants. Denote $\text{Cov}(\bm{u}_{i})$ as $\bm{\Sigma}\ \text{for all}\ i=1,\cdots,n$. There exists two positive constants $M_{1}$ and $M_{2}$ such that $\lambda_{\text{min}}\left(\bm{\Sigma}\right)\geq M_{1}$, $\mathop{\text{min}}_{s,t\in\{1,\cdots,p\}}\text{Var}\left(u_{is}u_{it}\right)\geq M_{1}$ and  $\left\|\bm{\Sigma}\right\|_{\infty}\leq M_{2}$.
\end{asmp}

\begin{asmp}\label{asmp:2}
	($\alpha$-Mixing Condition)
	Suppose $\mathcal{A}_{-\infty}^{0}$ and $\mathcal{A}_{n}^{\infty}$
	are $\sigma$-algebras generated by
	$\left\{\bm{f}_{i},\bm{u}_{i}\right\}_{-\infty}^{0}$ and  $\left\{\bm{f}_{i},\bm{u}_{i}\right\}_{n}^{\infty}$ respectively. We assume that there exists a constant $l_{3}$ such that $\displaystyle3/l_{1}+3/(2l_{2})+1/l_{3}>1$ and the following inequality holds:
	\begin{equation}
		\displaystyle\mathop{\text{sup}}_{E_{1}\in\mathcal{A}_{-\infty}^{0},E_{2}\in\mathcal{A}_{n}^{\infty}}\left|\mathbb{P}\left(E_{1}\right)\mathbb{P}\left(E_{2}\right)-\mathbb{P}\left(E_{1}E_{2}\right)\right|\leq\exp{\left(-an^{l_{3}}\right)},
		\nonumber
	\end{equation}
	where $a$ is a positive constant.
\end{asmp}

\begin{asmp}\label{asmp:3}
	There exists some constant $M_{0}$ such that $\left\|\bm{B}\right\|_{\text{max}}\leq M_{0}$, $\displaystyle \mathbb{E}[p^{-2}[\bm{u}_{i}^{\top}\bm{u}_{j}-\mathbb{E}(\bm{u}_{i}^{\top}\bm{u}_{j})] ]^{4}\leq M_{0}$ and all the eigenvalues of $p^{-1}\bm{B}^{T}\bm{B}$ are bounded away from $0$ and $\infty$. Furthermore, $\mathbb{E}\left\|p^{-1/2}\bm{B}^{T}\bm{u}_{i}\right\|_{2}^{4}\leq M_{0}$.
\end{asmp}

Assumption \ref{asmp:1} requires exponential-type tail probability for $\bbf_i$ and $\bu_i$, assuming they are uncorrelated. Moreover, $\left\|\bm{\Sigma}\right\|_{\infty}\leq M_{2}$ is imposed to ensure consistent estimation of the number of factors. Assumption \ref{asmp:2} relaxes the independence of $\{\bbf_i,\bu_i\}_{-\infty}^{\infty}$ to weak dependence. Combined with the exponential-type tail probability in Assumption \ref{asmp:1}, these conditions allow us to establish non-asymptotic bounds related to $\bbf_i$ and $\bu_i$. Assumption \ref{asmp:3} is a standard condition for deriving the consistency of factor loadings and factor scores with a fixed $r$, adapted from conditions in \cite{bai2003inferential}. Under these regular conditions, we derive the convergence rates of $\hat{\bm{F}},\hat{\bm{B}}$ and $\hat{\bm{U}}$, ensuring the efficiency of the plug-in estimations. In the following, we proceed to introduce some technical assumptions for subgroup identification.

\begin{asmp}\label{asmp:4}
	Suppose $\bm{\bepsilon}$ has tail probability of sub-Gaussian form, that is, $\mathbb{P}\left(\left|\bm{c}^{\top}\bm{\bepsilon}\right|\geq\left\|\bm{c}\right\|_{2}t\right)\leq2\exp{\left(-C_{0}t^{2}\right)}$ for any $\bm{c}\in \RR^{n}$, where $C_{0}$ is a positive constant.
\end{asmp}

\begin{asmp}\label{asmp:5}
	There exist two positive constants $C_{1}$ and $C_{2}$ such that 
	\begin{eqnarray}
		C_{1}\leq
		\displaystyle |\mathcal{G}_\text{{min}}|^{-1}\lambda_{\text{min}}\left[\text{diag}\left(    	\bm{\Omega}^{\top}\bm{\Omega},n\bm{I}_{r},n\bm{\Sigma}_{\mathcal{S}\mathcal{S}}\right)
		\right]\leq
		\displaystyle |\mathcal{G}_{\text{min}}|^{-1}\lambda_{\text{max}}\left[\text{diag}\left(	\bm{\Omega}^{\top}\bm{\Omega},n\bm{I}_{r},n\bm{\Sigma}_{\mathcal{S}\mathcal{S}}\right)\right]\leq C_{2},
		\nonumber
	\end{eqnarray}
	where $\cS$ stands for the support set of $\bbeta_0$. Additionally, we assume the irrepresentable condition holds for $\bm{\Sigma}$, i.e., there exist a positive value $\rho$ such that $\|\bm{\Sigma}_{\mathcal{S}^{c}\mathcal{S}}\bm{\Sigma}_{\mathcal{S}\mathcal{S}}^{-1}\|_{\infty}\leq 1-\rho$. Moreover, we assume $\|\bm{\Sigma}^{-1}\|_{\infty}$ is bounded by some positive constant $C_{4}$.
\end{asmp}

Assumption \ref{asmp:4} requires that the random noise in (\ref{equ:Subgroup2}) follows a sub-Gaussian distribution.
By the definition of $\bm{\Omega}$, we have $\bm{\Omega}^{\top}\bm{\Omega}=\text{diag}\left(\left|\mathcal{G}_{1}\right|,\cdots,\left|\mathcal{G}_{K}\right|\right)$. As $\lambda_{\text{max}}\left[\text{diag}\left(\bm{\Omega}^{\top}\bm{\Omega},n\bm{I}_{r},n\bm{\Sigma}_{\mathcal{SS}}\right)\right]\geq\lambda_{\text{max}}\left(\bm{\Omega}^{\top}\bm{\Omega}\right)$, Assumption \ref{asmp:5} implies a balanced group size: $C_{2}\left|\mathcal{G}_{\text{min}}\right|\geq\left|\mathcal{G}_{\text{max}}\right|$. The irrepresentable condition in Assumption \ref{asmp:5}, akin to that of \cite{zhao2006model}, ensures the sign consistency of $\hat{\bbeta}^{or}$. Moreover, the requirement that $\|\bm{\Sigma}^{-1}\|_{\infty}$ is bounded from above is equivalent to a weak correlation requirement of the idiosyncratic errors. 

With these assumptions, we then present the theoretical properties of the oracle estimators in the following theorem.
\begin{theorem}\label{thm:1}
	Let $\bm{\Theta}_{0}:=(\bm{\alpha}_{0}^{\top},(\bm{H}\bm{\theta}_{0})^{\top},\bm{\beta}_{0}^{\top})^{\top}$, where $\bm{H} = 1/n\bm{V}^{-1}\hat{\bm{F}}^{\top}\bm{F}\bm{B}^{\top}\bm{B}$ and $\bV$ denotes the diagonal matrix of the first $r$ largest eigenvalues of the sample covariance matrix, arranged in decreasing order.
	Suppose Assumptions \ref{asmp:1}--\ref{asmp:5} hold, $\lambda_{2}\geq (1/c)\|\nabla Z_{1}(\bm{\Theta}_{0})\|_{\infty}$, where $c\leq \rho/(2-\rho)$ and   $Z_{1}(\bm{\Theta})=1/2n\|\bm{Y}-\bm{\Omega}\bm{\gamma}-\hat{\bm{F}}\bm{\theta}-\hat{\bm{U}}\bm{\beta}\|_2^{2}$. If $p_{\mathcal{S}}(\log{p}/n+1/p)^{1/2}=o(1)$ with $p_{\mathcal{S}}=|\mathcal{S}|$, it holds
	$$\big\|(
	(\hat{\bm{\gamma}}^{or}-\bm{{\gamma}}_{0})^{\top},
	(\hat{\bm{\theta}}^{or}-\bm{H}\bm{\theta}_{0})^{\top},
	(\hat{\bm{\beta}}^{or}-\bm{\beta}_{0})^{\top})^{\top}\big\|_{\infty}
	=O_{\mathbb{P}}\left(p_{\mathcal{S}}(\sqrt{\log{p}/n}+1/\sqrt{p})\right).$$
\end{theorem}

Theorem \ref{thm:1} establishes the consistency and convergence rate of the oracle estimator when the number of subgroups is known in advance. The first term $O_{\mathbb{P}}\left(p_{\mathcal{S}}\sqrt{\log{p}/n}\right)$ corresponds to the rate of penalized regression, similar to the LASSO estimator. The second term $O_{\mathbb{P}}\left(1/\sqrt{p}\right)$ arises from the additional bias incurred due to the use of a plug-in estimate for the factor model. With further assumption, we can also obtain sign consistency, as stated in the following proposition.

\begin{proposition}\label{prop:1}
	Under the same conditions as in Theorem \ref{thm:1}, and further assuming $\text{min}(|(\bm{\Theta}_{0})_{\mathcal{S}}|) \geq \kappa\lambda_{2}$,
	where $\kappa$ is a sufficiently large positive value, then $\hat{\bm{\beta}}^{or}$ achieves sign consistency:
	$$\mathbb{P}(\text{sign}(\hat{\bm{\beta}}^{or})=\text{sign}(\bm{\beta}_{0}))\rightarrow1.$$
\end{proposition}

Next, we establish the asymptotic relationship between the oracle estimator and the local minimizer of the optimization problem (\ref{opt:CAFARM}) in the following theorem.
\begin{theorem}\label{thm:2}
	Let $\mathcal{M}(\lambda_1,\lambda_2)$ be the set of all local minima of the optimization problem (\ref{opt:CAFARM}) with $\ell_1$-type distance. Suppose the assumptions required in Theorem \ref{thm:1} hold. If $r_{n}\gg p_{\mathcal{S}}(\sqrt{\log{p}/n}+1/\sqrt{p})$ and $ \lambda_{1}\gg n^{-1}\text{max}\{\sqrt{\log{n}},p_{\mathcal{S}}^2(\sqrt{\log{p}}+\sqrt{n}/\sqrt{p})\}$, where $r_n= \mathop{\text{min}}_{i,j\in\{1,\cdots,K\}}\left|\gamma_{0i} -\gamma_{0j}\right|$ is the minimum gap between different group centroids with $\gamma_{0i}$ denoting the $i$-th element of $\bm{\gamma}_{0}$. Then, there exist a point $\hat{\bm{\Theta}}_{\mathcal{M}}\in\mathcal{M}(\lambda_{1},\lambda_{2})$ such that $\mathbb{P}(\hat{\bm{\Theta}}_{\mathcal{M}}=\hat{\bm{\Theta}}^{or})\rightarrow1$, where $\hat{\bm{\Theta}}^{or}=(\hat{\bm{\alpha}}^{or},\hat{\bm{\gamma}}^{or},\hat{\bm{\theta}}^{or},\hat{\bm{\beta}}^{or})$. 
\end{theorem}

Here is the refined text with some adjustments for clarity:

Theorem \ref{thm:2} asserts that the oracle estimator is a local minimizer of the optimization problem for the SILFS method with probability tending to 1. Combined with Theorem \ref{thm:1} and Proposition \ref{prop:1}, we conclude that there exists a local minimizer which achieves both estimation and sign consistency. Therefore, it is crucial to identify such a (local) minimizer for the proposed SILFS method. Fortunately, the proposed DC-ADMM algorithm ensures finite-step convergence to a local minimizer. In other words, with high probability, we can obtain consistent estimators provided we start with a ``good'' initial value.

\section{Simulation Study}\label{sec:4}

In this section, we conduct thorough simulation experiments to assess the finite sample behaviors of the proposed SILFS method. In Section \ref{sec:4.1}, we first compare the SILFS method with existing methods to evaluate empirical performance in terms of subgroup identification and variable selection. We also examine the computational time required for different methods. In Section \ref{sec:4.2}, we investigate the sensitivity of the SILFS method to various levels of collinearity.

In the simulation studies, we use the eigenvalue-ratio method to determine the  number of factors $r$. To elaborate, we denote
$\lambda_{i}(\bm{X}\bm{X}^{\top})$ as the $i$-th largest eigenvalue of $\bm{X}\bm{X}^{\top}$, and the estimator of the factor number is determined by the modified ER method proposed by \cite{chang2015high}:
$$\widehat{r} = \mathop{\text{argmin}}_{1\leq i\leq r^*}\frac{\lambda_{i}(\bm{X}\bm{X}^{\top})+C_{n,p}}{\lambda_{i+1}(\bm{X}\bm{X}^{\top})+C_{n,p}},$$
where $r^*$ is a positive integer larger than $r$, and $C_{n,p}$ is a constant that depends only on $n$ and $p$. When $\bm{X}$ itself is weakly correlated, one can directly set $r$ as 0.

\subsection{Subgroup Identification and Variable Selection}\label{sec:4.1}

In this section, we resort to Algorithm \ref{alg:1} to solve the optimization problem (\ref{opt:CAFARM}) with the $\ell_2$-type distance. To emphasize the advantage of CAR, we construct FA-PFP for comparison. FA-PFP initially considers the FARM and then replaces CAR with $\sum_{i<j}p(a_i-a_j)$ in the optimization problem (\ref{opt:CAFARM}), where $p(\cdot)$ is the SCAD penalty. Tuning parameters involved in FA-PFP is selected with the same strategy as that of the SILFS. Additionally, we compare SILFS with the standard center-augmented regularization model (S-CAR), which ignores the factor structure, to assess the impact of factor structure on subgroup identification and variable selection. Furthermore, we include the oracle estimator with known factors as a comparative benchmark for all models.

We adopt a data generation process similar to that of \cite{zhang2019robust} and \cite{fan2020factor}. Specifically, the model setting is defined as:
$$
Y_{i}=\alpha_{i}+\bm{X}_{i}^{\top}\bm{\beta}+\epsilon_{i},\quad \bm{X}_{i}=\bm{B}\bm{f}_{i}+\bm{u}_{i},\quad\ i=1,\cdots,n,
$$
where $\bbf_i\in\RR^r$ and $\bbeta\in\RR^p$.
We set $\epsilon_i$ are i.i.d. drawn from $\cN(0,0.1)$, and $\bm{\beta} =(\beta_{1},\dots,\beta_{5},\bm{0}^{\top}_{p-5})^{\top}$ with $\beta_{j}\sim \cU(0.8,1)$ for $j\le 5$. For the factor model of $\bX$, let $B_{ij}$ are i.i.d. drawn from $\cU(0,1)$.
Take the dependence of factor score into consideration, we employ a vector AR(1) model for $\bbf_i$: $\bm{f}_{i}=\bm{\Phi}\bm{f}_{i-1}+\bm{\xi}_{i}$, where the noise $\bxi_{i}$ are i.i.d. drawn from $\cN(\bm{0}_{r},0.1\bI_r)$ and $\Phi_{st}=(0.5)^{\mathbb{I}(s=t)}(0.3)^{{|s-t|}}$. The idiosyncratic errors $\bu_i$ are i.i.d. generated from $\cN(\bm{0}_{p},0.1\bI_p)$. As for group-specific parameters $\alpha_i$, the following two scenarios are considered.

\textbf {Scenario A} Set $K=2$ and $\mathbb{P}(\alpha_{i}=-a)=\mathbb{P}(\alpha_{i}=a)=1/2$, where $a\in\cbr{3,5}$.

\textbf {Scenario B} Set $K=3$ and $\mathbb{P}(\alpha_{i}=-a)=\mathbb{P}(\alpha_{i}=0)=\mathbb{P}(\alpha_{i}=a)=1/3$, where $a\in\cbr{3,5}$.

In the simulation studies, we fix the number of factors as $r=4$, set the sample size as $n=100$, and vary the dimensionality $p$ across $50$, $100$, and $150$. All simulation results in this paper are based on 100 replications. To evaluate the empirical performance of the estimators, we provide the following three  criteria:

(i) Estimation error indices, used by both \cite{zhang2019robust} and \cite{he2022center}, involve the Root Mean Squared Error (RMSE). Specifically, the RMSE for $\balpha$ and $\bbeta$ is denoted as $\text{RMSE}_{\bm{\alpha}}$ and $\text{RMSE}_{\bm{\bbeta}}$ respectively. The $\text{RMSE}_{\bm{\bbeta}}$ is computed as $(\sum_{i=1}^{N}||\widehat{\bbeta}^{(i)}-\bbeta||_2^2/(Np))^{1/2}$, where $\widehat{\bm{\beta}}^{(i)}$ represents the estimate of $\bm{\beta}$ in the $i$-th replication and $N$ is the number of replications. These indices evaluate the estimation accuracy, and a smaller $\text{RMSE}_{\bm{\alpha}}$ implies better subgroup identification.

(ii) Variable selection indices include the mean of Sensitivity and Specificity over 100 replications. These metrics are common and crucial for assessing feature selections, as illustrated in \cite{Chen2021Simultaneous}.

(iii) Subgroup identification indices include the Rand index (RI), a quantity between 0 and 1. It is commonly used to measure the performance of subgroup recovery, as discussed in \cite{rand1971objective} and \cite{zhang2019robust}. A higher Rand index value indicates better clustering performance. Additionally, we calculate the mean value of the estimated clusters, denoted as $\widehat{K}_{\text{mean}}$. The frequency of overestimation/underestimation of the group number is presented in the form $a|b$, where $a$ and $b$ represent the frequencies of overestimation and underestimation, respectively. These clustering-related indices are also adopted in \cite{liu2023simultaneous} to comprehensively evaluate clustering performance.

\begin{table}[h]
	\centering
	\renewcommand{\arraystretch}{1.5}
	\caption{Simulation results for Scenario A. The values in the parentheses denote standard deviation.}
	\label{tab:1}
	\scalebox{0.81}{
		\begin{tabular}{ccccccccc}
			\toprule[1pt]
			\multirow{2}{2cm}{\centering \textbf{Case} $(a,n,p,K)$}&\multirow{2}{2cm}{\centering \textbf{Method}}&\multicolumn{2}{c}{\centering \textbf{Estimation Error Indices} }&\multicolumn{3}{c}{\centering \textbf{Subgroup Identification Indices }}&\multicolumn{2}{c}{\centering \textbf{Variable Selection Indices}}\\
			\cmidrule(lr){3-4} \cmidrule(lr){5-7}
			\cmidrule(lr){8-9}
			&& $\text{RMSE}_{\bm{\alpha}}$ &  $\text{RMSE}_{\bm{\beta}}$ &$\widehat{K}_{mean}$  &Freq& RI & Sensitivity & Specificity  \\
			\hline
			\multirow{4}{2cm}{\centering $(3,100,50,2)$}
			& FA-PFP & 0.347 & 0.130 & 2.250(0.672) & $17 | 0$ & 0.997(0.007) & 0.992(0.039) & 0.979(0.025) \\
			& SILFS & 0.128 & 0.109 & 2.100(0.302) & $10 | 0$ & 0.985(0.045) & 1.000(0.000) & 0.986(0.023) \\
			& Oracle & 0.068 & 0.125 & NA & NA & NA & 0.996(0.040) & 0.996(0.009) \\
			& S-CAR & 0.313 & 0.212 & 2.430(0.820) & $27 | 0$ & 0.951(0.091) & 0.678(0.233) & 0.937(0.043) \\
			\hline
			\multirow{4}{2cm}{\centering $(3,100,100,2)$}
			& FA-PFP & 0.363 & 0.094 & 2.210(0.498) & $17 | 0$ & 0.997(0.006) & 0.990(0.044) & 0.978(0.019) \\
			& SILFS & 0.148 & 0.084 & 2.090(0.288) & $9 | 0$ & 0.987(0.042) & 0.982(0.076) & 0.990(0.013) \\
			& Oracle & 0.067 & 0.084 & NA & NA & NA & 1.000(0.000) & 0.997(0.005) \\
			& S-CAR & 0.343 & 0.158 & 2.590(0.954) & $33 | 0$ & 0.932(0.104) & 0.626(0.285) & 0.950(0.028) \\
			\hline
			\multirow{4}{2cm}{\centering $(3,100,150,2)$}
			& FA-PFP & 0.400 & 0.084 & 2.270(0.633) & $19 | 0$ & 0.996(0.013) & 0.962(0.120) & 0.976(0.021) \\
			& SILFS & 0.241 & 0.097 & 2.320(0.618) & $13 | 0$ & 0.960(0.076) & 0.907(0.215) & 0.999(0.003) \\
			& Oracle & 0.072 & 0.067 & NA & NA & NA & 1.000(0.000) & 0.997(0.004) \\
			& S-CAR & 0.334 & 0.126 & 2.550(0.936) & $31 | 0$ &0.936(0.102) & 0.656(0.280) & 0.965(0.021) \\
			\hline
			\multirow{4}{2cm}{\centering $(5,100,50,2)$}
			& FA-PFP & 0.558 & 0.157 & 2.160(0.507) & $11 | 0$ & 0.948(0.116) & 0.942(0.042) & 0.998(0.005) \\
			& SILFS & 0.168 & 0.156 & 2.090(0.288) & $9 | 0$ & 0.989(0.037) & 0.984(0.073) & 0.997(0.015) \\
			& Oracle & 0.081 & 0.120 & NA & NA & NA & 1.000(0.000) & 0.996(0.009) \\
			& S-CAR & 0.334 & 0.203 & 2.520(0.904) & $31 | 0$ & 0.942(0.097) & 0.702(0.270) & 0.937(0.043) \\
			\hline
			\multirow{4}{2cm}{\centering $(5,100,100,2)$}
			& FA-PFP & 0.608 & 0.118 & 2.270(0.709) & $16 | 0$ & 0.920(0.136) & 0.949(0.038) & 0.997(0.010) \\
			& SILFS & 0.137 & 0.105 & 2.070(0.326) & $5 | 0$ & 0.991(0.041) & 0.980(0.115) & 0.999(0.003) \\
			& Oracle & 0.079 & 0.085 & NA & NA & NA & 1.000(0.000) & 0.996(0.007) \\
			& S-CAR & 0.330 & 0.158 & 2.560(0.903) & $34 | 0$ & 0.935(0.100) & 0.612(0.278) & 0.955(0.024) \\
			\hline
			\multirow{4}{2cm}{\centering $(5,100,150,2)$}
			& FA-PFP & 0.560 & 0.098 & 2.170(0.551) & $12 | 0$ & 0.936(0.124) & 0.957(0.025) & 0.998(0.005) \\
			& SILFS & 0.241 & 0.095 & 2.230(0.468) & $21 | 0$ & 0.968(0.065) & 0.934(0.173) & 0.998(0.012) \\
			& Oracle & 0.073 & 0.071 & NA & NA & NA & 0.998(0.020) & 0.997(0.005) \\
			& S-CAR & 0.341 & 0.128 & 2.580(0.934) & $35 | 0$ & 0.935(0.098) & 0.618(0.296) & 0.968(0.018) \\
			\bottomrule[1pt]
		\end{tabular}
	}
\end{table}

\begin{table}[h]
	\centering
	\renewcommand{\arraystretch}{1.5}
	\caption{Simulation results for Scenario B. The values in the parentheses denote standard deviation.}
	\label{tab:2}
	\scalebox{0.81}{
		\begin{tabular}{ccccccccc}
			\toprule[1pt]
			\multirow{2}{2cm}{\centering \textbf{Case} $(a,n,p,K)$}&\multirow{2}{2cm}{\centering \textbf{Method}}&\multicolumn{2}{c}{\centering \textbf{Estimation Error Indices} }&\multicolumn{3}{c}{\centering \textbf{Subgroup Identification Indices }}&\multicolumn{2}{c}{\centering \textbf{Variable Selection Indices}}\\
			\cmidrule(lr){3-4} \cmidrule(lr){5-7}
			\cmidrule(lr){8-9}
			&& $\text{RMSE}_{\bm{\alpha}}$ &  $\text{RMSE}_{\bm{\beta}}$ &$\widehat{K}_{mean}$  &Freq& RI & Sensitivity & Specificity  \\
			\hline
			\multirow{4}{2cm}{\centering $(3,100,50,3)$}
			& FA-PFP & 0.689 & 0.218 & 4.140(1.181)  & $ 67 | 1 $& 0.932(0.048) & 0.778(0.244) & 0.914(0.058)  \\
			& SILFS & 0.455 & 0.184 & 3.000(0.000) &$ 0 | 0 $& 0.971(0.036) & 0.860(0.234) & 0.972(0.037) \\
			& Oracle & 0.080 & 0.113 &NA &NA& NA & 1.000(0.000) & 0.994(0.011) \\
			& S-CAR & 0.733 & 0.244 &3.100(0.560)&$21|11$& 0.917(0.079) &  0.550(0.252) & 0.894(0.060) \\
			\hline
			\multirow{4}{2cm}{\centering $(3,100,100,3)$}
			&FA-PFP &0.712 & 0.162 & 4.270(1.370) &$ 67 | 1 $&  0.930(0.051) & 0.752(0.248) & 0.920(0.042) \\
			&SILFS  & 0.526 & 0.146 & 3.000(0.000)&$ 0 | 0 $& 0.961(0.042) & 0.770(0.285) & 0.978(0.025) \\
			&Oracle & 0.090 & 0.081 &NA&NA& NA & 1.000(0.000) & 0.995(0.007)\\
			&S-CAR & 0.738 & 0.183 &3.170(0.551)&$25|8$&  0.915(0.074) & 0.456(0.272) & 0.931(0.033)\\
			\hline
			\multirow{4}{2cm}{\centering $(3,100,150,3)$}
			&FA-PFP & 0.742 & 0.142 & 4.130(1.292) & $ 62 | 2 $& 0.925(0.053) & 0.674(0.275) & 0.926(0.042) \\
			&SILFS  &0.599 & 0.122 & 3.000(0.000)&$ 0 | 0 $& 0.949(0.042) & 0.812(0.236) & 0.946(0.037) \\
			&Oracle & 0.076 & 0.063 & NA&NA&NA& 0.998(0.020) & 0.996(0.005)\\
			&S-CAR & 0.720 &  0.152 & 3.070(0.432)&$13|6$& 0.923(0.060) & 0.428(0.248) &0.942(0.027) \\
			\hline
			\multirow{4}{2cm}{\centering $(5,100,50,3)$}
			&FA-PFP & 0.684 & 0.192 & 6.580(3.075) & $ 86 | 0 $& 0.956(0.063) & 0.826(0.289) & 0.931(0.048) \\
			&SILFS  & 0.254 & 0.130 & 3.000(0.141)&$ 0 | 0 $& 0.997(0.014) & 0.976(0.104) & 0.986(0.044) \\
			&Oracle & 0.081 & 0.112 &NA&NA& NA & 1.000(0.000) & 0.995(0.011) \\
			&S-CAR & 0.423 & 0.216 & 3.310(0.581)& $25|0 $& 0.976(0.041) & 0.678(0.224) & 0.928(0.050) \\
			\hline
			\multirow{4}{2cm}{\centering $(5,100,100,3)$}
			&FA-PFP &0.644& 0.136 &6.520(2.761)& $95|0$ & 0.966(0.044) & 0.864(0.225) & 0.938(0.033) \\
			&SILFS  &0.346 & 0.101 & 3.010(0.100) &$ 1 | 0 $& 0.993(0.021) & 0.962(0.141) & 0.982(0.041) \\
			&Oracle &0.090 & 0.078 &NA&NA& NA & 1.000(0.000) & 0.995(0.008) \\
			&S-CAR &0.480 & 0.164 &3.310(0.506)& $31|1$ & 0.972(0.045) & 0.592(0.224) & 0.952(0.034)\\
			\hline
			\multirow{4}{2cm}{\centering $(5,100,150,3)$}
			&FA-PFP &0.713 & 0.118 & 6.750(3.010) &$ 93 | 0 $& 0.954(0.062) & 0.808(0.295) & 0.948(0.033) \\
			&SILFS  &0.423 & 0.088 & 3.000(0.000)&$ 0|0 $& 0.991(0.024) & 0.934(0.184) & 0.976(0.048) \\
			&Oracle & 0.083 & 0.063 &NA&NA& NA & 0.998(0.020) &0.996(0.005)\\
			&S-CAR & 0.531 & 0.140 & 3.330(0.533)&$ 28 | 1$ & 0.970(0.044) & 0.596(0.248) & 0.957(0.030) \\
			\bottomrule[1pt]
		\end{tabular}
	}
\end{table}
 
The simulation results for Scenario A and B are shown in Table \ref{tab:1} and \ref{tab:2} respectively.  There are four main takeaways from Table \ref{tab:1} for Scenario A. Firstly, though the S-CAR method proficiently identifies the number of subgroups, its Rand index is relatively lower compared with the  SILFS which incorporates the factor structure.
Secondly, SILFS outperforms both the FA-PFP and S-CAR methods in terms of variable selection and estimation accuracy. It also demonstrates comparable performance to the oracle method, highlighting the importance of overcoming feature dependence. Moreover, despite both FA-PFP and SILFS incorporating factor structures, FA-PFP exhibits poor performance in terms of RMSE. This can be attributed to its tendency to generate isolated and small-sized subgroups, leading to an overestimation of $K$. Lastly, as the group parameter $a$ increases, the performance of SILFS improves correspondingly with the increased distance between groups. Since the FA-PFP and SILFS is equipped with factor structures, the increase in dimensionality has no significant impact on the estimation errors. The results in Table \ref{tab:2} indicate that in the case of three groups, S-CAR performs worse in terms of subgroup identification and  completely loses power for variable selection. FA-PFP also exhibits a significant decline in terms of variable selection. In contrast, SILFS remains stable and continues to perform closely well with the oracle estimator. Overall, SILFS performs satisfactorily in various cases.

\begin{figure}[ht]
	\includegraphics[width=16cm]{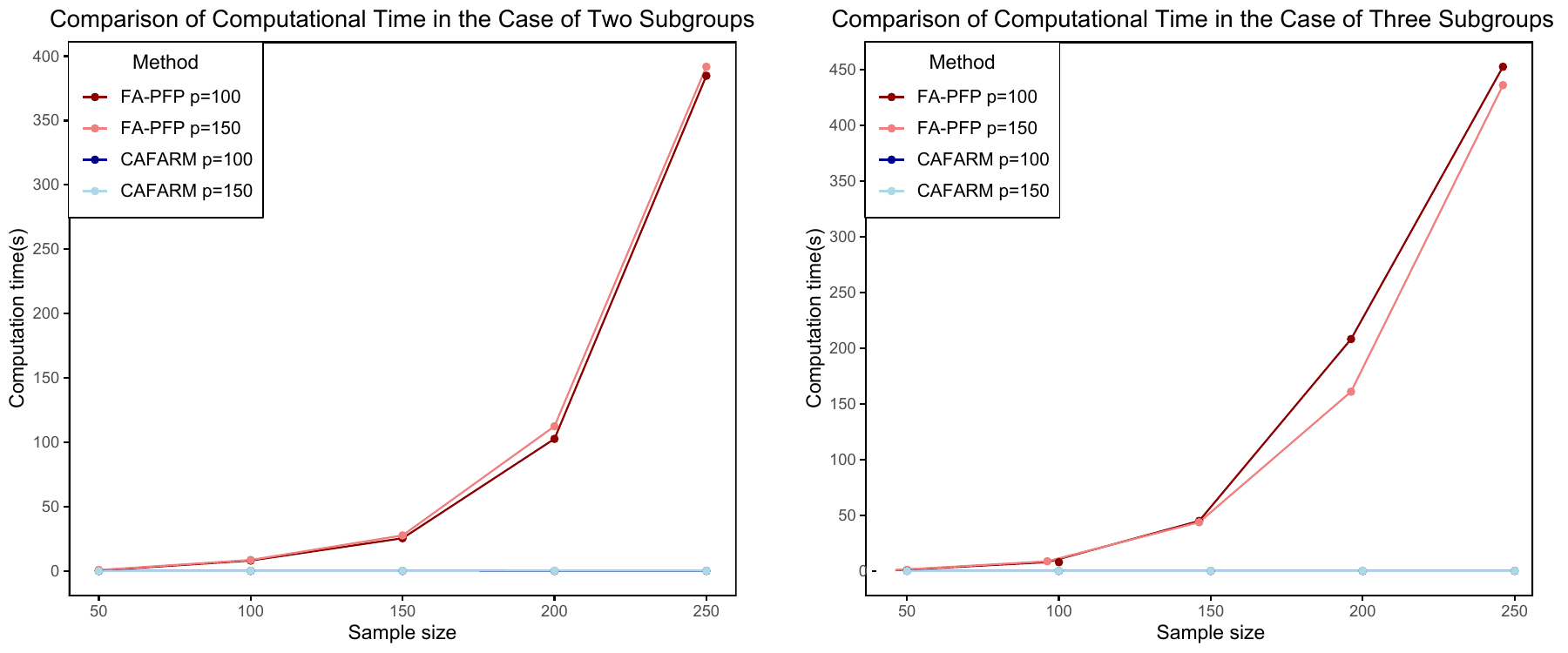}
	\caption{Comparison of Computing Time between FAPFP and SILFS Algorithms.}
	\label{fig:2}
\end{figure}

Next, we show the computing time of the SILFS and FA-PFP methods for Scenario A with $(\gamma_{1},\gamma_2) = (-3,3)$ and Scenario B with $(\gamma_{1},\gamma_2,\gamma_3) = (-3,0,3)$. For fair comparison, we use the same stopping criteria for SILFS and FA-PFP methods and only report the average time for 10 replications after selecting the optimal tuning parameters. The line charts of computing time are shown in Figure \ref{fig:2}. As the sample size increases, the computing time for FA-PFP increases rapidly, with a quadratic curve trend.
In contrast, the computational time required for SILFS increases steadily, almost linearly with respect to sample size.
 
In summary, based on the simulation results, we draw the following conclusions: (a) Existing methods in the literature show significantly reduced accuracy in variable selection and subgroup identification in the presence of covariate dependence. In contrast, the SILFS method performs well in both clustering and variable selection tasks. This is supported by high Rand Index and simplicity measures, as well as regression parameter estimators closely aligning with their true values; (b) Compared to the FA-PFP method, SILFS substantially reduces computational burdens, especially with large sample sizes and dimensions.

\subsection{Sensitivity to the Level of Collinearity}\label{sec:4.2}

In this section, we investigate the performance of SILFS in scenarios where the covariates  exhibit high collinearity but do not have a factor structure. We also consider the uncorrelated case. Specifically, we generate the data as follows:
$$
Y_{i}=\alpha_{i}+\bm{x}_{i}^{\top}\bm{\beta}+\epsilon_{i}, \ i=1,\dots, n.
$$
where $\bm{\beta} =(\beta_{1},\dots,\beta_{10},\bm{0}^{\top}_{p-10})^{\top}$ with $\beta_{j}\sim \cU(1,2)$ for $j\le 10$ and  $\mathbb{P}(\alpha_{i}=-3)=\mathbb{P}(\alpha_{i}=3)=1/2$. For the generation of covariate matrices, we consider the following cases:

\begin{itemize}
	\item \textbf {Collinearity Case:} We draw $\bm{x}_{i}$ i.i.d. from $\mathcal{N}(\bm{0}_{p},\bm{\Lambda})$, where $\bm{\Lambda} = \bm{\Gamma}\bm{\Gamma}^{\top}+\bm{I}_{p}$. As for $\bm{\Gamma}$, we first generate a $p$-dimensional random square matrix $\bA$, where $A_{ij}$ are i.i.d. from $\mathcal{U}(0,1)$. Then, by QR decomposition, we have $\bA = \bQ\bR$, where $\bQ=(\bq_1,\cdots,\bq_p)$ is an orthogonal matrix. Finally, we set $\bm{\Gamma} = 5(\bq_1,\cdots,\bq_s)$. When $s=p$, $\bm{\Gamma}\bm{\Gamma}^{\top}=25\bI_p$. In the simulation, we vary $s$ as 3, 4, and 5, respectively.
	\item \textbf {Uncorrelated Case:}
	We draw $\bm{x}_{i}$ i.i.d. from $\mathcal{N}(\bm{0}_{p},\bm{I}_{p})$.
\end{itemize}

\begin{table}[h]
	\centering
	\caption{SILFS estimation compared with S-CAR under the collinearity and uncorrelated cases. The values in parentheses denote the standard deviation.}
	\label{tab:3}
	\scalebox{0.77}{
		\begin{tabular}{@{}lcccccccccc@{}}
			\toprule
			& \multicolumn{5}{c}{\textbf{SILFS}} & \multicolumn{5}{c}{\textbf{S-CAR}} \\
			\cmidrule(lr){2-6} \cmidrule(l){7-11}
			& $\text{RMSE}_{\bm{\alpha}}$ & $\text{RMSE}_{\bm{\beta}}$ & RI & Sensitivity & Specificity & $\text{RMSE}_{\bm{\alpha}}$ & $\text{RMSE}_{\bm{\beta}}$ & RI & Sensitivity & Specificity\\
			\midrule
			\multicolumn{11}{c}{\textbf{Collinearity case with \( s = 3 \)}} \\
			\( p = 50 \) & 0.432 & 0.146 & 0.990(0.030) & 0.994(0.045) & 0.999(0.004) & 0.723 &  0.074 & 0.970(0.062)& 1.000(0.000) & 0.949(0.102) \\
			\( p = 100 \) & 0.530 & 0.104 & 0.985(0.033) & 0.996(0.028) & 0.998(0.007) & 0.605 & 0.055 & 0.979(0.055) & 0.998(0.020) & 0.984(0.042)\\
			\( p = 150 \) & 0.726 & 0.092 & 0.972(0.056) & 0.984(0.061) & 0.999(0.008) & 0.667 &0.053 & 0.975(0.060) & 1.000(0.000) & 0.987(0.029)\\
			\midrule
			\multicolumn{11}{c}{\textbf{Collinearity case with \( s = 4 \)}} \\
			\( p = 50 \) & 0.396 & 0.145 & 0.992(0.024) & 0.998(0.020) & 0.999(0.007) & 0.770 & 0.144 & 0.972(0.054) & 0.994(0.034) & 0.963(0.031)\\
			\( p = 100 \) & 0.675 & 0.114 & 0.976(0.050) & 0.990(0.052) & 0.997(0.011) & 0.849 & 0.110 & 0.965(0.055) & 0.994(0.045) & 0.987(0.015)\\
			\( p = 150 \) & 0.796 & 0.097 &  0.965(0.055) & 0.990(0.044) & 0.996(0.009) & 0.997 & 0.103 & 0.951(0.069) & 0.982(0.064) &  0.994(0.012)\\
			\midrule
			\multicolumn{11}{c}{\textbf{Collinearity case with \( s = 5 \)}} \\
			\( p = 50 \) & 0.531 & 0.162 & 0.985(0.032) & 0.996(0.028) & 0.998(0.008) & 0.888 & 0.161 & 0.964(0.060) & 0.992(0.039) & 0.960(0.034)\\
			\( p = 100 \) & 0.707 & 0.115 & 0.973(0.045) & 0.994(0.034) & 0.996(0.010) & 1.011 & 0.120 & 0.950(0.079) & 0.984(0.061) & 0.985(0.017)\\
			\( p = 150 \) & 0.909 & 0.101 & 0.956(0.069) & 0.978(0.069) & 0.995(0.011) & 1.005 & 0.103 &  0.952(0.080) & 0.984(0.068) & 0.991(0.013)\\
			\midrule
			\multicolumn{11}{c}{\textbf{Uncorrelated case}} \\
			\( p = 50 \) & 0.220 & 0.117 & 0.998(0.008) & 1.000(0.000) & 1.000(0.002) & 0.226 & 0.115 & 0.998(0.007) & 1.000(0.000) & 1.000(0.000)\\
			\( p = 100 \) & 0.551 & 0.097 & 0.984(0.044) & 0.992(0.049) & 0.999(0.006) & 0.471 & 0.093 & 0.988(0.033) & 0.998(0.020) & 0.998(0.008) \\
			\( p = 150 \) & 0.478 & 0.077 & 0.988(0.033) & 0.998(0.020) & 0.999(0.005) & 0.426 & 0.074 & 0.991(0.027) & 1.000(0.000)  & 0.999(0.002)\\
			\bottomrule
		\end{tabular}
	}
\end{table}

We set the sample size as $100$, while the dimensionality varies across 50, 100, and 150. All simulation results are based on 100 replications, and the evaluation metrics are consistent with those used in the last section. We present the simulation results in Table \ref{tab:3}. In the collinearity case, the covariance matrix exhibits a spike structure. Therefore, although there is no factor structure, the proposed SILFS can still achieve better performance in terms of subgroup identification and variable selection simultaneously. Additionally, as Table \ref{tab:3} illustrates,  for the cases where the covariates are uncorrelated, we find that the estimated $\hat{r}$ is quite large such that we believe there is no factor structures and directly set $\hat{r}=0$. Therefore, the performance of SILFS is comparable with that of SCAR. This indicates that  SILFS can be used as a safe replacement of the existing clustering methods, regardless of the level of collinearity among covariates.

\section{A Real Data Example}\label{sec:5}

In this section, we employ the proposed SILFS method  to explore the relationship between China's export value of commodities and exchange rates. The explanatory variables are sourced from the General Administration of Customs People's Republic of China and are available for download from \url{http://stats.customs.gov.cn/}. The raw datasets comprises panel data involving 219 trading partner countries (regions) of China and 61 non-industrial commodities. We focus on the top 50 countries (regions) with the highest total trade volume in 2019, resulting in a covariate dimension of $50\times 61$. The response variable is the Chinese exchange rates corresponding to these 50 countries (regions) in 2019. Our objective is to conduct subgroup analysis for this response variable while accounting for the effects of the covariates.

\begin{figure}[ht]
	\centering
	\begin{minipage}[b]{0.45\linewidth}
		\centering
		\includegraphics[height=7cm]{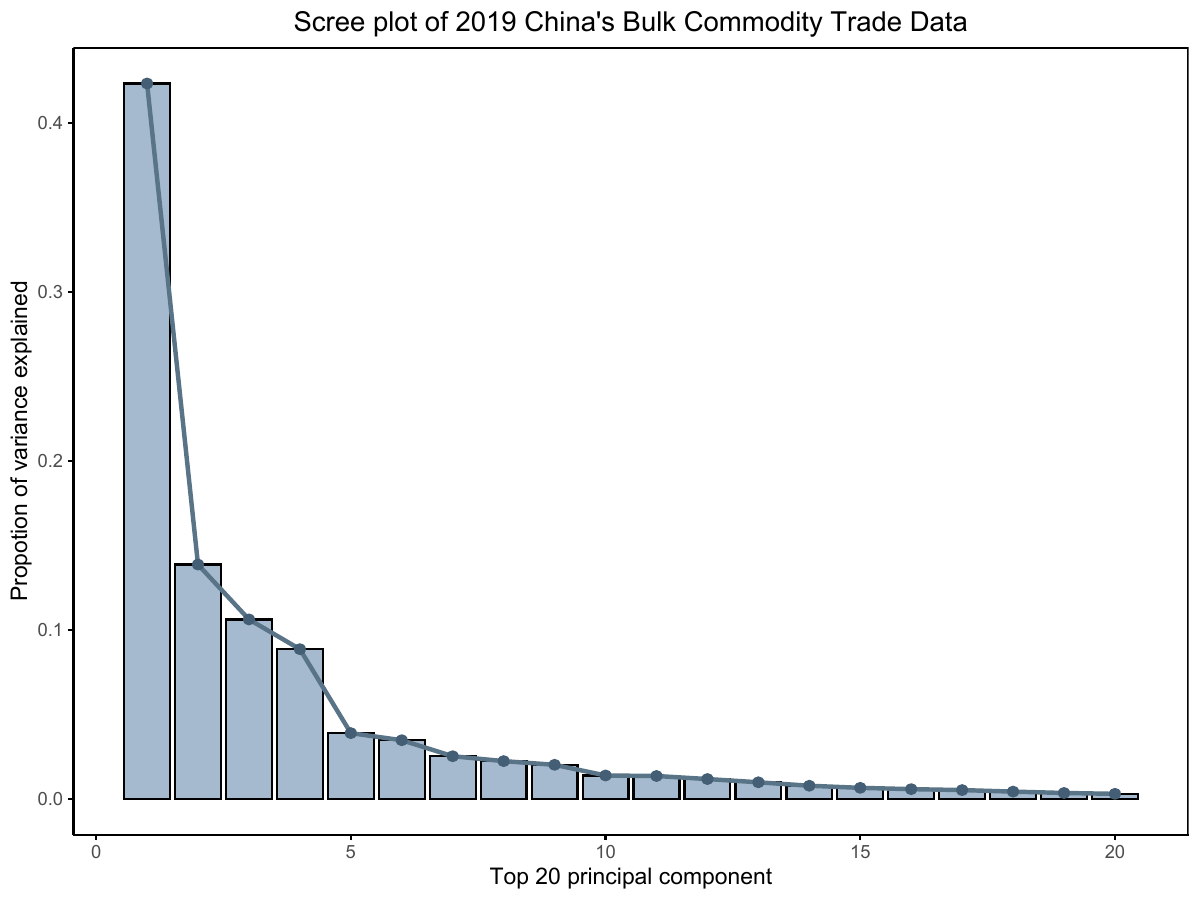}
	\end{minipage}
	\hfill
	\begin{minipage}[b]{0.45\linewidth}
		\centering
		\includegraphics[height=7cm]{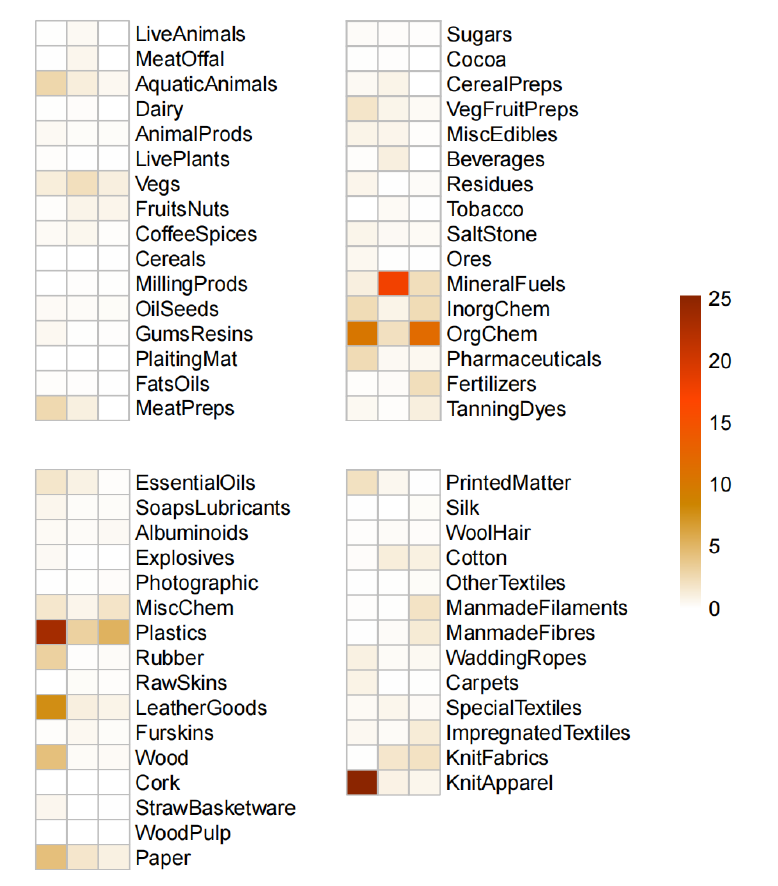}
		
	\end{minipage}
	\caption{Scree plot (left panel) of eigenvalues (dotted line) and proportions of variance explained (bar) by the top 20 principal components. Heatmap (right panel) of the absolute values of the factor loading matrix after orthogonal rotation.}
	\label{fig:3}
\end{figure}

Given the possibility for high collinearity among covariates due to commodity substitutability, we first conduct PCA on the covariates. The scree plot of the top 20 principal components, depicted in the left panel of Figure \ref{fig:3}, reveals that the first principal component explains 45\% of the total variance, while the first five components collectively account for 80\% of the variance. This underscores significant cross-sectional dependence among covariates, emphasizing the necessity of applying a factor model to mitigate the impact of collinearity.

To begin with, we determine the number of factors as $\widehat{r}=3$ using the eigenvalue-ratio method. To aid in interpreting these factors, we employ varimax orthogonal rotation through the \texttt{varimax} function in \texttt{R}. The heatmap in the right panel of Figure \ref{fig:3} displays the absolute values of the factor loading matrix after orthogonal rotation, where darker colors indicate larger values.
From Figure \ref{fig:3}, we conclude that the first factor is associated with variables such as ``Plastics," ``KnitApparel" (knitted or crocheted garments and clothing accessories), ``OrgChem" (Organic Chemicals), and ``Leather Goods", representing categories related to textiles, apparel, and accessories. This aligns with China's prominent role as an exporter of garments and textile products, as highlighted in studies such as \cite{altenburg2020exporting} and \cite{hussain2020non}. The second factor primarily relates to ``MineralFuels" (mineral fuels, mineral oils, products of their distillation; bituminous substances; mineral waxes), indicating its association with fuel and crude oil exports. In contrast, the third factor is influenced by variables such as ``OrgChem," ``Plastics," ``ManmadeFilaments" (synthetic filament; flat strips and similar forms of synthetic textile materials), and ``ManmadeFibres" (chemical fiber staple), suggesting a connection to light industry and manufacturing.
Therefore, we deduce that China's export structure is primarily driven by three latent factors, which we summarize as the ``Textiles and Apparel Category", ``Oil and Fuel Category", and ``Light Industry Category".

\begin{figure}[ht]
	\hspace{2.5cm}
	\includegraphics[width=11cm]{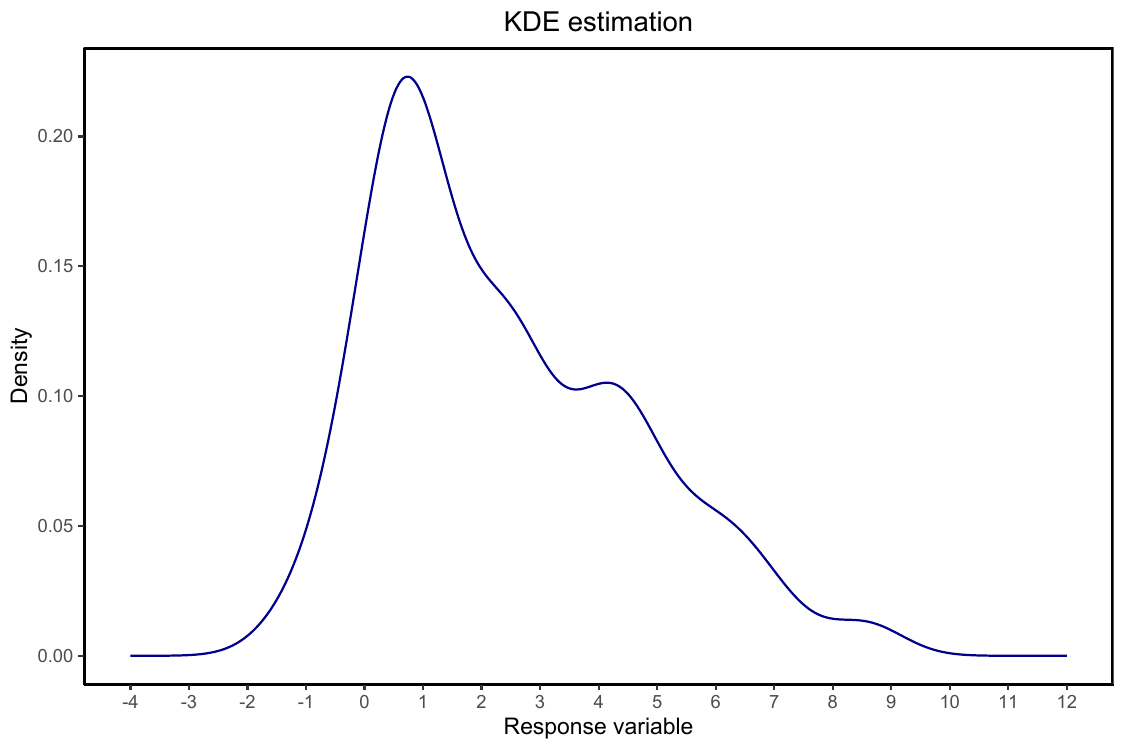}
	\caption{Density plot of the response variable after adjusting for the effects of  covariates in the empirical example.}
	\label{fig:4}
\end{figure}

Next, we apply the FARM model proposed by \cite{fan2020factor, fan2023latent} to estimate $\widehat{\bbeta}$. We further plot the kernel density estimate (KDE) of $Y_i - \bm{X}_{i}^{\top}\hat{\bm{\beta}}$ to assess its performance. According to Figure \ref{fig:4}, it is evident that even after adjusting for the influence of covariates, the distribution still exhibits multiple modes. This heterogeneity may stem from unobserved latent factors, such as subgroups. Therefore, employing SILFS for subgroup analysis appears to be a more appropriate approach.

We present the results of subgroup identification in Table \ref{tab:4}. Based on these findings, we cluster fifty countries (regions) into four groups. From the perspective of developmental status, the first two groups primarily comprise developed countries, including the United States, European countries, Australia and Singapore. Specifically, the first group exhibits greater concentration, while the second group displays variations in geographical location and industrial structure. The third group primarily consists of developing countries with significant growth potential. The last group mainly comprises countries with lower economic levels. Additionally, Hong Kong, China, and Taiwan, China, representing regions with smaller scales, are included in this group. From a geopolitical standpoint, the first group mainly comprises early-developed capitalist countries in Europe and America. The second group consists of countries from the Commonwealth. The third and fourth groups, which are less developed, are primarily situated in Asia and the Middle East.

\renewcommand{\arraystretch}{1.3}
\begin{table}[ht]
	\centering
	\caption{Subgroup identification of the top 50 countries (regions) from SILFS
.}
	\begin{tabular}{cc}
		\toprule
		Cluster & Countries (Regions)\\
		\midrule
 \multirow{2}{*}{C1}&United States; Germany; United Kingdom; Netherlands; Italy; Spain;\\
&France; Belgium; Poland; Panama; Greece.\\
\hline
\multirow{1}{*}{C2}&Australia; Canada; Singapore; New Zealand.\\
\hline
 \multirow{2}{*}{C3}& Japan; Malaysia; Brazil; United Arab Emirates; Pakistan; \\
 &Saudi Arabia; Israel; Peru.\\
 \hline
 \multirow{5}{*}{C4}&Hong Kong, China; Vietnam; South Korea; India; Indonesia; \\
 &Philippines; Thailand; Russia; Taiwan, China; Bangladesh; \\
 &Mexico; Nigeria; Turkey; Myanmar; South Africa; Cambodia; \\
 &Egypt; Chile; Iran; Kazakhstan; Colombia; Iraq; Kyrgyzstan;\\
 & Sri Lanka; Algeria;
 North Korea; Argentina.\\
		\bottomrule
	\end{tabular}
	
	\label{tab:4}
\end{table}

\section{Discussion}\label{sec:6}

Subgroup identification is crucial for characterizing heterogeneity in datasets from various fields such as biology, economics and finance. However, existing methods often rely on the computationally intensive pairwise fusion penalty (PFP) for group pursuit \citep{ma2017concave,zhang2019robust}. Moreover, high collinearity among covariates can lead to poor performance. To address these challenges, we propose SILFS to conduct variable selection and subgroup detection simultaneously. Essentially, we adopt a factor structure to represent collinearity and use the FARM framework \citep{fan2021augmented} in high dimensions for subgroup analysis. Furthermore, we introduce a novel Center-Augmented Regularization (CAR) method from \cite{he2022center} for clustering, significantly reducing computational complexity from $O(n^2)$ to $O(nK)$, where $n$ is the sample size and $K$ is the number of subgroups. We also investigate the corresponding algorithms and statistical properties of our proposed method. As a byproduct, an \texttt{R} package called \texttt{SILFS}, implementing the proposed method, is available on CRAN. Extensive simulations and real applications demonstrate the superiority of our proposed method.
 
Our study provides valuable insights into subgroup analysis in high-dimensional cases with collinearity among covariates. An important future research direction would be an in-depth exploration of the statistical properties of global minima and the theory of clustering consistency. Addressing these challenges will be essential for achieving a comprehensive understanding of the proposed approach. We leave this as future work.

\section*{Acknowledgements}
This work was supported in part by the National Natural Science Foundation of China (grant numbers: 12171282 and 12101412) and Qilu Young Scholar program of Shandong University.

\bibliography{ref}

\newpage
\begin{center}
APPENDIX
\end{center}
\begin{appendices}

\section{Technical Lemmas}
In this section, we introduce some useful technical lemmas. The asymptotic properties of factor model estimation
are frequently used in the following proof process. Hence, we summarize some useful results from \cite{fan2013large} in the following lemma. Besides, we also deliver some lemmas related to the subgroup identification.

\begin{lemma}$[$\cite{fan2013large} $]$ \label{lemma:prop1} Let $l = 3/l_1+3/(2l_2)+1/l_3+1$. Suppose $\log(p)=o(n^{l/6})$, $n=o(p^2)$
and Assumptions \ref{asmp:1}, \ref{asmp:2} and \ref{asmp:3} referred in the paper hold, we have:
\begin{enumerate}[(i)]
\item  $1/n\sum_{i=1}^{n}\|\widehat{\bm{f}}_{i}-\bm{H}\bm{f}_{i}\|_{2}^{2}=O_{\mathbb{P}}\left(1/n+1/p\right)$.
\item  $\bm{H}\bm{H}^{\top}=\bm{I}_{r}+O_{\mathbb{P}}\left(1/\sqrt{n}+1/\sqrt{p}\right)$.
\item For any subset of $\{1,2,\cdots,p\}$, denote as $\mathcal{D}$, we have $\mathop{\text{max}}_{j \in \mathcal{D}}\sum_{i=1}^{n}\rbr{\widehat{u}_{ij}-u_{ij}}^{2}=O_{\mathbb{P}}\left(\log{|\mathcal{D}|}+n/p\right)$.
\item $\mathop{\text{max}}_{i \in \{1,2,\cdots,p\}}\|\widehat{\bm{b}}_{i}-\bm{H}\bm{b}_{i}\|_{2}^{2}=O_{\mathbb{P}}\left(\log{p}/n+1/p\right)$.
\item $\mathop{\text{max}}_{i\in\{1,\cdots,r\};j\in\{1,\cdots,p\} }1/n|\sum_{t=1}^{n}f_{ti}u_{tj}|=O_{\mathbb{P}}(\sqrt{\log{p}/n})$.
\item
$\mathop{\text{max}}_{i\in\{1,\cdots,p\};j\in\{1,\cdots,p\} }1/n\left|\sum_{t=1}^{n}u_{ti}u_{tj}-\mathbb{E}(u_{ti}u_{tj})\right|=O_{\mathbb{P}}(\sqrt{\log{p}/n})$.

\end{enumerate}
\end{lemma}

The next lemma reveals the relationship of global optimum to the constrained optimum under some mild conditions.
\begin{lemma}\label{lemmaC1}$[$\cite{fan2020factor}$]$
Suppose $Z_{1}\left(\bm{\Theta}\right)$ is a convex, twice continuously differentiable function defined over $\mathbb{R}^{n}$. Let $g\left(\bm{\Theta}\right)\geq0$ be a convex and continuous penalty function satisfying the following  properties:
\begin{enumerate}[(i)]
	\item $g\left(\bm{\alpha}+\bm{\beta}\right)= g\left(\bm{\alpha}\right)+g\left(\bm{\beta}\right)$, for any $\bm{\alpha}\in\mathcal{U}$ and $\bm{\beta}\in\mathcal{U}^{\perp}$, where $\mathcal{U}$ is a subspace of $\mathbb{R}^{n}$ with orthogonal complement $\mathcal{U}^{\perp}$.
	\item There exists a continuous function $\widetilde{g}\left(\bm{\Theta}\right)$, such that $|\bm{\alpha}^{\top}\bm{\beta}|\leq g\left(\bm{\alpha}\right)\widetilde{g}\left(\bm{\beta}\right)$, for any $\bm{\alpha}\in\mathcal{U}^{\perp}$ and $\bm{\beta}\in\mathcal{R}^{n} $.								
\end{enumerate}	
Let $Z_{\lambda}\left(\bm{\Theta}\right)=Z_{1}\left(\bm{\Theta}\right)+\lambda g\left(\bm{\Theta}\right)$ where $\lambda \ge 0$ and denote the constrained minimization as
\begin{equation}\label{equ:resm}
\widehat{\bm{\Theta}}\in\mathop{\text{argmin}}_{\bm{\Theta}\in\mathcal{U}}Z_{\lambda}\left(\bm{\Theta}\right).
\end{equation}
If $\widetilde{g}(\nabla Z_{1}(\widehat{\bm{\Theta}}))<\lambda$ and $\bm{\Theta}^{\top}\nabla^{2}Z_{1}(\widehat{\bm{\Theta}})\bm{\Theta}>0$ for all $\bm{\Theta}\in\mathcal{U}$, then the constrained  minimization defined in \eqref{equ:resm} is the global minimizer of $Z_{\lambda}\left(\bm{\Theta}\right)$.
\end{lemma}

The following two lemmas provide the upper bound for invertible matrices.
\begin{lemma}\label{lemmaC2}
For two invertable $n\times n$ matrices $\bm{A}$ and $\bm{B}$, suppose $\|\bm{A}^{-1}\|\|\bm{A}-\bm{B}\|<1$ where $\|\cdot\|$ is an induced norm, we have
$$\|\bm{B}^{-1}\|\leq\frac{\|\bm{A}^{-1}\|}{1-\|\bm{A}^{-1}\|\|\bm{A}-\bm{B}\|}.$$
\end{lemma}

\begin{lemma}\label{lemmaC3}$[$\cite{fan2020factor}$]$
Let $\bm{C}$ be a $p\times q$ real matrix, and let $\bm{A}$ and $\bm{B}$ are two $q \times q$ symmetric matrices. Suppose $\bm{A}$ is nonsingular and $\|\bm{B}\bm{A}^{-1}\|<1$ where $\|\cdot\|$ is an induced norm, then
$$\Big\|\bm{C}\big[(\bm{A}+\bm{B})^{-1}-\bm{A}^{-1}\big]\Big\|\leq\frac{\left\|\bm{C}\bm{A}^{-1}\right\|\left\|\bm{B}\bm{A}^{-1}\right\|}{1-\left\|\bm{B}\bm{A}^{-1}\right\|}.$$
\end{lemma}

\begin{lemma}\label{lemmaC4}
Suppose Assumptions \ref{asmp:1}, \ref{asmp:2} and \ref{asmp:3} hold, for any vector $\bm{a}\in \mathbb{R}^r$ with $\|\bm{a}\|_{2}=1$, we have $$\left\|\widehat{\bm{U}}^{\top}\bm{F}\bm{a}\right\|_{\infty}=O_{\mathbb{P}}\left(\sqrt{n}+n/\sqrt{p}\right).$$
\end{lemma}

\begin{lemma}\label{lemmaC5}
Under the same assumptions as Lemma \ref{lemmaC4}, we have	
	$$\big\|\widehat{\bm{U}}^{\top}\widehat{\bm{U}}-\bm{U}^{\top}\bm{U}\big\|_{\text{max}}=O_{\mathbb{P}}\left(\log{p}+\sqrt{n}+n/\sqrt{p}\right).$$
\end{lemma}

The next lemma guarantees the separability of the oracle estimators.
\begin{lemma}\label{lemmaD1} $[$The separability of the oracle estimators.$]$
Under the same conditions required in Theorem \ref{thm:1},
for all $i$ and $k$, if the $i$-th sample is not from the $k$-th group, then
 $$\mathbb{P}\left(\left|\widehat{\alpha}_{i}^{or}-\widehat{\gamma}^{or}_{k}\right|\geq r_{n}\right)\rightarrow 1, $$
where $\widehat{\alpha}_{i}^{or}$ and $\widehat{\gamma}^{or}_{k}$ are the oracle estimators defined in (\ref{equ:oracle}).  $r_{n}$ is the minimum gap between different groups defined in Theorem \ref{thm:2}.
\end{lemma}

\section{Proof of Main Theorems}
\subsection{Proof of Theorem \ref{thm:A1}}
\begin{proof}
Without loss of generality, we consider the $m$-th iteration in the DC procedure. Recall that the optimization problem in the $m$-th iteration under $\ell_1$-type distance is:
\begin{equation}
\begin{aligned}\label{Aopt:Upper}
&(\widehat{\bm{\Theta}}^{(m)},\widehat{\bm{\delta}}^{(m)})=\argmin_{\bm{\Theta},\bm{\delta}}Z^{(m)}(\bm{\Theta},\bdelta)\\
\text{subject to}&\quad \delta_{ik}=\alpha_{i}-\gamma_{k}\quad i=1,\cdots,n, \   k=1,\cdots,K, \ \text{and}\ \gamma_{1}\leq\gamma_{2}\leq\cdots\leq\gamma_{K}.
	\end{aligned}
\end{equation}
where
\begin{align*}
Z^{(m)}(\bm{\Theta},\bdelta)=&\frac{1}{2n}\big\|\bm{Y}-\bm{\alpha}-\widehat{\bm{F}}\bm{\theta}-\widehat{\bm{U}}\bm{\beta}\big\|_2^{2} + \lambda_{1}\sum_{i=1}^{n}\sum_{k=1}^{K}|\delta_{ik}|-\lambda_1\sum_{i=1}^n\sum_{k=2}^{K}\max(|\widehat{\delta}_{i(k-1)}^{(m-1)}|,|\widehat{\delta}_{ik}^{(m-1)}|)\\
&-\lambda_1 g_{2}(\widehat{\bm{\delta}}^{(m-1)})-\lambda_1(\nabla g_{2}(\widehat{\bm{\delta}}^{(m-1)}))^{\top}(\bm{\delta}-\widehat{\bm{\delta}}^{(m-1)})+\lambda_{2}\|\bm{\beta}\|_{1}.
\end{align*}
Since $Z^{(m)}(\bm{\Theta},\bdelta)$ is a closed and proper convex function and the Lagragain function of (\ref{Aopt:Upper}) has saddle point by the saddle point theorem, then by the argument on convergence in the \cite{boyd2011distributed} and \cite{giesen2019combining}, the standard ADMM converges to a global minimizer. By some primary calculations, the following equalities hold:
$$Z\left(\widehat{\bm{\Theta}}^{(m-1)},\widehat{\bm{\delta}}^{(m-1)}\right)=Z^{\left(m\right)}\left(\widehat{\bm{\Theta}}^{(m-1)},\widehat{\bm{\delta}}^{(m-1)}\right),\quad Z\left(\widehat{\bm{\Theta}}^{(m)},\widehat{\bm{\delta}}^{(m)}\right)=Z^{\left(m+1\right)}\left(\widehat{\bm{\Theta}}^{(m)},\widehat{\bm{\delta}}^{(m)}\right).$$
By the construction of $Z^{(m)}(\bm{\Theta},\bm{\delta})$, for each $m\in N$,
\begin{eqnarray}
	Z^{\left(m\right)}\left(\widehat{\bm{\Theta}}^{(m-1)},\widehat{\bm{\delta}}^{(m-1)}\right) > Z^{\left(m\right)}\left(\widehat{\bm{\Theta}}^{(m)},\widehat{\bm{\delta}}^{(m)}\right) > Z^{\left(m+1\right)}\left(\widehat{\bm{\Theta}}^{(m)},\widehat{\bm{\delta}}^{(m)}\right),
\end{eqnarray}
implying that $Z\left(\widehat{\bm{\Theta}}^{(m-1)},\widehat{\bm{\delta}}^{(m-1)}\right)> Z\left(\widehat{\bm{\Theta}}^{(m)},\widehat{\bm{\delta}}^{(m)}\right)\geq0$.
Here the strict inequality always holds, otherwise the algorithm will be terminated when the equality is satisfied. By the monotone convergence theorem, we claim that $Z\left(\widehat{\bm{\Theta}}^{(m)},\widehat{\bm{\delta}}^{(m)}\right)$ converges as $m$ get larger.

Next, we illustrate the finite step convergence. Note that in each $Z^{(m)}(\bm{\Theta},\bdelta)$, $\bm{\delta}^{(m-1)}$ determines the upper bound function $Z^{(m)}(\bm{\Theta},\bm{\delta})$ only through sign function and indicator function. Thus $Z^{(m)}(\bm{\Theta},\bm{\delta})$ has only a finite set of possible options across all integer $m$. Therefore there exist a $m^{\star}$ such that for any $m\geq m^{\star}$, $\widehat{\bm{\Theta}}^{(m)}=\widehat{\bm{\Theta}}^{(m^{\star})}$.
Note that $(\widehat{\bm{\Theta}}^{(m^{\star})},\widehat{\bm{\delta}}^{(m^{\star})})$ is the global minimizer of $Z^{\left(m^{\star}\right)}\left(\bm{\Theta},\bm{\delta}\right)$ and $\nabla Z^{(m^{\star})}(\widehat{\bm{\Theta}}^{(m^{\star})},\widehat{\bm{\delta}}^{(m^{\star})})=\nabla Z(\widehat{\bm{\Theta}}^{(m^{\star})},\widehat{\bm{\delta}}^{(m^{\star})})$, thus we obtain that $\widehat{\bm{\Theta}}^{(m^{\star})}$ is a local minimizer of $Z(\widehat{\bm{\Theta}})$.
\end{proof}

\subsection{Proof of Theorem \ref{thm:1} and Proposition \ref{prop:1}}\label{subs:pTh2}
\begin{proof}
Recall that the oracle estimator is defined as:
\begin{equation*}
	(\hat{\bm{\gamma}}^{or},\hat{\bm{\theta}}^{or},\hat{\bm{\beta}}^{or})=\mathop{\text{argmin}}_{\bm{\Theta}}Z_{1}(\bm{\Theta})+\lambda_{2}\left\|\bm{\beta}\right\|_{1}.
\end{equation*}	
The proof proceeds in \textbf{three} steps.
The first step is to show that the constrained optimization estimator  converges to the oracle solution. In addition, the convergence rate is also provided.
In the second step, we prove that the constrained optimal is equivalent to the global optimal solution.
The third step proves the sign consistency of $\widehat{\bm \beta}$.

\textbf{Step 1.}
In this step,  we consider an optimization problem that limited to the subspace $\mathcal{U}:=\{\left(
\bm{\beta}^{\top},
\bm{\theta}^{\top},
\bm{\gamma}^{\top}
\right)^{\top}\in\mathbb{R}^{p+r+K}:\left(
\bm{\beta}^{\top}_{\mathcal{S}},
\bm{0}_{\mathcal{S}^{c}}^{\top},
\bm{\theta}^{\top},
\bm{\gamma}^{\top}
\right)^{\top}\}$. We define
\begin{equation*}
	\widehat{\bm{\Theta}}_{1}:=\left(
	(\widehat{\bm{\beta}}_{1})_{\mathcal{S}}^{\top},
	\bm{0}_{\mathcal{S}^{c}}^{\top},
	\widehat{\bm{\theta}}_{1}^{\top},
	\widehat{\bm{\gamma}}_{1}^{\top}
	\right)^{\top}=\mathop{\text{argmin}}_{\bm{\Theta}\in\mathcal{U}}\frac{1}{2n}\left\|\bm{Y}-\bm{\Omega}\bm{\gamma}-\widehat{\bm{F}}\bm{\theta}-\widehat{\bm{U}}\bm{\beta}\right\|_2^{2}+\lambda_{2}\|\bm{\beta}\|_{1}.
\end{equation*}
Note that this constrained optimization problem is equivalent to the following optimization problem,
\begin{equation}\label{equ:reso}
(\widehat{\bm{\Theta}}_{1})_{\mathcal{S}}:=\left(
(\widehat{\bm{\beta}}_{1})_{\mathcal{S}}^{\top},
\widehat{\bm{\theta}}_{1}^{\top},
\widehat{\bm{\gamma}}_{1}^{\top}
\right)^{\top}=\mathop{\text{argmin}}_{\bm{\beta}_{\mathcal{S}},\bm{\theta},\bm{\gamma}}\frac{1}{2n}\left\|\bm{Y}-\bm{\Omega}\bm{\gamma}-\widehat{\bm{F}}\bm{\theta}-\widehat{\bm{U}}_{\mathcal{S}}\bm{\beta}_{\mathcal{S}}\right\|_2^{2}+\lambda_{2}\|\bm{\beta}_{\mathcal{S}}\|_{1}.
\end{equation}
 The first order KKT conditions of problem \eqref{equ:reso} are
\begin{equation}
	\begin{aligned}
\partial Z_1[(\widehat{\bm{\Theta}}_{1})_{\mathcal{S}}]/\partial\bm{\beta}_{\mathcal{S}}&=-\frac{1}{n}\widehat{\bm{U}}_{\mathcal{S}}^{\top}\left(\bm{Y}-\bm{\Omega}\widehat{\bm{\gamma}}_{1}-\widehat{\bm{F}}\widehat{\bm{\theta}}_{1}-\widehat{\bm{U}}_{\mathcal{S}}(\widehat{\bm{\beta}}_{1})_{\mathcal{S}}\right)\in-\lambda_{2}\partial\big\|(\widehat{\bm{\beta}}_{1})_{\mathcal{S}}\big\|_{1},
\\
\partial Z_1[(\widehat{\bm{\Theta}}_{1})_{\mathcal{S}}]/\partial\bm{\theta}&=-\frac{1}{n}\widehat{\bm{F}}^{\top}\left(\bm{Y}-\bm{\Omega}\widehat{\bm{\gamma}}_{1}-\widehat{\bm{F}}\widehat{\bm{\theta}}_{1}-\widehat{\bm{U}}_{\mathcal{S}}(\widehat{\bm{\beta}}_{1})_{\mathcal{S}}\right)=\bm{0},
\\
\partial Z_1[(\widehat{\bm{\Theta}}_{1})_{\mathcal{S}}]/\partial\bm{\gamma}&=-\frac{1}{n}\bm{\Omega}^{\top}\left(\bm{Y}-\bm{\Omega}\widehat{\bm{\gamma}}_{1}-\widehat{\bm{F}}\widehat{\bm{\theta}}_{1}-\widehat{\bm{U}}_{\mathcal{S}}(\widehat{\bm{\beta}}_{1})_{\mathcal{S}}\right)=\bm{0},
	\end{aligned}
\end{equation}
where $\partial\| \bm{\Theta}\|_{1}$ is the sub-gradient of $\|\bm{\Theta}\|_{1}$.
The above KKT condition implies $\|\nabla_{\mathcal{S}}Z_{1}(\widehat{\bm{\Theta}}_{1})\|_{\infty}\leq\lambda_{2}$. We consider the difference between the first order differential of the $Z_1\left(\bm{\Theta}\right)$ at $\bm{\Theta}_{0}$ and $\widehat{\bm{\Theta}}_{1}$.
\begin{equation}\label{equ:mainres}
 \nabla_{\mathcal{S}}Z_1(\widehat{\bm{\Theta}}_{1})- \nabla_{\mathcal{S}}Z_1(\bm{\Theta}_{0})=\frac{1}{n}\left(\bm{\Omega},\widehat{\bm{F}},\widehat{\bm{U}}_{ \mathcal{S}}\right)^{\top}\left(\bm{\Omega},\widehat{\bm{F}},\widehat{\bm{U}}_ \mathcal{S}\right)\left(
\begin{array}{c}
	\widehat{\bm{\gamma}}_{1}-\bm{\gamma}_{0}\\
	\widehat{\bm{\theta}}_{1}-\bm{H}\bm{\theta}_{0}\\
	(\widehat{\bm{\beta}}_{1})_{\mathcal{S}}-(\bm{\beta}_{0})_{\mathcal{S}}
\end{array}
\right).
\end{equation}
Denote
\begin{equation}\label{equ:Adef}
\bm{A}:=\frac{1}{n}\left(\bm{\Omega},\widehat{\bm{F}},\widehat{\bm{U}}_{ \mathcal{S}}\right)^{\top}\left(\bm{\Omega},\widehat{\bm{F}},\widehat{\bm{U}}_ \mathcal{S}\right).
\end{equation}
If $\bm{A}$ is invertible, then left multiply both sides by matrix $\bA^{-1}$ and take the infinite norm on both sides, we have
\begin{equation}\label{equ:main1}
\left\|(\widehat{\bm{\Theta}}_{1})_{\mathcal{S}}-(\bm{\Theta}_{0})_{\mathcal{S}}\right\|_{\infty}\leq
\left\|\left[\frac{1}{n}\left(\bm{\Omega},\widehat{\bm{F}},\widehat{\bm{U}}_{ \mathcal{S}}\right)^{\top}\left(\bm{\Omega},\widehat{\bm{F}},\widehat{\bm{U}}_ \mathcal{S}\right)\right]^{-1}\right\|_{\infty}
{\left\|\nabla_{\mathcal{S}}Z_1(\widehat{\bm{\Theta}}_{1})- \nabla_{\mathcal{S}}Z_1(\bm{\Theta}_{0})\right\|_{\infty}}.
\end{equation}
In the following, we need to illustrate the $\bm{A}$ is invertible with probability tending to 1 and focus on the upper bounds for the two infinity norms on the right-hand side of the inequality \eqref{equ:main1}.

To show $\bm{A}$ is invertible, we set $\bm{D}=\text{diag}(\bm{\Omega}^{\top}\bm{\Omega}/n,\bm{I}_{r},\bm{\Sigma}_{\mathcal{SS}})$ and $\bm{\Delta}=\bm{A}-\bm{D}$. By Weyl theorem, we have $|\lambda_{\text{min}}(\bm{A})-\lambda_{\text{min}}(\bm{D})|\leq \|\bm{\Delta}\|_{2}\leq \left\|\bm{ \Delta}\right\|_{\infty}$. Thus, it is sufficient to show $\left\|\bm{ \Delta}\right\|_{\infty}$ is $o_{\mathbb{P}}(1)$. On the other hand, $\|\bm{A}^{-1}\|_{\infty}$ can be bounded by Lemma \ref{lemmaC2}
\begin{equation}\label{equ:Ainv}
	\left\|\bm{A}^{-1}\right\|_{\infty}\leq\frac{\left\|\bm{D}^{-1}\right\|_{\infty}}{1-\left\|\bm{D}^{-1}\right\|_{\infty}\left\|\bm{\Delta}\right\|_{\infty}}.
\end{equation}
It is not hard to verify that $\left\|\bm{D}^{-1}\right\|_{\infty}$ is bounded under Assumption \ref{asmp:5} and we assume it is smaller than $C$. In conclusion, we should find the convergence rate of  $\left\|\bm{ \Delta}\right\|_{\infty}$ to show $\bm{A}$ is invertible and then bound $\bm{A}^{-1}$ with infinite norm.

To analyze the gap between $\bm{A}$ and $\bm{D}$, an ``intermediary matrix" of them are introduced as:
$$
\bm{W}:=\frac{1}{n}\begin{pmatrix}
	\bm{\Omega}^{\top}\bm{\Omega}&\bm{\Omega}^{\top}\bm{F}\bm{H}^{\top}&\bm{\Omega}^{\top}\bm{U}_{\mathcal{S}}\\
	\bm{H}\bm{F}^{\top}\bm{\Omega}&n\bm{I}_{r}&\bm{0}\\
	\bm{U}_{\mathcal{S}}^{\top}\bm{\Omega}&\bm{0}&\bm{U}_{\mathcal{S}}^{\top}\bm{U}_{\mathcal{S}}\end{pmatrix}.$$
By triangle inequality, we have
$$\left\|\bm{\Delta}\right\|_{\infty} = \left\|\bm{D} - \bm{W} + \bm{W} - \bm{A}\right\|_{\infty}\leq\left\|\bm{A} - \bm{W}\right\|_{\infty}+\left\|\bm{D} - \bm{W}\right\|_{\infty} =:\left\|\bm{\Delta}_{1}\right\|_{\infty} + \left\|\bm{\Delta}_{2}\right\|_{\infty}.$$

By the property of infinite norm, we have
\begin{eqnarray}\label{equ:delta1}
	\begin{aligned}
		\left\|\bm{\Delta}_{1}\right\|_{\infty}\leq
		&\frac{1}{n}\text{max}\Big(\Big\|\left(\widehat{\bm{F}}-\bm{F}\bm{H}^{\top}\right)^{\top}\bm{\Omega}\Big\|_{\infty},\Big\|\left(\widehat{\bm{U}}_{\mathcal{S}}-\bm{U}_{\mathcal{S}}\right)^{\top}\bm{\Omega}\Big\|_{\infty}\Big)+\frac{1}{n}\Big\|\bm{\Omega}^{\top}\left(\widehat{\bm{F}}-\bm{F}\bm{H}^{\top}\right)\Big\|_{\infty}\\
	&	+\frac{1}{n}\text{max}\Big(\Big\|\bm{\Omega}^{\top}\left(\widehat{\bm{U}}_{\mathcal{S}}-\bm{U}_{\mathcal{S}}\right)\Big\|_{\infty},\Big\|\widehat{\bm{U}}_{\mathcal{S}}^{\top}\widehat{\bm{U}}_{\mathcal{S}}-\bm{U}_{\mathcal{S}}^{\top}\bm{U}_{\mathcal{S}}\Big\|_{\infty}\Big).
	\end{aligned}
\end{eqnarray}
Recalling the properties of the factor model estimators listed in Lemma \ref{lemma:prop1}, we obtain that
$$\Big\|\left(\widehat{\bm{F}}-\bm{F}\bm{H}^{\top}\right)^{\top}\bm{\Omega}\Big\|_{\infty}\leq\sqrt{K}\Big\|\bm{\Omega}\Big\|_{\mathbb{F}}\Big\|\widehat{\bm{F}}-\bm{F}\bm{H}^{\top}\Big\|_{\mathbb{F}}=O_{\mathbb{P}}\left(\sqrt{n}+n/\sqrt{p}\right).$$
$$\Big\|\bm{\Omega}^{\top}\left(\widehat{\bm{F}}-\bm{F}\bm{H}^{\top}\right)\Big\|_{\infty}\leq\sqrt{r}\Big\|\bm{\Omega}\Big\|_{\mathbb{F}}\Big\|\widehat{\bm{F}}-\bm{F}\bm{H}^{\top}\Big\|_{\mathbb{F}}=O_{\mathbb{P}}\left(\sqrt{n}+n/\sqrt{p}\right).$$
Similarly, we also have
$$\Big\|\left(\widehat{\bm{U}}_{\mathcal{S}}-\bm{U}_{\mathcal{S}}\right)^{\top}\bm{\Omega}\Big\|_{\infty}=O_{\mathbb{P}}\Big(\left(np_{\mathcal{S}}\log{p_{\mathcal{S}}}+p_{\mathcal{S}}n^{2}/p\right)^{1/2}\Big).$$ $$\Big\|\bm{\Omega}^{\top}\left(\widehat{\bm{U}}_{\mathcal{S}}-\bm{U}_{\mathcal{S}}\right)\Big\|_{\infty}=O_{\mathbb{P}}\Big(p_{\mathcal{S}}\left(n\log{p_{\mathcal{S}}}+n^{2}/p\right)^{1/2}\Big).$$
According to the result of Lemma \ref{lemmaC5}, it can be obtained that
$$\Big\|\widehat{\bm{U}}_{\mathcal{S}}^{\top}\widehat{\bm{U}}_{\mathcal{S}}-\bm{U}_{\mathcal{S}}^{\top}\bm{U}_{\mathcal{S}}\Big\|_{\infty}\leq p_{\mathcal{S}}\Big\|\widehat{\bm{U}}_{\mathcal{S}}^{\top}\widehat{\bm{U}}_{\mathcal{S}}-\bm{U}_{\mathcal{S}}^{\top}\bm{U}_{\mathcal{S}}\Big\|_{\text{max}}=O_{\mathbb{P}}\Big(p_{\mathcal{S}}\Big(\log{p_{\mathcal{S}}}+\sqrt{n}+n/\sqrt{p}\Big)\Big).$$
Combining all these results and the decomposition in \eqref{equ:delta1} , it's obvious that
\begin{equation}\label{equ:delta1r}
	\left\|\bm{\Delta}_{1}\right\|_{\infty}=O_{\mathbb{P}}(p_{\mathcal{S}}(\sqrt{\log{p_{\mathcal{S}}}/n}+1/\sqrt{p})).
\end{equation}

With similar arguments, we have
\begin{equation}\label{equ:delta2}
		\left\|\bm{\Delta}_{2}\right\|_{\infty}\leq \frac{1}{n}\text{max}\Big(\Big\|\bm{H}\bm{F}^{\top}\bm{\Omega}\Big\|_{\infty},\Big\|\bm{U}_{\mathcal{S}}^{\top}\bm{\Omega}\Big\|_{\infty}\Big)+\frac{1}{n}\Big\|\bm{\Omega}^{\top}\bm{F}\bm{H}^{\top}\Big\|_{\infty}
		+\frac{1}{n}\text{max}\Big(\Big\|\bm{\Omega}^{\top}\bm{U}_{\mathcal{S}}\Big\|_{\infty},\Big\|\bm{U}_{\mathcal{S}}^{\top}\bm{U}_{\mathcal{S}}-n\bm{\Sigma}_{\mathcal{S}}\Big\|_{\infty}\Big).
		\nonumber
\end{equation}
According to the concentration inequality in  \cite{merlevede2011bernstein}, we have
\begin{equation}
		\mathbb{P}\Big(\frac{1}{|\mathcal{G}_{k}|}\Big|\sum_{\mathcal{G}_{k}}f_{ij}\Big|\geq t\Big)
		\leq |\mathcal{G}_{k}|\exp{\Big(-\frac{\left(|\mathcal{G}_{k}|t\right)^{l}}{V_{1}}\Big)}+\exp{\Big(-\frac{\left(|\mathcal{G}_{k}|t\right)^{2}}{|\mathcal{G}_{k}|V_{2}}\Big)}
		+\exp{\Big(-\frac{\left(|\mathcal{G}_{k}|t\right)^{2}}{|\mathcal{G}_{k}|V_{3}}\exp{\Big(\frac{\left(|\mathcal{G}_{k}|t\right)^{l\left(1-l\right)}}{V_{4}\log^{l}{\left(nt\right)}}\Big)}\Big)},
		\nonumber
\end{equation}
which implies  $|\sum_{\mathcal{G}_{k}}f_{ij}|=O_{\mathbb{P}}(\sqrt{\left|\mathcal{G}_{k}\right|})$.
Hence, we have
\begin{equation*}
\left\|\bm{F}^{\top}\bm{\Omega}\right\|_{\infty}=O_{\mathbb{P}}(\sqrt{\left|\mathcal{G}_{\text{max}}\right|})  \quad \quad \text{and} \quad \quad \left\|\bm{\Omega}^{\top}\bm{F}\right\|_{\infty}=O_{\mathbb{P}}(\sqrt{\left|\mathcal{G}_{\text{max}}\right|}).
\end{equation*}
With similar arguments, we have
\begin{equation*}
   \left\|\bm{U}_{\mathcal{S}}^{\top}\bm{\Omega}\right\|_{\infty}=O_{\mathbb{P}}(\sqrt{\left|\mathcal{G}_{\text{max}}\right|\log{p_{\mathcal{S}}}})
   \quad \text{and} \quad  \left\|\bm{\Omega}^{\top}\bm{U}_{\mathcal{S}}\right\|_{\infty}=O_{\mathbb{P}}(p_{\mathcal{S}}\sqrt{\left|\mathcal{G}_{\text{max}}\right|\log{p_{\mathcal{S}}}}).
\end{equation*}
Further, by Lemma \ref{lemma:prop1},
$$\Big\|\bm{U}_{\mathcal{S}}^{\top}\bm{U}_{\mathcal{S}}-n\bm{\Sigma}_{\mathcal{S}}\Big\|_{\infty}\leq p_{\mathcal{S}}\Big\|\bm{U}_{\mathcal{S}}^{\top}\bm{U}_{\mathcal{S}}-n\bm{\Sigma}_{\mathcal{S}}\Big\|_{\text{max}}=O_{\mathbb{P}}\left(p_{\mathcal{S}}\sqrt{n\log{p_{\mathcal{S}}}}\right).$$
Thus, we obtain that $\left\|\bm{\Delta}_{2}\right\|_{\infty}=O_{\mathbb{P}}(p_{\mathcal{S}}\sqrt{\log{p_{\mathcal{S}}}/n})$. Combining the upper bound of $\left\|\bm{\Delta}_{1}\right\|_{\infty}$ in equation \eqref{equ:delta1r}, we obtain
 \begin{equation}\label{equ:deltar}
 \left\|\bm{\Delta}\right\|_{\infty}=O_{\mathbb{P}}(p_{\mathcal{S}}(\sqrt{\log{p_{\mathcal{S}}}/n}+1/\sqrt{p})).
 \end{equation}
It implies that $\left\|\bm{\Delta}\right\|_{\infty}\leq1/(2C)$ holds with probability tending to 1. Combined with the assumptions in Theorem \ref{thm:1}, we get $p_{\mathcal{S}}(\sqrt{\log{p_{\mathcal{S}}}/n}+1/\sqrt{p})\to 0$. Thus, we obtain that $\bA$ is invertible with probability tending to 1. Plugging this results in the inequality \eqref{equ:Ainv} we have
\begin{equation}\label{equ:main1r}
\left\|\Big[\frac{1}{n}\left(\bm{\Omega},\widehat{\bm{F}},\widehat{\bm{U}}_{ \mathcal{S}}\right)^{\top}\left(\bm{\Omega},\widehat{\bm{F}},\widehat{\bm{U}}_ \mathcal{S}\right)\Big]^{-1}\right\|_{\infty}\leq 2C
\end{equation}
holds with probability approaching to 1.

According to \eqref{equ:main1}, in order to bound $\left\|(\widehat{\bm{\Theta}}_{1})_{\mathcal{S}}-(\bm{\Theta}_{0})_{\mathcal{S}}\right\|_{\infty}$, we still need to find the upper bound of $\|\nabla_{\mathcal{S}}Z_1(\bm{\Theta}_{0})\|_{\infty}$. Recall that  $\bm{X}=\bm{F}\bm{B}^{\top}+\bm{U}=\widehat{\bm{F}}\widehat{\bm{B}}^{\top}+\widehat{\bm{U}}$, we have
$$\bm{Y}-\bm{\Omega}\bm{\gamma}_{0}-\widehat{\bm{F}}\bm{H}\bm{\theta}_{0}-\widehat{\bm{U}}_{\mathcal{S}}(\bm{\beta}_{0})_{\mathcal{S}}=\bepsilon+\widehat{\bm{F}}\left(\widehat{\bm{B}}^{\top}-\bm{H}\bm{B}^{\top}\right)\bm{\beta}_{0}.$$
Then, we can write $\nabla Z_1(\bm{\Theta}_{0})$ as
\begin{equation}\label{equ:main12}
\nabla Z_1(\bm{\Theta}_{0})=-\frac{1}{n}\left(\bm{\Omega},\widehat{\bm{F}},\widehat{\bm{U}}\right)^{\top}
\left(\bepsilon+\widehat{\bm{F}}\big(\widehat{\bm{B}}^{\top}-\bm{H}\bm{B}^{\top}\big)\bm{\beta}_{0}\right).
\end{equation}
We control the two parts of the right hand side of (\ref{equ:main12}) separately.
The first part is
\begin{eqnarray}\label{equ:main12f}
	\begin{aligned}
		\big\|\big(\bm{\Omega},\widehat{\bm{F}},\widehat{\bm{U}}\big)^{\top}\bm{\epsilon}\big\|_{\infty}\leq\text{max}(\left\|\bm{\Omega}^{\top}\bm{\epsilon}\right\|_{\infty},\big\|\widehat{\bm{F}}^{\top}\bm{\epsilon}\big\|_{\infty},\big\|\widehat{\bm{U}}^{\top}\bm{\epsilon}\big\|_{\infty}).
	\end{aligned}
\end{eqnarray}

Note that $\bm{\Omega}^{\top}\bm{\epsilon} = (\sum_{ \mathcal{G}_{1}}\epsilon_{i},\cdots,\sum_{\mathcal{G}_{K}}\epsilon_{i})$, the concentration inequality of sub-Gaussian random variable leads to $\|\bm{\Omega}^{\top}\bm{\epsilon}\|_{\infty} = O_{\mathbb{P}}\left(\sqrt{\left|\mathcal{G}_{\text{max}}\right|}\right)$. For $\|\widehat{\bm{F}}^{\top}\bm{\epsilon}\|_{\infty}$, we have
$$\big\|\widehat{\bm{F}}^{\top}\bm{\epsilon}\big\|_{\infty}\leq\big\|\bm{H}\bm{F}^{\top}\bm{\epsilon}\big\|_{\infty}+\big\|\big(\widehat{\bm{F}}^{\top}-\bm{H}\bm{F}^{\top}\big)\bm{\epsilon}\big\|_{2}\leq\big\|\bm{H}\big\|_{\infty}\big\|\bm{F}^{\top}\bm{\epsilon}\big\|_{\infty}+\big\|\big(\widehat{\bm{F}}^{\top}-\bm{H}\bm{F}^{\top}\big)\bm{\epsilon}\big\|_{2}.$$
Using the concentration inequality in \cite{merlevede2011bernstein} again, we have
\begin{eqnarray}
	\begin{aligned}
\mathbb{P}\Big(\frac{1}{n}\Big|\sum_{i=1}^{n}f_{ij}\epsilon_{i}\Big|\geq t\Big)\leq n\exp{\Big(-\frac{\left(nt\right)^{l}}{V_{1}}\Big)}+\exp{\Big(-\frac{\left(nt\right)^{2}}{nV_{2}}\Big)}+\exp{\Big(-\frac{\left(nt\right)^{2}}{nV_{3}}\exp{\Big(\frac{\left(nt\right)^{l\left(1-l\right)}}{V_{4}\log^{l}{\left(nt\right)}}\Big)}\Big)}.
		\nonumber
	\end{aligned}
\end{eqnarray}
Thus, we know $\|\bm{F}^{\top}\bm{\epsilon}\|_{\infty}\leq\|\bm{F}^{\top}\bm{\epsilon}\|_{2}=O_{\mathbb{P}}\left(\sqrt{n}\right)$.
By Cauchy-Schwarz inequality,
\begin{equation}
	\begin{aligned}
\|\widehat{\bm{U}}^{\top}\bm{\epsilon}\|_{\infty}
&\leq\|\bm{U}^{\top}\bm{\epsilon}\|_{\infty}+\mathop{\text{max}}_{j\in\left\{1\cdots p\right\}}\big|\sum_{i=1}^{n}(\widehat{u}_{ij}-u_{ij})\epsilon_{i}\big|
\\ &\leq
\|\bm{U}^{\top}\bm{\epsilon}\|_{\infty}+\mathop{\text{max}}_{j\in\left\{1\cdots p\right\}}\big(\sum_{i=1}^{n}(\widehat{u}_{ij}-u_{ij})^{2}\big)^{1/2}\big(\sum_{i=1}^{n}\epsilon_{i}^{2}\big)^{1/2}.
	\end{aligned}
\end{equation}
With similar arguments of $\|\bm{F}^{\top}\bm{\epsilon}\|_{\infty}$, one can obtain $\|\bm{U}^{\top}\bm{\epsilon}\|_{\infty}=O_{\mathbb{P}}(\sqrt{n\log{p}})$.
Further, combining the convergence rate of factor models in the Lemma \ref{lemma:prop1}, we obtain that
\begin{equation}
\big\|\widehat{\bm{F}}^{\top}\bm{\epsilon}\big\|_{\infty}=O_{\mathbb{P}}\left(\sqrt{n}+n/\sqrt{p}\right), \quad\text{and}\quad  \big\|\widehat{\bm{U}}^{\top}\bm{\epsilon}\big\|_{\infty}=O_{\mathbb{P}}\Big(\left(n\log{p}+n^{2}/p\right)^{1/2}\Big).
\end{equation}
Recall the decomposition in \eqref{equ:main12f}, we have
\begin{eqnarray}\label{equ:main12fr}
	\begin{aligned}
		\big\|\big(\bm{\Omega},\widehat{\bm{F}},\widehat{\bm{U}}\big)^{\top}\bm{\epsilon}\big\|_{\infty}=O_{\mathbb{P}}(\left(n\log{p}+n^{2}/p\right)^{1/2})
		.
	\end{aligned}
\end{eqnarray}

Next, we bound the second part. According to the maximum inequality we have
\begin{eqnarray}\label{equ:main12t}
	\begin{aligned}
		\big\|\big(\bm{\Omega},\widehat{\bm{F}},\widehat{\bm{U}}\big)^{\top}\widehat{\bm{F}}\big(\widehat{\bm{B}}^{\top}-\bm{H}\bm{B}^{\top}\big)\bm{\beta}_{0}\big\|_{\infty}
		\leq\text{max}\bigg(\left
		\|\bm{\Omega}^{\top}\widehat{\bm{F}}\big(\widehat{\bm{B}}^{\top}-\bm{H}\bm{B}^{\top}\big)\bm{\beta}_{0}\right\|_{\infty},
	\\
		\left\|\widehat{\bm{F}}^{\top}\widehat{\bm{F}}\big(\widehat{\bm{B}}^{\top}-\bm{H}\bm{B}^{\top}\big)\bm{\beta}_{0}\right\|_{\infty},
		\left\|\widehat{\bm{U}}^{\top}\widehat{\bm{F}}\big(\widehat{\bm{B}}^{\top}-\bm{H}\bm{B}^{\top}\big)\bm{\beta}_{0}\right\|_{\infty}
		\bigg).
	\end{aligned}
\end{eqnarray}
Note that  $\|\bm{\Omega}^{\top}\widehat{\bm{F}}\|_{\infty}\leq \|\bm{\Omega}^{\top}\widehat{\bm{F}}\|_{2}\leq\|\bm{\Omega}\|_{2}\|\widehat{\bm{F}}\|_{2}=O_{\mathbb{P}}(n)$.
Then we conclude
$$\big\|\bm{\Omega}^{\top}\widehat{\bm{F}}\big(\widehat{\bm{B}}^{\top}-\bm{H}\bm{B}^{\top}\big)\bm{\beta}_{0}\big\|_{\infty}\leq n\big\|\big(\widehat{\bm{B}}^{\top}-\bm{H}\bm{B}^{\top}\big)\bm{\beta}_{0}\big\|_{\infty}\leq n\big\|\big(\widehat{\bm{B}}^{\top}-\bm{H}\bm{B}^{\top}\big)\bm{\beta}_{0}\big\|_{2}.$$
According to the Lemma \ref{lemma:prop1}, the following inequality holds
$$\big\|\big(\widehat{\bm{B}}^{\top}-\bm{H}\bm{B}^{T}\big)\bm{\beta}_{0}\big\|_{2}\leq\mathop{\text{max}}_{j\in\mathcal{S}}\big\|\widehat{\bm{b}}_{j}-\bm{H}\bm{b}_{j}\big\|_{2}\left\|\bm{\beta}_{0}\right\|_{1}=O_{\mathbb{P}}\big(p_{\mathcal{S}}(\log{p_{\mathcal{S}}}/n+1/p)^{1/2}\big).$$
Therefore,
$$\big\|\bm{\Omega}^{\top}\widehat{\bm{F}}\big(\widehat{\bm{B}}^{\top}-\bm{H}\bm{B}^{\top}\big)\bm{\beta}_{0}\big\|_{\infty}=O_{\mathbb{P}}\big(p_{\mathcal{S}}(n\log{p_{\mathcal{S}}}+n^2/p)^{1/2}\big).$$
	Note that
 $\widehat{\bm{F}}^{\top}\widehat{\bm{F}}=n \bm{I}_{r}$, we have
\begin{equation*}
	\left\|\widehat{\bm{F}}^{\top}\widehat{\bm{F}}\big(\widehat{\bm{B}}^{\top}-\bm{H}\bm{B}^{\top}\big)\bm{\beta}_{0}\right\|_{\infty}
	\leq n\left\|\big(\widehat{\bm{B}}^{\top}-\bm{H}\bm{B}^{\top}\big)\bm{\beta}_{0}\right\|_{\infty}
	=O_{\mathbb{P}}\big(p_{\mathcal{S}}(n\log{p_{\mathcal{S}}}+n^2/p)^{1/2}\big).
\end{equation*}
Recall that estimation of factor model produce
$\widehat{\bm{U}}^{\top}\widehat{\bm{F}}=\bm{0}.$ We have
\begin{eqnarray}\label{equ:main12tr}
	\begin{aligned}
		\big\|\big(\bm{\Omega},\widehat{\bm{F}},\widehat{\bm{U}}\big)^{\top}\widehat{\bm{F}}\big(\widehat{\bm{B}}^{\top}-\bm{H}\bm{B}^{\top}\big)\bm{\beta}_{0}\big\|_{\infty}=O_{\mathbb{P}}\big(p_{\mathcal{S}}(n\log{p}+n^2/p)^{1/2}\big).
	\end{aligned}
\end{eqnarray}
Therefore, combine the results for \eqref{equ:main12},\eqref{equ:main12fr} and \eqref{equ:main12tr}, we get
\begin{eqnarray}\label{equ:main12tr1}
	\begin{aligned}
 \left\|\nabla Z_{1}(\bm{\Theta}_{0})\right\|_{\infty}=O_{\mathbb{P}}\big(p_{\mathcal{S}}(\log{p}/n+1/p)^{1/2}\big).
	\end{aligned}
\end{eqnarray}
WLOG, we let $c\lambda_{2}=\left\|\nabla Z_{1}(\bm{\Theta}_{0})\right\|_{\infty}$.
Combining the inequality in \eqref{equ:main1} and the convergence rate in \eqref{equ:main1r}, we obtain
\begin{eqnarray}\label{equ:main12tr2}
	\begin{aligned}
\left\|(\widehat{\bm{\Theta}}_{1})_{\mathcal{S}}-(\bm{\Theta}_{0})_{\mathcal{S}}\right\|_{\infty}
&\leq 2C\Big\|\nabla_{\mathcal{S}}Z_{1}(\widehat{\bm{\Theta}}_{1})- \nabla_{\mathcal{S}}Z_{1}(\bm{\Theta}_{0})\Big\|_{\infty}\\
&\leq
2C\Big(\Big\|\nabla_{\mathcal{S}}Z_{1}(\widehat{\bm{\Theta}}_{1})\Big\|_{\infty}+\Big\|\nabla_{\mathcal{S}}Z_{1}(\bm{\Theta}_{0})\Big\|_{\infty}\Big)\leq2C(1+c)\lambda_{2}.
\end{aligned}
\end{eqnarray}

\textbf{Step 2.} In this step, we show that the constrained minima $\widehat{\bm{\Theta}}_{1}$ is the global minimum  of (\ref{equ:oracle}).
According to Lemma \ref{lemmaC1}, we only need to show $\|\nabla_{\mathcal{S}^c}Z_{1}(\widehat{\bm{\Theta}}_{1})\|_{\infty}\leq\lambda_{2}$. By Lagrange mean value theorem, we have
$$\nabla_{\mathcal{S}^{c}}Z_{1}(\widehat{\bm{\Theta}}_{1})- \nabla_{\mathcal{S}^{c}}Z_{1}(\bm{\Theta}_{0})=\bm{L}\big((\widehat{\bm{\Theta}}_{1})_{\mathcal{S}}-(\bm{\Theta}_{0})_{\mathcal{S}}\big),$$
where $\bm{L}=1/n(\widehat{\bm{U}}_{\mathcal{S}^{c}}^{\top}\bm{\Omega},\bm{0},\widehat{\bm{U}}_{\mathcal{S}^{c}}^{\top}\widehat{\bm{U}}_{\mathcal{S}})$ is the sub-matrix of Hessian matrix of $Z_{1}(\bm{\Theta})$. By norm inequality we get
$$
\left\|\nabla_{\mathcal{S}^c}Z_{1}(\widehat{\bm{\Theta}}_{1})\right\|_{\infty}\leq\left\|\nabla_{\mathcal{S}^c}Z_{1}(\bm{\Theta}_{0})\right\|_{\infty}+\left\|\bm{L}\bm{A}^{-1}\right\|_{\infty}\left\|\bm{A}\big((\widehat{\bm{\Theta}}_{1})_{\mathcal{S}}-(\bm{\Theta}_{0})_{\mathcal{S}}\big)\right\|_{\infty},
$$
where $\bm{A}$ is defined in \eqref{equ:Adef} and $\bm{A}=\bm{\Delta}+\bm{D}$.
According to Lemma \ref{lemmaC3},
$$
\left\|\bm{L}\left[\bm{A}^{-1}-\bm{D}^{-1}\right]\right\|_{\infty}\leq\frac{\left\|\bm{L}\bm{D}^{-1}\right\|_{\infty}\left\|\bm{\Delta}\bm{D}^{-1}\right\|_{\infty}}{1-\left\|\bm{\Delta}\bm{D}^{-1}\right\|_{\infty}}.
$$
Combining the norm inequality and the irrepresentable condition, we have
$$
\Big\|\bm{L}\bm{A}^{-1}-\big(\bm{0}_{|\mathcal{S}^{c}|\times K},\bm{0}_{|\mathcal{S}^{c}|\times r},\bm{\Sigma}_{\mathcal{S}^{c}\mathcal{S}}\big)\bm{D}^{-1}\Big\|_{\infty}=O_{\mathbb{P}}\big(p_{\mathcal{S}}(\log{p}/n+1/p)^{1/2}\big).
$$
 Hence,  $\|\bm{L}\bm{A}^{-1}\|_{\infty}\leq 1-\rho$ holds with probability approaching to 1.
Combined with equation \eqref{equ:mainres} and the definition of $\bm{A}$, we have
$$
\big\|\bm{A}\big((\widehat{\bm{\Theta}}_{1})_{\mathcal{S}}-(\bm{\Theta}_{0})_{\mathcal{S}}\big)\big\|_{\infty}= \Big\|\nabla_{\mathcal{S}}Z_{1}(\widehat{\bm{\Theta}}_{1})- \nabla_{\mathcal{S}}Z_{1}(\bm{\Theta}_{0})\Big\|_{\infty}\leq\Big\|\nabla_{\mathcal{S}}Z_{1}(\widehat{\bm{\Theta}}_{1})\Big\|_{\infty}+\Big\|\nabla_{\mathcal{S}}Z_{1}(\bm{\Theta}_{0})\Big\|_{\infty}\leq\lambda_{2}+c\lambda_{2}.
$$
Thus, recall $c<\rho/(2-\rho)$, we obtain
$$\Big\|\nabla_{\mathcal{S}^c}Z_{1}(\widehat{\bm{\Theta}}_{1})\Big\|_{\infty}\leq c\lambda_{2}+(1-\rho)(\lambda_{2}+c\lambda_{2})<\lambda_{2}.$$
This implies $\widehat{\bm{\Theta}}_{1}$ is the unique global minima of the objective function, and
$$
\|(\widehat{\bm{\Theta}}_{1})_{\mathcal{S}}-(\bm{\Theta}_{0})_{\mathcal{S}}\|_{\infty}=O_{\mathbb{P}}(p_{\mathcal{S}}(\sqrt{\log{p}/n}+1/\sqrt{p}))=o_{\mathbb{P}}\left(1\right).
$$
Thus Theorem \ref{thm:1} is proved.

\textbf{Step 3.} In this step, we proof the proposition \ref{prop:1}. Note that
$$\mathbb{P}\big(\text{sign}(\widehat{\bm{\beta}}^{or})=\text{sign}(\bm{\beta}_{0})\big)\geq\mathbb{P}\big(\mathcal{A}_{n}\cap \mathcal{B}_{n}\big),$$
where
$$
\mathcal{A}_{n}=\left\{\left|\bm{A}^{-1}\frac{1}{\sqrt{n}}\left(\bm{\Omega},\widehat{\bm{F}},\widehat{\bm{U}}_{ \mathcal{S}}\right)^{\top}\bm{\epsilon}\right|\leq\sqrt{n}\left(\left|(\bm{\Theta}_{0})_{\mathcal{S}}\right|-\frac{\lambda_{2}}{2}\left|\bm{A}^{-1}\left(\bm{0}_{|\mathcal{S}^{c}|\times K}, \bm{0}_{|\mathcal{S}^{c}|\times r},\text{sign}[(\bm{\beta}_{0})_{\mathcal{S}}]\right)^{\top}\right|\right)\right\}
$$
and
$$
\mathcal{B}_{n}=\left\{\Big|(\widehat{\bm{U}}_{\mathcal{S}^{c}}^{\top}\bm{\Omega},\bm{0}_{|\mathcal{S}^{c}|\times r},\widehat{\bm{U}}_{\mathcal{S}^{c}}^{\top}\widehat{\bm{U}}_{\mathcal{S}})\bm{A}^{-1}\frac{1}{\sqrt{n}}(\bm{\Omega},\widehat{\bm{F}},\widehat{\bm{U}}_{ \mathcal{S}})^{\top}\bm{\epsilon}-\frac{1}{\sqrt{n}}\widehat{\bm{U}}_{ \mathcal{S}^{c}}^{\top}\bm{\epsilon}\Big|\leq\sqrt{n}\lambda_{2}\rho\bm{1}\right\},
$$
where the inequality holds element-wise. We consider the complement of $\mathcal{A}_{n}$ and $\mathcal{B}_{n}$ respectively.
$$
\mathbb{P}\left(\mathcal{A}_{n}^{c}\right)\leq\mathbb{P}\left( \bigcup_{i=1}^{p_{\mathcal{S}}+r+K}\left\{|\bm{s}_{i}^{\top}\bm{\varepsilon}|\geq\sqrt{n}\left(\min\left(|(\bm{\Theta}_{0})_{\mathcal{S}}|\right)-\frac{\lambda_{2}}{2}\left\|\bm{A}^{-1}\left(\bm{0}_{|\mathcal{S}^{c}|\times K}, \bm{0}_{|\mathcal{S}^{c}|\times r},\text{sign}[(\bm{\beta}_{0})_{\mathcal{S}}]\right)^{\top}\right\|_{\infty}\right)\right\}\right),
$$
where $\bm{s}_{j}$ is the $j$-th row of  $\bm{A}^{-1}(\bm{\Omega},\widehat{\bm{F}},\widehat{\bm{U}}_{ \mathcal{S}})^{\top}/\sqrt{n}$. Hence, it's easy to show that $\bm{s}_{i}^{\top}\bm{s}_{j}=\bm{A}^{-1}$. According to \eqref{equ:main1r},  $\|\bm{A}^{-1}\|_{2}\leq 2C$.
Thus, $\|(\bm{s}_{i}^{T}\bm{s}_{j})\|_{2}$ is bounded with high probability. There exists a constant $M$ such that
$$
\|\bm{s}_{i}\|_{2}\leq M \quad \text{for all}\quad  i=1,\cdots,k+r+p_{\mathcal{S}}.
$$
Similarly, we can prove $\|(\widehat{\bm{U}}_{ \mathcal{S}}^{\top}\widehat{\bm{U}}_{ \mathcal{S}}/n)^{-1}\|_{\infty}$ has finite upper bound $\kappa$.
\\
\\
By concentration inequality of sub-Gaussian random variable, we obtain

\begin{equation*}
	\begin{aligned}
\mathbb{P}\big(\mathcal{A}_{n}^{c}\big)\leq\mathbb{P}\Big(\bigcup_{i=1}^{p_{\mathcal{S}}+r+K}\big\{\frac{|\bm{s}_{i}^{\top}\bm{\epsilon}|}{\|\bm{s}_{i}\|}\geq\frac{1}{M}\sqrt{n}\big(\min(|(\bm{\Theta}_{0})_{\mathcal{S}}|)-\frac{\lambda_{2}}{2}\big\|\bm{A}^{-1}\big(\bm{0}_{|\mathcal{S}^{c}|\times K}, \bm{0}_{|\mathcal{S}^{c}|\times r},\text{sign}[(\bm{\beta}_{0})_{\mathcal{S}}]\big)^{\top}\big\|_{\infty})\big\}\Big)\\
\leq\sum_{i=1}^{p_{\mathcal{S}}+r+K}\mathbb{P}\Big(\frac{|\bm{s}_{i}^{\top}\bm{\epsilon}|}{\|\bm{s}_{i}\|}\geq\frac{1}{M}\sqrt{n}\frac{\lambda_{2}\kappa}{2}\Big)\leq 2p_{\mathcal{S}}\exp{\big(-B_{1}p_{\mathcal{S}}^{2}\log{p_{\mathcal{S}}}\big)}\leq \frac{\widetilde{B}_{1}}{p_{\mathcal{S}}^{p_{\mathcal{S}}^{2}-1}},
\end{aligned}
\end{equation*}
where $B_{1}$ and $\widetilde{B}_{1}$ are two positive values. Denote $\widetilde{\bm{s}}_{j}$ as the $j$-th row of
$$
\widetilde{\bm{S}}:=(\widehat{\bm{U}}_{\mathcal{S}^{c}}^{\top}\bm{\Omega},\bm{0},\widehat{\bm{U}}_{\mathcal{S}^{c}}^{\top}\widehat{\bm{U}}_{\mathcal{S}})\bm{A}^{-1}(\bm{\Omega},\widehat{\bm{F}},\widehat{\bm{U}}_{ \mathcal{S}})^{\top}/\sqrt{n}-\widehat{\bm{U}}_{\mathcal{S}^{c}}^{\top}/\sqrt{n}.
$$
Note that $\widetilde{\bm{S}}\widetilde{\bm{S}}^{\top}=\widehat{\bm{U}}_{ \mathcal{S}^{c}}^{\top}\bm{P}\widehat{\bm{U}}_{ \mathcal{S}^{c}}/n$, where $\bm{P}$ is a projection matrix. Hence, we obtain that $\|\widetilde{\bm{s}}_{j}\|_{2}\leq \widetilde{M}$ for all $j\in\mathcal{S}^{c}$. Further, we have
$$
\mathbb{P}\big(\mathcal{B}_{n}^{c}\big)\leq\sum_{i=1}^{p-p_{\mathcal{S}}}\mathbb{P}\Big(\frac{|\widetilde{\bm{s}}_{i}^{\top}\bm{\epsilon}|}{\|\widetilde{\bm{s}}_{i}\|}\geq\frac{1}{\widetilde{M}}\sqrt{n}\lambda_{2}\rho\Big)\leq 2(p-p_{\mathcal{S}})\exp{\big(-B_{2}p_{\mathcal{S}}^{2}\log{p_{\mathcal{S}}}\big)}\leq \frac{\widetilde{B}_{2}(p-p_{\mathcal{S}})}{p_{\mathcal{S}}^{p_{\mathcal{S}}^{2}}},
$$
where $B_{2}$ and $\widetilde{B}_{2}$ are two positive values. Therefore, the estimators has sign consistency, i.e.
$\mathbb{P}(\text{sign}(\widehat{\bm{\beta}})=\text{sign}(\bm{\beta}_{0}))\rightarrow1$.

\end{proof}

\subsection{Proof of Theorem \ref{thm:2}}\label{subs:pTh3.4}
\begin{proof}
Before starting the proof, we define a function to facilitate the subsequent proof process. For any vector $\bm{\alpha}\in\mathcal{V}_{\mathcal{G}}$, where $\mathcal{V}_{\mathcal{G}}=\{\bm{\alpha},\bm{\alpha}=\bm{\Omega}\bm{\gamma},\bm{\gamma}\in \mathbb{R}^{K}\}$, $T(\bm{\alpha})$ is a $K$-dimensional vector, i.e.,
$$
T:\mathcal{V}_{\mathcal{G}}\rightarrow \mathbb{R}^{K}.
$$
Specifically, the $k$-th coordinate of $T(\bm{\alpha})$ is the common value of $\alpha_{i}$ for $i\in \mathcal{G}_{k}$. According to this definition of $T$, $T^{-1}(\bm{\gamma})$ represents the $n$-dimensional vector obtained by restoring $\bm{\gamma}$ to the corresponding the true grouping structure.

With a slight abuse of notations, we define $\bTheta=\rbr{\balpha^{\top},\bgamma^{\top},\btheta^{\top},\bbeta^{\top}}^{\top}$, the corresponding oracle estimator can be defined as $\widehat{\bm{\Theta}}^{or}=((\widehat{\bm{\alpha}}^{or})^{\top},(\widehat{\bm{\gamma}}^{or})^{\top},(\widehat{\bm{\theta}}^{or})^{\top},(\widehat{\bm{\beta}}^{or})^{\top})^{\top}$
and
\begin{equation}\label{equ:Z0}
Z(\bm{\Theta}) = 1/2n\|\bm{Y}-\bm{\alpha}-\widehat{\bm{F}}\bm{\theta}-\widehat{\bm{U}}\bm{\beta}\|_{2}^{2} + \lambda_{1}\sum_{i=1}^{n}\text{min}\{d\left(\alpha_{i},\gamma_{1}\right),\cdots,d\left(\alpha_{i},\gamma_{K}\right)\}+\lambda_{2}\left\|\bm{\beta}\right\|_{1}.
\end{equation}
We choose an open set of $\bTheta$ related to $t_{n}$, denoted as
$$
\mathcal{N}_{n}=\left\{\bm{\Theta}:\big\|\bm{\Theta}-\widehat{\bm{\Theta}}^{or}\big\|_{\infty}< t_{n}\right\},
$$
where $t_{n}$ is a real value sequence satisfying $t_{n}=o(r_{n})$ and hence we have $t_{n}=o\big(p_{\mathcal{S}}(\sqrt{\log{p}/n}+1/\sqrt{p})\big)$.

In the neighborhood of $\widehat{\bm{\Theta}}^{or}$, we show that  $\widehat{\bm{\Theta}}^{or}$ is the local minima of $Z$ with high  probability in the following
\textbf{two steps}. The first step is to prove that for any parameter $\bm{\Theta}$ belonging to $\mathcal{N}_{n}$ satisfy $Z(
\widehat{\bm{\alpha}}^{or},
\widehat{\bm{\gamma}}^{or},
\widehat{\bm{\theta}}^{or},
\widehat{\bm{\beta}}^{or}
)< Z(T^{-1}(\bm{\gamma}),\bm{\gamma},\bm{\theta},\bm{\beta})$. In the second step, we show that $Z(T^{-1}(\bm{\gamma}),\bm{\gamma},\bm{\theta},\bm{\beta})\leq Z(\bm{\alpha},\bm{\gamma},\bm{\theta},\bm{\beta})$ holds on $\mathcal{N}_{n}$. Hence, $\widehat{\bm{\Theta}}^{or}$ is the local minima of the objective function.

\textbf{Step 1.} In this step,  wo focus on show that  $Z(
\widehat{\bm{\alpha}}^{or},
\widehat{\bm{\gamma}}^{or},
\widehat{\bm{\theta}}^{or},
\widehat{\bm{\beta}}^{or}
)< Z(T^{-1}(\bm{\gamma}),\bm{\gamma},\bm{\theta},\bm{\beta})$ holds for any parameter $\bm{\Theta}\in \mathcal{N}_{n}$.

As $\widehat{\bm{\Theta}}^{or}$ is the oracle estimator, taking it into \eqref{equ:Z0}, we have
$$
\sum_{i=1}^{n}\text{min}\{d\left(\widehat{\alpha}_{i}^{or},\widehat{\gamma}_{1}^{or}\right),\cdots,d\left(\widehat{\alpha}_{i}^{or},\widehat{\gamma}_{K}^{or}\right)\}=0.
$$
Hence, the oracle estimator of \eqref{equ:Z0} is,
$$
Z(\widehat{\bm{\alpha}}^{or},
\widehat{\bm{\gamma}}^{or},
\widehat{\bm{\theta}}^{or},
\widehat{\bm{\beta}}^{or})=1/2n\|\bm{Y}-\widehat{\bm{\alpha}}^{or}-\widehat{\bm{F}}\widehat{\bm{\theta}}^{or}-\widehat{\bm{U}}\widehat{\bm{\beta}}^{or}\|_{2}^{2} +\lambda_{2}\|\widehat{\bm{\beta}}^{or}\|_{1}.
$$
Similarly, by taking $(T^{-1}(\bm{\gamma}),\bm{\gamma},\bm{\theta},\bm{\beta})^{\top}$ into the objective function and using the property of $T^{-1}(\bm{\gamma}) $, we have
$$
Z(T^{-1}(\bm{\gamma}),\bm{\gamma},\bm{\theta},\bm{\beta})=1/2n\|\bm{Y}-T^{-1}(\bm{\gamma})-\widehat{\bm{F}}\bm{\theta}-\widehat{\bm{U}}\bm{\beta}\|_{2}^{2} +\lambda_{2}\|\bm{\beta}\|_{1}.
$$
As $\widehat{\bm{\alpha}}^{or}$ and $T^{-1}(\bm{\gamma})$ has the same grouping structure as the $\bm{\alpha}_{0}$, we denote them as $\bm{\Omega}\widehat{\bm{\gamma}}^{or}$ and $\bm{\Omega}\bm{\gamma}$. The objective function can be rewritten as
$$Z(\widehat{\bm{\alpha}}^{or},
\widehat{\bm{\gamma}}^{or},
\widehat{\bm{\theta}}^{or},
\widehat{\bm{\beta}}^{or})=\frac{1}{2n}\|\bm{Y}-\bm{\Omega}\widehat{\bm{\gamma}}^{or}-\widehat{\bm{F}}\widehat{\bm{\theta}}^{or}-\widehat{\bm{U}}\widehat{\bm{\beta}}^{or}\|_{2}^{2} +\lambda_{2}\|\widehat{\bm{\beta}}^{or}\|_{1}$$
and
$$Z(T^{-1}(\bm{\gamma}),\bm{\gamma},\bm{\theta},\bm{\beta})=\frac{1}{2n}\|\bm{Y}-\bm{\Omega}\bm{\gamma}-\widehat{\bm{F}}\bm{\theta}-\widehat{\bm{U}}\bm{\beta}\|_{2}^{2} +\lambda_{2}\|\bm{\beta}\|_{1}.$$
By the definition of the oracle estimator, we get
$$
Z(
\widehat{\bm{\alpha}}^{or},
\widehat{\bm{\gamma}}^{or},
\widehat{\bm{\theta}}^{or},
\widehat{\bm{\beta}}^{or}
)< Z(T^{-1}(\bm{\gamma}),\bm{\gamma},\bm{\theta},\bm{\beta}).
$$

\textbf{Step 2.} In this step, we prove that $Z(T^{-1}(\bm{\gamma}),\bm{\gamma},\bm{\theta},\bm{\beta})\leq Z(\bm{\alpha},\bm{\gamma},\bm{\theta},\bm{\beta})$ holds for any $(\bm{\alpha},\bm{\gamma},\bm{\theta},\bm{\beta})^{\top}\in\mathcal{N}_{n}$.

If sample $i$ belongs to the $k$-th subgroup, we have
$$
\left|\alpha_{i}-\gamma_{k}\right|\leq\left|\alpha_{i}-\widehat{\alpha}_{i}^{or}\right|+\left|\widehat{\gamma}_{k}^{or}-\gamma_{k}\right|\leq2t_{n}.
$$
If sample $i$ does not belong to the $k$-th subgroup, by the Lemma \ref{lemmaD1}, we have
$$
\left|\alpha_{i}-\gamma_{k}\right|=\left|\widehat{\alpha}_{i}^{or}-\widehat{\gamma}_{k}^{or}\right|-|(\widehat{\alpha}_{i}^{or}-\widehat{\gamma}_{k}^{or})-(\alpha_{i}-\gamma_{k})|\geq r_{n}/2-o(r_{n})\geq 2t_{n}
$$
holds with probability approaching to 1.
Therefore, for a given $\alpha_{i}$, we define $\gamma_{K(i)}$ as
$$
\min_{k=1,\cdots,K}|\alpha_{i}-\gamma_{k}|=|\alpha_{i}-\gamma_{K(i)}|,
$$
and we account the $i$-th sample belongs to $K(i)$-th group. Further, the objective function in the \eqref{equ:Z0} can be written as
$$
Z(\bm{\alpha},\bm{\gamma},\bm{\theta},\bm{\beta}) = \frac{1}{2n}\|\bm{Y}-\bm{\alpha}-\widehat{\bm{F}}\bm{\theta}-\widehat{\bm{U}}\bm{\beta}\|_{2}^{2} + \lambda_{1}\sum_{i=1}^{n}|\alpha_{i}-\gamma_{K(i)}|+\lambda_{2}\left\|\bm{\beta}\right\|_{1}.
$$
By Lagrange mean value theorem, we have:
\begin{equation}
\begin{aligned}
	&Z(\bm{\alpha},\bm{\gamma},\bm{\theta},\bm{\beta})-Z(T^{-1}(\bm{\gamma}),\bm{\gamma},\bm{\theta},\bm{\beta})\\
	=&\lambda_{1}\sum_{i=1}^{n}|\alpha_{i}-\gamma_{K(i)}|
	-\frac{1}{n}(\bm{Y}-\bm{\alpha}^{m}-\widehat{\bm{F}}\bm{\theta}-\widehat{\bm{U}}\bm{\beta})^{\top}(\bm{\alpha}-T^{-1}(\bm{\gamma}))
	\\
	=&\sum_{i=1}^{n}\left[\lambda_{1}|\alpha_{i}-\gamma_{K(i)}|-1/n(Y_{i}-\alpha^{m}_{i}-\widehat{\bm{f}}_{i}^{\top}\bm{\theta}-\widehat{\bm{u}}_{i}^{\top}\bm{\beta})(\alpha_{i}-\gamma_{K(i)})\right]
	.
\end{aligned}
\end{equation}
where $\bm{\alpha}^{m}=\varsigma\bm{\alpha}+(1-\varsigma)T^{-1}(\bm{\gamma}),\ \varsigma\in[-1,1]$. Thus, it is sufficient to show
$$
\sum_{i=1}^{n}\big(\lambda_{1}-\frac{1}{n}|Y_{i}-\alpha^{m}_{i}-\widehat{\bm{f}}_{i}^{\top}\bm{\theta}-\widehat{\bm{u}}_{i}^{\top}\bm{\beta}|\big)|\alpha_{i}-\gamma_{K(i)}|\geq 0.
$$
Next, we aim to show that each term of the summation is positive.
By triangular inequality and the property of  $\widehat{\bm{\beta}}^{or}$, we have:
\begin{eqnarray}
	\begin{aligned}
		\|\bm{Y}-\bm{\alpha}^{m}-\widehat{\bm{F}}\bm{\theta}- \widehat{\bm{U}}\bm{\beta}\|_{\infty}\leq&\left\|\bm{\epsilon}\right\|_{\infty}+\left\|\bm{\alpha}^{m}-\bm{\alpha}_{0}\right\|_{\infty}+\big\|\widehat{\bm{F}}\big\|_{\infty}\big\|\bm{\theta}-\widehat{\bm{B}}^{\top}\bm{\beta}_{0}\big\|_{\infty}+\big\|\widehat{\bm{U}}(\bm{\beta}-\bm{\beta}_{0})\big\|_{\infty}\\
		\leq& \left\|\bm{\epsilon}\right\|_{\infty}+\left\|\bm{\alpha}^{m}-\bm{\alpha}_{0}\right\|_{\infty}+\big\|\widehat{\bm{F}}\big\|_{\infty}\big\|\bm{\theta}-\widehat{\bm{B}}^{\top}\bm{\beta}_{0}\big\|_{\infty}\\
		&+\big\|\widehat{\bm{U}}\big\|_{\infty}\big\|(\bm{\beta}-\widehat{\bm{\beta}}^{or})\big\|_{\infty}+\big\|\widehat{\bm{U}}_{\mathcal{S}}\big\|_{\infty}\big\|(\widehat{\bm{\beta}}_{\mathcal{S}}^{or}-(\bm{\beta}_{0})_{\mathcal{S}})\big\|_{\infty}.
		\nonumber
	\end{aligned}
\end{eqnarray}
Note that there exist a sufficient large $C_{1}$ such that
$$
\mathbb{P}\big(\|\bm{\epsilon}\|_{\infty}\geq C_{1}\sqrt{\log{n}}\big)\leq2\sum_{i=1}^{n}\exp{\big(-C_{2}\log{n}\big)}\leq 2/n^{C_{2}-1},
$$
which implies that $\|\bm{\epsilon}\|_{\infty}=O_{\mathbb{P}}\big(\sqrt{\log{n}}\big)$.
By triangle inequality and Theorem \ref{thm:1}, we have
\begin{equation}
	\begin{aligned}
\|\bm{\alpha}^{m}-\bm{\alpha}_{0}\|_{\infty}&\leq\|\bm{\alpha}^{m}-\widehat{\bm{\alpha}}^{or}\|_{\infty}+\|\widehat{\bm{\alpha}}^{or}-\bm{\alpha}_{0}\|_{\infty}\\
&\leq t_{n}+O_{\mathbb{P}}\big(p_{\mathcal{S}}(\sqrt{\log{p}/n}+1/\sqrt{p})\big)=O_{\mathbb{P}}\big(p_{\mathcal{S}}(\sqrt{\log{p}/n}+1/\sqrt{p})\big).
\nonumber
\end{aligned}
\end{equation}
Similarly, we have
$$ \big\|\bm{\theta}-\widehat{\bm{B}}^{\top}\bm{\beta}_{0}\big\|_{\infty}\leq\big\|\bm{\theta}-\widehat{\bm{\theta}}^{or}\big\|_{\infty}+\big\|\widehat{\bm{\theta}}^{or}-\widehat{\bm{B}}^{\top}\bm{\beta}_{0}\big\|_{\infty}.
$$
Note that $$
\|\widehat{\bm{\theta}}_{1}^{or}-\widehat{\bm{B}}^{\top}\bm{\beta}_{0}\|_{\infty}\leq\|\widehat{\bm{\theta}}_{1}^{or}-\bm{H}\bm{\theta}_{0}\|_{\infty}+\|(\bm{H}\bm{B}^{\top}-\widehat{\bm{B}}^{\top})\bm{\beta}_{0}\|_{\infty}=O_{\mathbb{P}}\big(p_{\mathcal{S}}(\sqrt{\log{p}/n}+1/\sqrt{p})\big),
$$ thus $\big\|\bm{\theta}-\widehat{\bm{B}}^{\top}\bm{\beta}_{0}\big\|_{\infty}=O_{\mathbb{P}}\big(p_{\mathcal{S}}(\sqrt{\log{p}}/n+1/\sqrt{p})\big)$. By the concentration inequality and Lemma \ref{lemma:prop1}, we obtain $\|\widehat{\bm{F}}\|_{\infty}=O_{\mathbb{P}}\big(\sqrt{n}\big)$.
By norm inequality, we have,
\begin{eqnarray}
	\begin{aligned}
		\|\widehat{\bm{U}}\|_{\infty}&\leq\|\bm{U}\|_{\infty}+\|\widehat{\bm{U}}-\bm{U}\|_{\infty}\\
		&\leq\sqrt{p}\|\bm{U}\|_{2}+\sqrt{p}\|\widehat{\bm{U}}-\bm{U}\|_{2}\\
		&\leq p\big(\|\bm{U}^{\top}\bm{U}\|_{\text{max}}\big)^{1/2}+p\big(\|\big(\widehat{\bm{U}}-\bm{U}\big)^{\top}\big(\widehat{\bm{U}}-\bm{U}\big)\|_{\text{max}}\big)^{1/2}.
\end{aligned}
\end{eqnarray}
By the proposed assumption and Lemma \ref{lemmaC5}, we have
 $$
 \|\bm{U}^{\top}\bm{U}\|_{\text{max}}\leq n\|\bm{\Sigma}\|_{\text{max}}+\|\bm{U}^{\top}\bm{U}-n\bm{\Sigma}\|_{\text{max}}=O_{\mathbb{P}}\big(n\big)+O_{\mathbb{P}}\big(\sqrt{n\log{p}}\big)=O_{\mathbb{P}}\big(n\big).
 $$
	Thus, $\|\widehat{\bm{U}}\|_{\infty}=O_{\mathbb{P}}\big(p\sqrt{n}\big)$ and similarly $\|\widehat{\bm{U}}_{\mathcal{S}}\|_{\infty}=O_{\mathbb{P}}\big(p_{\mathcal{S}}\sqrt{n}\big)$. Recall that $t_n=o\big(p_{\mathcal{S}}(\sqrt{\log{p}/n}+1/\sqrt{p})\big)$, we obtain that
$$ \|\bm{Y}-\bm{\alpha}^{m}-\widehat{\bm{F}}\bm{\theta}- \widehat{\bm{U}}\bm{\beta}\|_{\infty}=O_{\mathbb{P}}\big(\sqrt{\log{n}}+p_{\mathcal{S}}^2(\sqrt{\log{p}}+\sqrt{n}/\sqrt{p})+p\sqrt{n}t_{n}\big).
$$
By setting $t_{n}$ to satisfy  $pt_{n}=o\big(p_{\mathcal{S}}^2(\sqrt{\log{p}/n}+1/\sqrt{p})\big)$, we have
$$
\|\bm{Y}-\bm{\alpha}^{m}-\widehat{\bm{F}}\bm{\theta}- \widehat{\bm{U}}\bm{\beta}\|_{\infty}=O_{\mathbb{P}}\big(\sqrt{\log{n}}+p_{\mathcal{S}}^2(\sqrt{\log{p}}+\sqrt{n}/\sqrt{p})\big).
$$
If $\lambda_{1}$ satisfy $\lambda_{1}\gg n^{-1}\text{max}(\sqrt{\log{n}},p_{\mathcal{S}}^2(\sqrt{\log{p}}+\sqrt{n}/\sqrt{p})),$ then  $$\sum_{i=1}^{n}\big(\lambda_{1}-\frac{1}{n}(Y_{i}-\alpha^{m}_{i}-\widehat{\bm{f}}_{i}^{\top}\bm{\theta}-\widehat{\bm{u}}_{i}^{\top}\bm{\beta})\big)|\alpha_{i}-\gamma_{K(i)}|\geq0,$$ with high probability, i.e., $Z(\bm{\alpha},\bm{\gamma},\bm{\theta},\bm{\beta})-Z(T^{-1}(\bm{\gamma}),\bm{\gamma},\bm{\theta},\bm{\beta})\geq0$ is proved.

Combining with the inequality from the first step, we get $$Z(\bm{\alpha},\bm{\gamma},\bm{\theta},\bm{\beta})-Z(\widehat{\bm{\alpha}}^{or},
\widehat{\bm{\gamma}}^{or},
\widehat{\bm{\theta}}^{or},
\widehat{\bm{\beta}}^{or})\geq0 \quad \text{for all }\bTheta\in\mathcal{N}_{n},$$
which implies the $\widehat{\bm{\Theta}}^{or}$ is a local minima of $Z(\bm{\Theta})$.

\end{proof}

\section{Proof of Technical Lemmas}
\subsection{Proof of Lemma \ref{lemmaC4}}
\begin{proof} From Lemma \ref{lemma:prop1}, we know $\bm{H}$ is invertible with the probability approaching to 1 and $\|\bm{H}^{-1}\|_{2}=O_{\mathbb{P}}(1)$, where the inverse matrix of $\bm{H}$ in sense of probability denoted as $\bm{H}^{-1}$.
Let $\bm{e}_{j}$ be the unit vector whose $j$-th element is 1. According to the estimation process, it is clear that $\widehat{\bm{F}}^{\top}\widehat{\bm{U}}=0$.
Combined with the triangle inequality, we have
\begin{equation}\label{equ:l51}
\begin{aligned}
\|\widehat{\bm{U}}^{\top}\bm{F}\bm{a}\|_{\infty}
&=\|\widehat{\bm{U}}^{\top}(\bm{F}-\widehat{\bm{F}}\bm{H}^{-1})\bm{a}\|_{\infty}\\
&\leq\big\|(\widehat{\bm{U}}^{\top}-\bm{U}^{\top})(\bm{F}-\widehat{\bm{F}}\bm{H}^{-1})\bm{a}\big\|_{\infty}+\big\|\bm{U}^{\top}(\bm{F}-\widehat{\bm{F}}\bm{H}^{-1})\bm{a}\big\|_{\infty}.
\end{aligned}
\end{equation}
By the Cauchy-Schwarz inequality and Lemma \ref{lemma:prop1}, we have
\begin{eqnarray}\label{equ:l511}
\begin{aligned}
	\big\|(\widehat{\bm{U}}^{\top}-\bm{U}^{\top})(\bm{F}-\widehat{\bm{F}}\bm{H}^{-1})\bm{a}\big\|_{\infty}
	&=\mathop{\text{max}}_{j\in\left\{1,\cdots,p\right\}}|\bm{e}_{j}^{\top}(\widehat{\bm{U}}^{\top}-\bm{U}^{\top})(\bm{F}-\widehat{\bm{F}}\bm{H}^{-1})\bm{a}|\\
	&\leq
	\Big(\mathop{\text{max}}_{j\in \left\{1,\cdots,p\right\}}\sum_{i=1}^{n}\left(\widehat{u}_{ij}-u_{ij}\right)^{2}\Big)^{1/2}\big\|\bm{F}\bm{H}^{\top}-\widehat{\bm{F}}\big\|_{\mathbb{F}}\big\|\bm{H}^{-1}\bm{a}\big\|_{2}\\
	&=O_{\mathbb{P}}\Big(\big[(\log{p}+\frac{n}{p})(1+\frac{n}{p})\big]^{1/2}\Big).
\end{aligned}
\end{eqnarray}
As $\bH$ is invertible with probability tending to 1,
\begin{equation}\label{equ:l512}
\|\bm{U}^{\top}(\bm{F}-\widehat{\bm{F}}\bm{H}^{-1})\bm{a}\|_{\infty}=\|\bm{U}^{\top}(\bm{F}\bm{H}^{\top}-\widehat{\bm{F}})\bm{H}^{-1}\bm{a}\|_{\infty}.
\end{equation}
Recall the defination of $\bm{H}$ in Theorem \ref{thm:1}, we have
$$\widehat{\bm{F}}-\bm{F}\bm{H}^{\top}=\frac{1}{n}\bm{U}\bm{U}^{\top}\widehat{\bm{F}}\bm{V}^{-1}+\frac{1}{n}\bm{U}\bm{B}\bm{F}^{\top}\widehat{\bm{F}}\bm{V}^{-1}+\frac{1}{n}\bm{F}\bm{B}^{\top}\bm{U}^{\top}\widehat{\bm{F}}\bm{V}^{-1}.$$
Plug it back into \eqref{equ:l512}, we have
\begin{eqnarray}\label{equ:l5123}
	\begin{aligned}
		\|\bm{U}^{\top}(\bm{F}-\widehat{\bm{F}}\bm{H}^{-1})\bm{a}\|_{\infty}\leq &
		\underbrace{\frac{1}{n}\big\|\bm{U}^{\top}\bm{U}\bm{U}^{\top}\widehat{\bm{F}}\bm{V}^{-1}\bm{H}^{-1}\bm{a}\big\|_{\infty}}_{S_1}
+\underbrace{\frac{1}{n}\big\|\bm{U}^{\top}\bm{U}\bm{B}\bm{F}^{\top}\widehat{\bm{F}}\bm{V}^{-1}\bm{H}^{-1}\bm{a}\big\|_{\infty}}_{S_2}\\
&+\underbrace{\frac{1}{n}\big\|\bm{U}^{\top}\bm{F}\bm{B}^{\top}\bm{U}^{\top}\widehat{\bm{F}}\bm{V}^{-1}\bm{H}^{-1}\bm{a}\big\|_{\infty}}_{S_3}.
	\end{aligned}
\end{eqnarray}

Next, we will bound each term separately.
For $S_1$, according to the triangle inequality, we have
\begin{eqnarray}\label{equ:S1}
		S_{1}\leq
		\underbrace{
			\frac{1}{n}\big\|\bm{U}^{\top}\bm{U}\bm{U}^{\top}\bm{F}\bm{H}^{\top}\bm{V}^{-1}\bm{H}^{-1}\bm{a}\big\|_{\infty}
		}_{S_{11}}+
		\underbrace{
		\frac{1}{n}\big\|\bm{U}^{\top}\bm{U}\bm{U}^{\top}(\widehat{\bm{F}}-\bm{F}\bm{H}^{\top})\bm{V}^{-1}\bm{H}^{-1}\bm{a}\big\|_{\infty}
		}_{S_{12}}.
\end{eqnarray}
For $S_{11}$, according to the definition of the infinity norm of the matrix, and using the Cauchy-Schwarz inequality, we have
\begin{eqnarray*}
		S_{11} \!=\! \frac{1}{n}\mathop{\text{max}}_{j\in \left\{1,\cdots,p\right\}}\left|\bm{e}_{j}^{\top}\bm{U}^{\top}\bm{U}\bm{U}^{\top}\bm{F}\bm{H}^{\top}\bm{V}^{\!-\!1}\bm{H}^{-1}\bm{a}\right|
		\leq \frac{1}{n}\mathop{\text{max}}_{j\in
			\left\{1,\cdots,p\right\}}\left\|\bm{e}_{j}^{\top}\bm{U}^{\top}\bm{U}\bm{U}^{\top}\bm{F}\right\|_{2}\big\|\bm{V}^{\!-\!1}\bm{H}^{-1}\big\|_{2}.
\end{eqnarray*}
Note that
\begin{eqnarray}
	\begin{aligned}
		\bm{e}_{j}^{\top}\bm{U}^{\top}\bm{U}\bm{U}^{\top}\bm{F}=\Big(\sum_{t=1}^{n}u_{tj}\displaystyle\sum_{s=1}^{n}\bm{u}_{t}^{\top}\bm{u}_{s}f_{s1},\cdots,\sum_{t=1}^{n}u_{tj}\displaystyle\sum_{s=1}^{n}\bm{u}_{t}^{\top}\bm{u}_{s}f_{sr}\Big)=\sum_{t=1}^{n}u_{tj}\displaystyle\sum_{s=1}^{n}\bm{u}_{t}^{\top}\bm{u}_{s}\bm{f}_{s}^{\top}.
		\nonumber
	\end{aligned}
\end{eqnarray}
After centralization, according to the triangle inequality and the Cauchy-Schwarz inequality, we have,
\begin{eqnarray}\label{equ:S11}
	\begin{aligned}
		S_{11} \leq \frac{1}{n}\mathop{\text{max}}_{j\in \left\{1,\cdots,p\right\}}\left(\sum_{t=1}^{n}u_{tj}^{2}\right)^{1/2}\Bigg[
		&\left(\sum_{t=1}^{n}\big\|\sum_{s=1}^{n}\left(\bm{u}_{t}^{\top}\bm{u}_{s}-\mathbb{E}\left(\bm{u}_{t}^{\top}\bm{u}_{s}\right)\right)\bm{f}_{s}\big\|_{2}^{2}\right)^{1/2}\\
		&+\left(\sum_{t=1}^{n}\big\|\sum_{s=1}^{n}\mathbb{E}\left(\bm{u}_{t}^{\top}\bm{u}_{s}\right)\bm{f}_{s}\big\|_{2}^{2}\right)^{1/2}\Bigg]\big\|\bm{V}^{-1}\|_2\|\bm{H}^{-1}\big\|_{2}.
	\end{aligned}
\end{eqnarray}
By the similar technique used in the proof of Theorem \ref{thm:1}, we have
\begin{eqnarray}
	\begin{aligned}
		\mathbb{P}\left(\frac{1}{n}\left|\sum_{i=1}^{n}u_{ij}^{2}-\sum_{i=1}^{n}\mathbb{E}u_{ij}^{2}\right|\geq t\right)\leq n\exp{\left(-\frac{\left(nt\right)^{l}}{V_{1}}\right)}+\exp{\left(-\frac{\left(nt\right)^{2}}{nV_{2}}\right)}+\exp{\left(-\frac{\left(nt\right)^{2}}{nV_{3}}\exp{\Big(-\frac{\left(nt\right)^{l\left(1-l\right)}}{V_{4}\left(\log{nt}\right)^{l}}\Big)}\right)},
		\nonumber
	\end{aligned}
\end{eqnarray}
which implies
$
	\mathop{\text{max}}_{j\in \left\{1,\cdots,p\right\}}(\sum_{t=1}^{n}u_{tj}^{2})=O_{\mathbb{P}}(n+\sqrt{n\log{p}})=O_{\mathbb{P}}(n).
$
Combined with similar arguments of Lemma 8 in \cite{fan2013large}, we have
\begin{equation*}
	\begin{aligned}
		\sum_{t=1}^{n}\left\|\sum_{s=1}^{n}\left(\bm{u}_{t}^{\top}\bm{u}_{s}-\mathbb{E}\left(\bm{u}_{t}^{\top}\bm{u}_{s}\right)\right)\bm{f}_{s}\right\|_{2}^{2}=O_{\mathbb{P}}\left(n^{3}p\right)\quad\sum_{t=1}^{n}\left\|\sum_{s=1}^{n}\mathbb{E}\left(\bm{u}_{t}^{\top}\bm{u}_{s}\right)\bm{f}_{s}\right\|_{2}^{2}=O_{\mathbb{P}}\left(n^{2}p^{2}\right).
		\nonumber
	\end{aligned}
\end{equation*}
According to the Lemma 5 in the \cite{fan2013large}, it's obvious that $\|\bm{V}^{-1}\|_{2}=O_{\mathbb{P}}(1/p)$.
According to the decomposition of $S_{11}$ in equation \eqref{equ:S11} and above results, we obtain
\begin{equation*}
S_{11}=O_{\mathbb{P}}\left(\sqrt{n}+\frac{n}{\sqrt{p}}\right).	
\end{equation*}
With similar arguments, we have $S_{12}=O_{\mathbb{P}}\left(n/p+\sqrt{n}/\sqrt{p}\right)$.
Therefore, reference to the decomposition of $S_1$ in \eqref{equ:S1}, we get
\begin{equation}
	S_{1}=O_{\mathbb{P}}\left(\sqrt{n}+\frac{n}{\sqrt{p}}\right).	
\end{equation}

We apply Cauchy-Schwarz inequality to bound $S_2$ and consider the defination of the infinite norm, the following inequality holds
\begin{eqnarray}
	\begin{aligned}
		S_{2}=&\frac{1}{n}\mathop{\text{max}}_{j\in
			\left\{1,\cdots,p\right\}}\left|\bm{e}_{j}^{\top}\bm{U}^{\top}\bm{U}\bm{B}\bm{F}^{\top}\widehat{\bm{F}}\bm{V}^{-1}\widetilde{\bm{H}}^{\top}\bm{a}\right|\\
		=&\frac{1}{n}\mathop{\text{max}}_{j\in
			\left\{1,\cdots,p\right\}}\left|\sum_{i=1}^{n}u_{ij}\bm{u}_{i}^{\top}\bm{B}\sum_{s=1}^{n}\bm{f}_{s}\widehat{\bm{f}}_{s}^{\top}\bm{V}^{-1}\widetilde{\bm{H}}^{\top}\bm{a} \right| \\
		\leq & \frac{1}{n}\mathop{\text{max}}_{j\in
			\left\{1,\cdots,p\right\}}\left\|\sum_{i=1}^{n}u_{ij}\bm{u}_{i}^{\top}\bm{B}\right\|_{2}\left\|\sum_{s=1}^{n}\bm{f}_{s}\widehat{\bm{f}}_{s}^{\top}\right\|_{2}\left\|\bm{V}^{-1}\widetilde{\bm{H}}^{\top}\right\|_{2}.
		\nonumber
	\end{aligned}
\end{eqnarray}
By simple triangle inequality, we have
\begin{eqnarray}
	\begin{aligned}
		\mathop{\text{max}}_{j\in
		\left\{1,\cdots,p\right\}}\Big\|\sum_{i=1}^{n}u_{ij}\bm{u}_{i}^{\top}\bm{B}\Big\|_{2}&\leq\mathop{\text{max}}_{j\in
		\left\{1,\cdots,p\right\}}
	\Bigg[
	\Big\|\sum_{i=1}^{n}\mathbb{E}\left(u_{ij}\bm{u}_{i}^{\top}\bm{B}\right)\Big\|_{2}+
	\Big\|\sum_{i=1}^{n}\left(u_{ij}\bm{u}_{i}^{\top}\bm{B}-\mathbb{E}\left(u_{ij}\bm{u}_{i}^{\top}\bm{B}\right)\right)\Big\|_{2}
	\Bigg].
		\nonumber
	\end{aligned}
\end{eqnarray}
By the distribution assumption of $\bm{u}_{i}$, we know the $\bm{b}_{k}^{\top}\bm{u}_{i}u_{ij}/\left\|\bm{b}_{k}\right\|_{2}$ also has the exponential-type tail,
where $\bm{b}_{s}$ is the $ s$-th column of the matrix $\bm{B}$.
Applying the concentration inequality in \cite{merlevede2011bernstein} again, we derive
$$\mathop{\text{max}}_{j\in
	\left\{1,\cdots,p\right\}}\left\|\sum_{i=1}^{n}\left(\bm{b}_{k}^{\top}\bm{u}_{i}u_{ij}-\mathbb{E}\bm{b}_{k}^{\top}\bm{u}_{i}u_{ij}\right)\right\|_{2}=O_{\mathbb{P}}\left(\left\|\bm{b}_{k}\right\|_{2}\sqrt{n\log{p}}\right)=O_{\mathbb{P}}\left(\sqrt{np\log{p}}\right).$$
As $\bm{B}\in \mathbb{R}^{p\times r}$ and $r$ is finite, thus
\begin{equation*}
	\mathop{\text{max}}_{j\in
		\left\{1,\cdots,p\right\}}\|\sum_{i=1}^{n}\left(u_{ij}\bm{u}_{i}^{\top}\bm{B}-\mathbb{E}\left(u_{ij}\bm{u}_{i}^{\top}\bm{B}\right)\right)\|_{2}=O_{\mathbb{P}}\left(\sqrt{np\log{p}}\right)
\end{equation*}
and
\begin{equation}
	\mathop{\text{max}}_{j\in
		\left\{1,\cdots,p\right\}}\Big\|\sum_{i=1}^{n}u_{ij}\bm{u}_{i}^{\top}\bm{B}\Big\|_{2}=O_{\mathbb{P}}\left(n+\sqrt{np\log{p}}\right).
\end{equation}
Combining these results, we finally obtain that
\begin{equation}\label{equ:S2}
	S_{2}=O_{\mathbb{P}}\left(1+\sqrt{p\log{p}/n}\right).
\end{equation}
Recall the definition of  $S_3$ in \eqref{equ:l5123}, an upper bound of it is given by
$$S_{3}\leq\mathop{\text{max}}_{j\in
	\left\{1,\cdots,p\right\}}1/n\left\|\bm{e}_{j}^{\top}\bm{U}^{\top}\bm{F}\right\|_{2}\left\|\bm{B}^{\top}\bm{U}^{\top}\right\|_{\mathbb{F}}\big\|\widehat{\bm{F}}\big\|_{2}\big\|\bm{V}^{-1}\bm{H}^{-1}\big\|_{2}.$$
According to the results in \cite{fan2013large}, we  have $\mathop{\text{max}}_{j\in
	\left\{1,\cdots,p\right\}}\left\|\bm{e}_{j}^{\top}\bm{U}^{\top}\bm{F}\right\|_{2}=O_{\mathbb{P}}\left(\sqrt{n\log{p}}\right)$
and
$
\|\bm{V}^{-1}\bm{H}^{-1}\|_{2}=O({1}/{\sqrt{p}}).
$
\\

Since $\mathbb{E}\|\bm{B}^{\top}\bm{U}^{\top}\|_{\mathbb{F}}^{2}=n\text{tr}(\bm{B}^{\top}\bm{\Sigma}\bm{B})=O_{\mathbb{P}}\left(np\right)$, we conclude that $S_{3}=O_{\mathbb{P}}(\sqrt{n\log{p}/p})$. Combining the equation \eqref{equ:l5123},\eqref{equ:S1} and \eqref{equ:S2}, we have $\|\bm{U}^{\top}(\bm{F}-\widehat{\bm{F}}\bm{H}^{-1})\bm{a}\|_{\infty}=O_{\mathbb{P}}\left(\sqrt{n}+n/\sqrt{p} \right)$.
Recall the decomposition in \eqref{equ:l51} and results in \eqref{equ:l511}, the final result is as follow
\begin{equation*}
   \|\widehat{\bm{U}}^{\top}\bm{F}\bm{a}\|_{\infty}= O_{\mathbb{P}}\left(\sqrt{n}+n/\sqrt{p}\right).
\end{equation*}

\end{proof}

\subsection{Proof of Lemma \ref{lemmaC5}}
\begin{proof}
By the triangle inequality, we have
	\begin{eqnarray}
	\begin{aligned}
	\left\|\widehat{\bm{U}}^{\top}\widehat{\bm{U}}-\bm{U}^{\top}\bm{U}\right\|_{\text{max}}=&\left\|\widehat{\bm{U}}^{\top}\left(\widehat{\bm{U}}-\bm{U}\right)+\left(\widehat{\bm{U}}-\bm{U}\right)^{\top}\widehat{\bm{U}}-\left(\widehat{\bm{U}}-\bm{U}\right)^{\top}\left(\widehat{\bm{U}}-\bm{U}\right)\right\|_{\text{max}}\\
	\leq & 2\left\|\widehat{\bm{U}}^{\top}\left(\widehat{\bm{U}}-\bm{U}\right)\right\|_{\text{max}}+\left\|\left(\widehat{\bm{U}}-\bm{U}\right)^{\top}\left(\widehat{\bm{U}}-\bm{U}\right)\right\|_{\text{max}}.
	\end{aligned}
	\nonumber
	\end{eqnarray}
According to the definition of factor models and the estimation process, it is clear that
\begin{equation*}
\bm{X}=\bm{F}\bm{B}^{\top}+\bm{U}=\widehat{\bm{F}}\widehat{\bm{B}}^{\top}+\widehat{\bm{U}} \quad \quad \text{and} \quad \quad \widehat{\bm{U}}^{\top}\widehat{\bm{F}}=\bm{0}.
\end{equation*}
Combining the result of Lemma \ref{lemmaC5}, it is not hard to verify that
\begin{equation}\label{equ:Uhat}
\begin{aligned}
\big\|\widehat{\bm{U}}^{\top}\big(\widehat{\bm{U}}-\bm{U}\big)\big\|_{\text{max}}&=\big\|\widehat{\bm{U}}^{\top}\left(\bm{F}\bm{B}^{\top}-\widehat{\bm{F}}\widehat{\bm{B}}^{\top}\right)
\big\|_{\text{max}}
=\big\|\widehat{\bm{U}}^{\top}\bm{F}\bm{B}^{\top}\big\|_{\text{max}}
\\
&=\mathop{\text{max}}_{j\in\left\{1,\cdots,p\right\}}\big\|\widehat{\bm{U}}^{\top}\bm{F}\widetilde{\bm{b}}_{j}\big\|_{\infty}=O_{\mathbb{P}}\left(\sqrt{n}+n/\sqrt{p}\right),
\end{aligned}
\end{equation}
where $\widetilde{\bm{b}}_{j}^{\top}$ is the $j$-th row of the matrix $\bm{B}$.
Applying the Lemma \ref{lemma:prop1}, we have
\begin{equation}\label{equ:Uhat2}
\big\|\big(\widehat{\bm{U}}-\bm{U}\big)^{\top}\big(\widehat{\bm{U}}-\bm{U}\big)\big\|_{\text{max}}\leq\mathop{\text{max}}_{j\in \{1,\cdots,p\}}\sum_{i=1}^{n}\left(\widehat{u}_{ij}-u_{ij}\right)^{2}=O_{\mathbb{P}}\big(\log{p}+n/p\big).
\end{equation}
Hence, combining the results of \eqref{equ:Uhat} and \eqref{equ:Uhat2},
\begin{equation*}
	\big\|\widehat{\bm{U}}^{\top}\widehat{\bm{U}}-\bm{U}^{\top}\bm{U}\big\|_{\text{max}}=O_{\mathbb{P}}\left(\log{p}+\sqrt{n}+\frac{n}{\sqrt{p}}\right).
\end{equation*}

\end{proof}

\subsection{Proof of Lemma \ref{lemmaD1}}
\begin{proof}
	The gap of the oracle estimators can be bounded from below by
	\begin{equation}\label{equ:orf}
	\mathop{\text{min}}_{i,k}\left|\widehat{\alpha}_{i}^{or}-\widehat{\gamma}_{k}^{or}\right|\geq\mathop{\text{min}}_{i,k}\left|\alpha_{0i}-\gamma_{0k}\right|-\mathop{\text{max}}_{i,k}\left|\left(\widehat{\alpha}_{i}^{or}-\widehat{\gamma}_{k}^{or}\right)-\left(\alpha_{0i}-\gamma_{0k}\right)\right|.
	\end{equation}
Recall that $r_{n}=\mathop{\text{min}}_{i,j\in1,\cdots,K}\left|\gamma_{0i}-\gamma_{0j}\right|$, we have
\begin{equation}\label{equ:min0}
\mathop{\text{min}}_{i,k}\left|\alpha_{0i}-\gamma_{0k}\right|\geq r_{n}.
\end{equation}
Combined with the Theorem \ref{thm:1} and the condition $r_{n}\gg p_{\mathcal{S}}(\sqrt{\log{p}/n}+1/\sqrt{p})$, we have
	\begin{equation*}
	\begin{aligned}
	\mathop{\text{max}}_{i,k}\left|\left(\widehat{\alpha}_{i}^{or}-\widehat{\gamma}_{k}^{or}\right)-\left(\alpha_{0i}-\gamma_{0k}\right)\right|\leq& \mathop{\text{max}}_{i}\left|\widehat{\alpha}_{i}^{or}-\alpha_{0i}\right|+\mathop{\text{max}}_{k}\left|\widehat{\gamma}_{k}^{or}-\gamma_{0k}\right|\\
	\leq &2\left\|\widehat{\bm{\gamma}}^{or}-\bm{\gamma}_{0}\right\|_{\infty}=o_{\mathbb{P}}\left(r_{n}\right).\\
	\end{aligned}
	\end{equation*}
Hence, we get
		\begin{equation*}
		\mathbb{P}\left(\left|\widehat{\alpha}_{i}^{or}-\widehat{\gamma}^{or}_{k}\right|\geq r_{n}\right)\rightarrow 1.
		\end{equation*}
		
	\end{proof}

\end{appendices}
\end{document}